\documentclass[aps,nofootinbib,floatfix]{revtex4}

\usepackage{epsfig}
\usepackage{latexsym}
\usepackage{graphicx}
\usepackage{amsmath}
\usepackage{amsfonts}   
\usepackage{amssymb}    

\newcommand{\newc}{\newcommand}

\newc{\hone}{h_1} \newc{\mhone}{m_{\hone}}
\newc{\htwo}{h_2} \newc{\mhtwo}{m_{\htwo}}
\newc{\hthree}{h_3} \newc{\mhthree}{m_{\hthree}}
\newc{\aone}{a_1} \newc{\maone}{m_{\aone}}
\newc{\atwo}{a_2} \newc{\matwo}{m_{\atwo}}

\newc{\msew}{m_{S}}
\newc{\mssq}{m_{S}^{2}}
\newc{\alambda}{A_{\lamda}} \newc{\aappa}{A_{\kappa}}

\def\eq#1{Eq.~(\ref{#1})}

\def\e3{$\epsilon_3$}

\def\ch2{$\chi^2$}

\def\co#1{{\ifmmode{\cal O}_{#1}\else${\cal O}_{#1}$\fi}}

\newdimen\unit
\def\point#1 #2 #3{\vbox to0pt{\kern-#2\unit
  \hbox{\kern#1\unit#3}\vss}
 \nointerlineskip}

\newcommand{\be}{\begin{equation}}
\newcommand{\ee}{\end{equation}}
\newcommand{\bea}{\begin{eqnarray}}
\newcommand{\eea}{\end{eqnarray}}

\newcommand{\gev}{\mbox{ GeV}}
\newcommand{\tev}{\mbox{ TeV}}

\newcommand{\cl}{\text{CL}}

\newcommand{\alphaem}{\alpha_{\text{em}}}

\newcommand{\alphas}{\alpha_s(M_Z)^{\overline{MS}}}

\newcount\hour
\newcount\minute
\newtoks\amorpm
\hour=\time\divide\hour by60 \minute=\time{\multiply\hour by60
\global\advance\minute by- \hour}
\edef\standardtime{{\ifnum\hour<12 \global\amorpm={am}%
    \else\global\amorpm={pm}\advance\hour by-12 \fi
    \ifnum\hour=0 \hour=12 \fi
    \number\hour:\ifnum\minute<100\fi\number\minute\the\amorpm}}
\edef\militarytime{\number\hour:\ifnum\minute<100\fi\number\minute}
\def\bold#1{\setbox0=\hbox{$#1$}%
     \kern-.025em\copy0\kern-\wd0
     \kern.05em\copy0\kern-\wd0
     \kern-.025em\raise.0433em\box0 }

\newc\eg{{\rm {e.g.}}}  \newc\etal{{\rm {et al.}}} \newc\ie{{\rm i.e.}}
\newc\etc{{\rm {etc}}}
\newcommand\lsim{\mathrel{\rlap{\lower4pt\hbox{\hskip1pt$\sim$}}
    \raise1pt\hbox{$<$}}}
\newcommand\gsim{\mathrel{\rlap{\lower4pt\hbox{\hskip1pt$\sim$}}
    \raise1pt\hbox{$>$}}}

\newc{\mhalf}{m_{1/2}}      \newc{\mzero}{m_0}
\newc{\tanb}{\tan\beta}
\newc{\azero}{A_0}
\newc{\at}{A_t} \newc{\ab}{A_b} \newc{\atau}{A_\tau}
\newc{\bmu}{B\mu}           \newc{\sgn}{{\rm sgn}}
\newc{\mone}{M_1}           \newc{\mtwo}{M_2}

 \newc{\hu}{H_u}       \newc{\hd}{H_d}
 \newc{\mhu}{m_{H_u}}       \newc{\mhd}{m_{H_d}}
 \newc{\mhuew}{m^{\ast}_{H_u}}       \newc{\mhdew}{m^{\ast}_{H_d}}
 \newc{\mhuewsq}{m^{\ast\, 2}_{H_u}}       \newc{\mhdewsq}{m^{\ast\, 2}_{H_d}}
 \newc{\mhuast}{m^{\ast}_{H_u}}       \newc{\mhdast}{m^{\ast}_{H_d}}

\newc{\charone}{\chi_1^\pm} \newc{\mcharone}{m_{\chi_1^\pm}}

\newc{\hl}{h}               \newc{\mhl}{m_{\hl}}   \newc{\gammahl}{\Gamma_{\hl}}
\newc{\hh}{H}               \newc{\mhh}{m_{\hh}}   \newc{\gammahh}{\Gamma_{\hh}}
\newc{\ha}{A}               \newc{\mha}{m_{\ha}}   \newc{\gammaha}{\Gamma_{\ha}}
\newc{\hpm}{H^{\pm}}        \newc{\mhpm}{m_{\hpm}} \newc{\gammahpm}{\Gamma_{\hpm}}
\newc{\hp}{H^{+}} \newc{\mhp}{m_{\hp}} \newc{\hm}{H^{-}}
\newc{\mhm}{m_{\hm}}
\newc{\xt}{X_{t}}           \newc{\xb}{X_{b}}

\newc{\qzero}{Q_0}          \newc{\qstop}{Q_{\widetilde t}}
\newc{\amu}{a_{\mu}}        \newc{\amususy}{a_{\mu}^{\text{SUSY}}}
\newc{\amuexpt}{a_{\mu}^{\text{expt}}}        \newc{\amusm}{a_{\mu}^{\text{SM}}}
\newc{\deltaamususy}{\delta a_{\mu}^{\text{SUSY}}}
\newc\gmtwo{(g-2)_{\mu}} 
\newc\deltagmtwo{\delta (g-2)_{\mu}} 
\newc\deltaamu{\Delta a_{\mu}}
\newc{\msbar}{\overline{MS}} \newc{\drbar}{\overline{DR}}
\newc{\yt}{h_t} \newc{\yb}{h_b} \newc{\ytau}{h_{\tau}}

\newc{\mtop}{m_t}               \newc{\mtpole}{M_t}
\newc{\mtaupole}{m_{\tau}^{\text{pole}}}
\newc{\mtmtsmmsbar}{m_t(m_t)^{\msbar}_{{\text{SM}}}}
\newc{\mtmtsmdrbar}{m_t(m_t)^{\drbar}_{{\text{SM}}}}
\newc{\mtmtmssmdrbar}{m_t(m_t)^{\drbar}_{{\text{SUSY}}}}

\newc{\mbmbmsbar}{m_b(m_b)^{\msbar} }

\newc{\mbmbsmmsbar}{m_b(m_b)^{\msbar}_{{\text{SM}}}}
\newc{\mbmzsmmsbar}{m_b(\mz)^{\msbar}_{{\text{SM}}}}
\newc{\mbmzsmdrbar}{m_b(\mz)^{\drbar}_{{\text{SM}}}}
\newc{\mbmzmssmdrbar}{m_b(\mz)^{\drbar}_{{\text{SUSY}}}}

\newc{\mtaumzsmmsbar}{m_{\tau}(\mz)^{\msbar}_{{\text{SM}}}}
\newc{\mtaumzsmdrbar}{m_{\tau}(\mz)^{\drbar}_{{\text{SM}}}}
\newc{\mtaumzmssmdrbar}{m_{\tau}(\mz)^{\drbar}_{{\text{SUSY}}}}

\newc{\mgut}{M_{\rm GUT}}
\newc{\mplanck}{M_{\rm P}}      \newc{\mpl}{M_{\text{Pl}}}
\newc{\msusy}{M_{\rm SUSY}}      \newc{\ms}{M_{\text{S}}}
\newc{\jxf}{J({\xf})}
\newc{\jxfexact}{J_{\rm exact}({\xf})}  \newc{\jxfexp}{J_{\rm exp}({\xf})}
\newc{\VEV}[1]{\langle #1 \rangle}
\newc{\xf}{x_f}
\newc\vrel{v_{\rm rel}}
\newcommand\mchi{m_{\chi}}              
\newc\sell{{\widetilde e}_L}      \newc\msell{m_{\sell}}
\newc\selr{{\widetilde e}_R}      \newc\mselr{m_{\selr}}
\newc\snue{{\widetilde \nu}_e}      \newc\msnue{m_{\snue}}
\newc\snutau{{\widetilde \nu}_\tau}      \newc\msnutau{m_{\snutau}}
\newc\supl{{\widetilde u}_L}      \newc\msupl{m_{\supl}}
\newc\supr{{\widetilde u}_R}      \newc\msupr{m_{\supr}}
\newc\sdl{{\widetilde d}_L}      \newc\msdl{m_{\sdl}}
\newc\sdr{{\widetilde d}_R}      \newc\msdr{m_{\sdr}}

\newcommand\stopone{{\widetilde t}_1}

\newcommand\stauone{{\widetilde \tau}_1}

\newcommand\mgluino{m_{\widetilde g}}

\newc\sfermion{\tilde f}  \newc\msfermion{m_{\sfermion}}
\newc\cmeter{{\rm cm}} \newc\meter{{\rm m}} \newc\kmeter{{\rm km}}
\newc\second{{\rm sec}}

\newc\sr{{\rm sr}}

\newc{\gstar}{g_\ast}           \newc{\gsstar}{g_{s\ast}}
\newc{\geff}{g_{\rm eff}}
\newcommand\mz{m_{Z}}

\newc{\sthw}{\sin\theta_W}              \newc{\cthw}{\cos\theta_W}
\newc{\bino}{\widetilde B}              \newc{\wino}{\widetilde W_30}
\newc{\higgsinob}{{\widetilde H}^0_b}   \newc{\higgsinot}{{\widetilde H}^0_t}
\newc{\abund}{\Omega h^2}
\newc{\abundchi}{\Omega_\chi h^2}
\newc{\abundcdm}{\Omega_{\text{CDM}} h^2}
\newc{\omegam}{\Omega_{M}}       \newc{\abundm}{\Omega_{M} h^2}
\newc{\omegab}{\Omega_{b}}       \newc{\abundb}{\Omega_{b} h^2}
\newc{\omegacdm}{\Omega_{CDM}}
\newc{\omegatot}{\Omega_{TOT}}
\newc{\rhocrit}{\rho_{crit}}
\newc{\rhochi}{\rho_{\chi}}
\newcommand\pb{\,\mbox{pb}} 

\newc\pc{\,\mbox{pc}} \newc\kpc{\,\mbox{kpc}}
\newc\mpc{\,\mbox{Mpc}} \newc\gpc{\,\mbox{Gpc}}

\newc\BR{BR}
\newc\bsgamma{b\rightarrow s \gamma }
\newc\bxsgamma{\overline{B}\rightarrow X_{s}\gamma}
\newc\brbsgamma{\BR(\overline{B}\rightarrow X_s\gamma)}

\newcommand\brbsmumu{\BR(\overline{B}_s\to\mu^+\mu^-)}

\newcommand\brbtaunu{\BR(\overline{B}_u\to \tau \nu)}

\newc{\beq}{\begin{equation}}
\newc{\eeq}{\end{equation}}

\newcommand\vs{{\it {vs.}}}

\newc\stoponetwo{{\widetilde t}_{1,2}}
\newc\sbotonetwo{{\widetilde b}_{1,2}}
\newc\stauonetwo{{\widetilde \tau}_{1,2}}

\newc{\sigsip}{\sigma^{SI}_{p}} \newc{\sigsin}{\sigma^{SI}_{n}}
\newc{\sigsiN}{\sigma^{SI}_{N}}
\newc{\sigsdp}{\sigma^{SD}_{p}} \newc{\sigsdn}{\sigma^{SD}_{n}}
\newc{\sigsiA}{\sigma^{SI}_{A}}

\newc{\pbar}{\bar{p}}

\newc{\egamma}{E_{\gamma}}
\newc{\flux}[1]{\Phi_{#1}}
\newc{\dfluxde}[1]{\frac{d\Phi_{#1}}{d E_{#1}}}

\newc{\fluxg}{\Phi_{\gamma}}
\newc{\dfluxgde}{\frac{d\Phi_{\gamma}}{d\egamma}}
\newc{\dfluxgdetext}{ d\Phi_{\gamma} / d\egamma}

\newc{\eplus}{e^+}
\newc{\epos}{E_{\eplus}}
\newc{\eps}{\varepsilon}

\newc{\npos}{n_{\eplus}} \newc{\Npos}{N_{\eplus}}
\newc{\dnposde}{\frac{d n_{\eplus}}{d\epos}}
\newc{\dnposdeps}{\frac{d n_{\eplus}}{d\eps\phantom{_{\eplus}}}}
\newc{\dnposdepstext}{ d n_{\eplus} / d\eps}
\newc{\fluxpos}{\Phi_{\eplus}}  \newc{\fluxelec}{\Phi_{e^{-}}}
\newc{\dfluxposde}{\frac{d\Phi_{\eplus}}{d\epos}}
\newc{\dfluxposdetext}{ d\Phi_{\eplus} / d\epos}

\newc{\nfwc}{{\text{NFW+ac}}} \newc{\moorec}{{\text{Moore+ac}}}

\newc{\chisq}{\chi^2}  \newc{\chisqred}{\chi^2_{\text{red}}}

\newc\xilim{\xi_{\rm lim}} 
\newc\tlim{t_{\rm lim}} 
\newc\zetalim{\zeta_{\rm lim}} 

\newc\zetah{\zeta_h}
\newc{\relprobone}[1]{p({#1} \vert d)}
\newc{\relprobtwo}[2]{p({#1},{#2} \vert d)}

\long\def\begincomment#1\endcomment{%
        \begingroup\sf\baselineskip12pt#1\endgroup}

\newcommand{\squishlist}{
   \begin{list}{$\bullet$}
    { \setlength{\itemsep}{0pt}      \setlength{\parsep}{3pt}
      \setlength{\topsep}{3pt}       \setlength{\partopsep}{0pt}
      \setlength{\leftmargin}{1.em} \setlength{\labelwidth}{1em}
      \setlength{\labelsep}{0.5em} } }
\newcommand{\squishend}{
    \end{list}  }
\newcommand\mZ{m_{Z}}        \newcommand\mW{m_{W}}

\def    \be            {\begin{equation}}
\def    \ee            {\end{equation}}
\def    \bea           {\begin{eqnarray}}
\def    \eea           {\end{eqnarray}}

\def \ie{{\it i.e.}}
\def \eg{{\it e.g.}}
\def \etal{{\it et al.}}

\newcommand{\data}{d}
\newcommand{\nuis}{\psi}
\newcommand{\params}{\theta}
\newcommand{\basis}{m}
\newcommand{\derived}{\xi}

\def\jhep{J. High Energy Phys. }
\def\ds@jhep{\def\@journal{jhep}}

\def\ds@jhep{\def\@journal{jcap}}

\def\plb{Phys. Lett. B}
\def\ds@plb{\def\@journal{plb}}

\def\prep{Phys. Rep.}
\def\ds@prep{\def\@journal{prep}}

\def\ds@cpc{\def\@journal{cpc}}

\def\ds@ijmpa{\def\@journal{ijmpa}}

\def\hepph{hep-ph/}
\def\ds@hepph{\def\@journal{hepph}}

\def\app{Astropart. Phys.}
\def\ds@app{\def\@journal{app}}

\begin{document}

\title{Some novel features of the Non-Universal Higgs Model}

\preprint{
\rightline{ 25 June 2009 LR, v5}
}

\title{A Bayesian Analysis of the Constrained NMSSM}

\author{Daniel E. L\'opez-Fogliani$^{1}$, Leszek Roszkowski$^{1,2}$,
  Roberto Ruiz de Austri$^3$, and Tom A. Varley$^{1}$}
\affiliation{
$^1$Department of Physics and Astronomy, The University of Sheffield,
Sheffield S3 7RH, England\\
$^2$The Andrzej Soltan Institute for Nuclear Studies, Warsaw, Poland\\
$^3$Instituto de F\'isica Corpuscular, IFIC-UV/CSIC, Valencia, Spain\\
}

\begin{abstract}
We perform a first global exploration of the Constrained
  Next-to-Minimal Supersymmetric Standard Model using Bayesian
  statistics. We derive several global features of the model and find
  that, in some contrast to initial expectations, they closely
  resemble the Constrained MSSM. This remains true even away from the
  decoupling limit which is nevertheless strongly preferred. We present
  ensuing implications for several key observables, including collider
  signatures and predictions for direct detection of dark matter.
 \end{abstract}

\keywords{Supersymmetric Effective Theories, Cosmology of  Theories beyond the SM, Dark Matter}

\maketitle

\section{Introduction}\label{sec:intro}

Effective low-energy supersymmetry (SUSY) has many attractive features
and is widely expected to provide a more complete description of
phenomena at and above the electroweak scale than the Standard Model
(SM) of electroweak and strong interactions. In many SUSY models,
gauge coupling unification can easily be achieved, unlike in the SM or
non-SUSY versions of those models. Moreover, SUSY, when supplemented
by R-parity (or matter parity), offers an attractive candidate for
resolving the dark matter (DM) problem in the Universe.

On the other hand, SUSY, being a global symmetry, allows for
a whole multitude of possible effective models which otherwise would
suffer from the well-known hierarchy and fine-tuning problems. In the past,
most phenomenological studies focused on the Minimal Supersymmetric Standard Model
(MSSM) -- a supersymmetrized version of the
SM~\cite{susy-reviews}. More recently, a constrained version of the
MSSM (CMSSM)~\cite{kkrw94}, 
which includes a minimal supergravity (mSUGRA) model, has become more
popular by virtue of its relative simplicity and a small number of
free parameters. This is achieved by relating at the unification scale
the soft masses of the MSSM gauginos to a common value $\mhalf$, those
of the scalar partners of SM fermions to $\mzero$, the tri-linear
terms to $\azero$, in addition to $\tanb$ - the ratio of v.e.v.'s of
the neutral components of the two Higgs doublets. 

One puzzling and unsatisfactory feature of the MSSM is the so-called
$\mu$-problem~\cite{mupb}: the Higgs/higgsino mass parameter is SUSY-preserving
but, on phenomenological grounds, it is expected to be of the same
order as soft SUSY breaking masses, $\mu\sim\msusy\simeq 1\tev$. Various solutions
have been suggested, for example~\cite{gm88}.

A model that solves the $\mu$-problem of the MSSM in a simple way is
the Next-to-Minimal Supersymmetric Standard Model (NMSSM)~\cite{NMSSM1}. In the
NMSSM one adds a singlet chiral superfield S. The
explicit $\mu$ term is absent and the superpotential has only
dimensionless parameters and therefore the only new effective scale is the
one of the soft breaking terms $\msusy$. The $\mu$ parameter is
generated dynamically through the v.e.v. of the spin-0 component of
the singlet superfield.

At the phenomenological level, the presence of additional fields,
namely an extra CP-even and CP-odd neutral Higgs bosons, as well as a
singlino component of a neutralino, leads in general to a richer and more complex
phenomenology~\cite{NMSSM2,NLEP,NLHC,NHIGGS,NMHDECAY}, as well as cosmology,
in particular with respect to the domain wall problem~\cite{Abel1, Abel2, tadp}.

In analogy with the CMSSM, successful gauge coupling unification in
SUSY has provided motivation for considering a constrained version of
the NMSSM (CNMSSM)~\cite{CNMSSM1, NMSPEC}, which we will define below. Extensive
phenomenological investigations of the CNMSSM have been carried out
in~\cite{NMSSM2}. 

On the other hand, the enlarged set of parameters makes a full exploration of the CNMSSM
even more challenging than in the case of the CMSSM. Traditional techniques of
sampling slices of the parameter space provide limited information and
are inadequate in a number of other aspects, for example in fixing
relevant SM parameters which may have much impact on the outcome, as
shown in~\cite{rtr1,rrt2,rrt3}. 

More recently, it has been demonstrated that a Markov Chain Monte
Carlo (MCMC), or some other, scanning technique, coupled with Bayesian
statistics, can very efficiently
probe the whole parameter space and thus allow one to derive global
properties of the model under
investigation~\cite{al05,rtr1,rrt2,rrt3}.

Over the last few years, several studies using the Bayesian approach
have been performed of the CMSSM~\cite{al05,rtr1,rrt2,rrt3}, the
Non-Universal Higgs Mass Model (NUHM)~\cite{nuhm1}, the
MSSM~\cite{allanach+quevedo,hewett+rizzo} and large volume string
\cite{lvstring}.

One advantage of the Bayesian approach is that it allows quantitative
model comparison, using so called Bayes factors to see which model
best fits the data, be it selecting the sign of $\mu$ in the CMSSM to
picking a class of SUSY breaking. 

In this paper we perform a Bayesian analysis of the CNMSSM. We explore
very broad ranges of the CNMSSM parameter and apply all most important
experimental and cosmological constraints, including all collider
limits, the branching ratio of $\bsgamma$, the difference
$\deltagmtwo$ between the experimental and SM values of the magnetic
moment of the muon, the LEP limits on sparticle and Higgs masses and
the 5 year WMAP limit on the relic abundance $\abundchi$ of the
lightest neutralino assumed to be the dark matter. 
A full list
of constraints used and the exact numbers used in the analysis will be
given below.  

Our main finding is that, somewhat contrary to initial expectations,
from the statistical point of view, most phenomenological and dark
matter features of the CNMSSM of crucial importance for experimental
tests closely resemble those of the CMSSM. In particular,
singlino-dominated LSP only appears in a very limited number of cases
that are not yet excluded by experimental bounds on the parameter
space.  Clearly, this will make it very challenging, although not
impossible, to distinguish the models at the LHC and in DM searches,
as we discuss below.

The paper is organized as follows: in Sec.~\ref{sec:overNMSSM} we
define and overview the NMSSM and the CNMSSM. In Sec.~\ref{sec:Bayes}
we describe our the statistical approach.  The results are presented in
Sec.~\ref{sec:CNMSSMpars} and we finish with our conclusions in
Sec.~\ref{sec:summary}.

\section{The NMSSM and the CNMSSM}\label{sec:overNMSSM}

The NMSSM superpotential contains a new superfield $S$ which is a singlet under the
SM gauge group $SU(3)_c \times SU(2)_L \times U(1)_Y$. (We use the
same notation for superfields and their respective spin-0 component fields for simplicity.)
\begin{equation}\label{2:Wnmssm}
W=
\epsilon_{ij} \left(
Y_u \, H_u^j\, Q^i \, u +
Y_d \, H_d^i\, Q^j \, d +
Y_e \, H_d^i\, L^j \, e \right)
- \epsilon_{ij} \lambda \,S \,H_d^i H_u^j +\frac{1}{3} \kappa S^3\,,
\end{equation}
where $H_d^T=(H_d^0, H_d^-)$, $H_u^T=(H_u^+, H_u^0)$, $i,j$ are
$SU(2)$ indices with $\epsilon_{12}=1$, while $\lambda$ and $\kappa$
are dimensionless couplings in the enlarged Higgs sector.

The superpotential in Eq.~(\ref{2:Wnmssm})
is scale invariant, and the EW scale will only appear through the soft SUSY
breaking terms in $\mathcal{L}_{\text{soft}}$, which in our conventions is
given by
\begin{align}\label{2:Vsoft}
-\mathcal{L}_{\text{soft}}=&\,
 {m^2_{\tilde{Q}}} \, \tilde{Q}^* \, \tilde{Q}
+{m^2_{\tilde{U}}} \, \tilde{u}^* \, \tilde{u}
+{m^2_{\tilde{D}}} \, \tilde{d}^* \, \tilde{d}
+{m^2_{\tilde{L}}} \, \tilde{L}^* \, \tilde{L}
+{m^2_{\tilde{E}}} \, \tilde{e}^* \, \tilde{e}
 \nonumber \\
&
+m_{H_d}^2 \,H_d^*\,H_d + m_{H_u}^2 \,H_u^* H_u +
m_{S}^2 \,S^* S \nonumber \\
&
+\epsilon_{ij}\, \left(
A_u \, Y_u \, H_u^j \, \tilde{Q}^i \, \tilde{u} +
A_d \, Y_d \, H_d^i \, \tilde{Q}^j \, \tilde{d} +
A_e \, Y_e \, H_d^i \, \tilde{L}^j \, \tilde{e} + \text{H.c.}
\right) \nonumber \\
&
+ \left( -\epsilon_{ij} \lambda\, A_\lambda S H_d^i H_u^j +
\frac{1}{3} \kappa \,A_\kappa\,S^3 + \text{H.c.} \right)\nonumber \\
&
- \frac{1}{2}\, \left(M_3\, \lambda_3\, \lambda_3+M_2\, \lambda_2\, \lambda_2
+M_1\, \lambda_1 \, \lambda_1 + \text{H.c.} \right) \,.
\end{align}
When the scalar component of $S$ acquires a VEV, $s=\langle S
\rangle$, an effective interaction $\mu H_d H_u$ is generated, with
$\mu \equiv \lambda s$.

In addition to terms from $\mathcal{L}_{\text{soft}}$, the
tree-level scalar Higgs potential receives the usual $D$ and $F$ term
contributions:
\begin{align}\label{2:Vfd}
V_D = & \, \,\frac{g_1^2+g_2^2}{8} \left( |H_d|^2 - |H_u|^2 \right)^2 +
\frac{g_2^2}{2} |H_d^\dagger H_u|^2 \, , \nonumber \\
V_F = & \, \,|\lambda|^2
\left( |H_d|^2 |S|^2 + |H_u|^2 |S|^2 + |\epsilon_{ij} H_d^i H_u^j|^2 \right)
+ |\kappa|^2 |S|^4
\nonumber \\
&
-\left( \epsilon_{ij} \lambda \kappa^* H_d^{i} H_u^{j}S^{*2} + \mathrm{H.c.}
\right) \,.
\end{align}

Using the minimization equations we can re-express the soft breaking
Higgs masses in terms of $\lambda$, $\kappa$, $A_\lambda$, $A_\kappa$,
$v_d=\langle H_d^0 \rangle$, $v_u=\langle H_u^0 \rangle$ (with
$\tanb=v_u/v_d$), and $s$: 
 \begin{align}
m_{H_d}^2 = & -\lambda^2 \left( s^2 + v^2\sin^2\beta \right)
- \frac{1}{2} M_Z^2 \cos 2\beta
+\lambda s \tan \beta \left(\kappa s +A_\lambda \right) \,,
\label{2:minima:mh1} 
\\
m_{H_u}^2 = & -\lambda^2 \left( s^2 + v^2\cos^2\beta \right)
+\frac{1}{2} M_Z^2 \cos 2\beta
+\lambda s \cot \beta \left(\kappa s +A_\lambda \right) \,,
\label{2:minima:mh2} 
\\
\label{2:minima:ms} 
m_{S}^2 = & -\lambda^2 v^2 - 2\kappa^2 s^2 + \lambda \kappa v^2
\sin 2\beta + \frac{\lambda A_\lambda v^2}{2s} \sin 2\beta -
\kappa A_\kappa s\,,
\end{align}

The boundary conditions at the grand unification scale $\mgut\simeq
2\times 10^{16}\gev$ are analogous to those of the CMSSM, with the exception of $m_S$. 
$\kappa$, $M_s$ and $s$ are fixed
by the minimization
equations~(\ref{2:minima:mh1})-(\ref{2:minima:ms}) which leads to five
continuous free parameters of the CNMSSM: $\mhalf$, $\mzero$,
$\azero$, $\tanb$ and $\lambda$, in addition to $\sgn(\mu)$. 
%
%

The feature of not unifying $m_s$ with all the other soft scalar
masses at $\mzero$ gives one the necessary freedom to obtain, in the
limit $\lambda \to 0$, with $\lambda s$ fixed, the CMSSM plus a
singlet and a singlino field that both decouple from the spectrum, as
discussed in~\cite{NMSPEC}. For cosmological analyzes those extra
particles can still play an important role but from the particle
phenomenology point of view the model becomes indistinguishable from the
CMSSM. Also in this limit a singlino LSP is excluded since, being
completed decoupled, it can not annihilate into SM particles.

We also present the neutralino sector since the lightest neutralino
will, by assumption, play the r\^{o}le of dark matter. The mass term in
the Lagrangian is given by
\begin{equation}
\mathcal{L}_{\mathrm{mass}}^{\chi^0} =
-\frac{1}{2} (\Psi^0)^T \mathcal{M}_{\tilde \chi^0} \Psi^0 + \mathrm{H.c.}\,,
\end{equation}
with $\mathcal{M}_{\tilde \chi^0}$ given by a $5 \times 5$ matrix,
{\footnotesize \begin{equation}
  \mathcal{M}_{\chi^0} = \left(
    \begin{array}{ccccc}
      M_1 & 0 & -M_Z \sin \theta_W \cos \beta &
      M_Z \sin \theta_W \sin \beta & 0 \\
      0 & M_2 & M_Z \cos \theta_W \cos \beta &
      -M_Z \cos \theta_W \sin \beta & 0 \\
      -M_Z \sin \theta_W \cos \beta &
      M_Z \cos \theta_W \cos \beta &
      0 & -\lambda s & -\lambda v_u \\
      M_Z \sin \theta_W \sin \beta &
      -M_Z \cos \theta_W \sin \beta &
      -\lambda s &0 & -\lambda v_d \\
      0 & 0 & -\lambda v_u & -\lambda v_d & 2 \kappa s
    \end{array} \right),
  \label{neumatrix}
\end{equation}}
where $\mone$ ($\mtwo$) denotes soft the mass of the bino (wino) and
$\theta_W$ denotes the weak mixing angle.

\section{Outline of the method}\label{sec:Bayes}

Following the discussion of the previous Section, in the CNMSSM the 
free parameters are given by
\beq
\theta = (\mhalf, \mzero,\azero, \tanb,\lambda),
\label{eq:cmssm}
\eeq
while we fix $\sgn(\mu)$= +1, which implies $s>0$. Furthermore,
without loss of generality we choose
$\lambda>0$~\cite{Cerdeno:2004xw}.

As the values of relevant SM
parameters, when varied over their experimental constraints, have an
impact on the observable quantities, fixing them would lead to
inaccurate results. Instead, here we incorporate them explicitly as
free parameters (which are then constrained using their measured
values), which we call {\em nuisance parameters} $\psi$, where
\beq
\psi = (\mtpole,m_b(m_b)^{\msbar}, \alpha_{s}(\mz)^{\msbar}).
\label{eq:nuis}
\eeq
In \eq{eq:nuis} $\mtpole$ denotes the pole top quark mass, while the
other two parameters: $\mbmbmsbar$ -- the bottom quark mass evaluated
at $m_b$ and $\alphas$ -- the strong coupling constant evaluated at
the $Z$ pole mass $\mZ$ - are all computed in the $\msbar$
scheme. Note that, in contrast to recent analyzes of the
CMSSM~\cite{al05,rtr1,rrt2}, we do not include among the nuisance
parameters the fine structure constant. This is because here we use
the Fermi constant, $\mZ$ and $\mW$ as input parameters, yielding
$\alphaem$ as output.

Using notation consistent with previous analyzes we define our  eight dimensional
{\em basis  parameter} set as 
\beq
\basis=(\theta,\psi)
\label{eq:basis}
\eeq
which we will be scanning simultaneously over. For
each choice of $\basis$ a number of colliders or cosmological observables
are calculated. These derived variables are denoted by
$\derived=(\xi_1,\xi_2,\ldots)$, which are then compared with the
relevant measured data $\data$.

The quantity we are interested in is
the {\em posterior probability density function}, (or simply
posterior) $p(\basis|\data)$ which gives the probability of the
parameters after the constraints coming from the data have been
applied. The posterior follows from Bayes' theorem,
\beq 
p(\basis|\data)=\frac{p(\data|\derived) \pi(\basis)}{p(\data)}
\eeq
where $p(\data|\derived)$, taken as a function of $\derived$ for {\em
fixed data} $\data$, is called the {\em likelihood} (where the
dependence of $\derived(\basis)$ is understood).  The likelihood is
the quantity that compares the data with the derived
observables. $\pi(m)$ is the {\em prior} which encodes our state of
knowledge of the parameters before comparison with the data. This
state of knowledge is then updated by the likelihood to give us the
posterior. $p(d)$ is called the {\em evidence} or {\em model
likelihood}, and in our analysis can be treated as a normalization
factor and hence is ignored subsequently for an example of how the
evidence can be used for model comparison purposes.

\begin{table}
\centering
\begin{tabular}{|c |}
 \hline
{CNMSSM parameters $\params$}       \\ \hline
 $50 < \mhalf < 4 \tev$ \\ 
 $50 < \mzero < 4 \tev$ \\
 $|\azero| <  7\tev$    \\
$2 < \tanb < 65$       \\ 
 $10^{-3}< \lambda <0.7$  \\ \hline\hline
{SM (nuisance) parameters $\nuis$}       \\
\hline
{$160 < \mtpole < 190 \gev$}\\
{$4 < m_b(m_b)^{\overline{MS}} < 5 \gev$} \\
{$0.10 < \alphas  < 0.13$} \\ \hline
\end{tabular}
\caption{Initial ranges for our basis parameters
  $\basis=(\params,\nuis)$. }
\label{table:prior}
\end{table}

As our main prior we take very wide ranges of the CNMSSM parameters as
 given in Table~\ref{table:prior}, although we have performed a number
 of additional scans which will be discussed below. We adopt a flat
 prior in $\log \mhalf$, $\log \mzero$, $\azero$, $\tanb$ and
 $\lambda$.  Following Ref.~\cite{nuhm1}, we call this choice a {\em
 log prior}, as opposed to a completely {\em flat prior} used in some
 of our earlier analyzes where all the basis parameters are scanned
 with a flat measure. 

As before~\cite{tfhrr1,nuhm1}, our rationale for this choice of priors
is that they are distinctively different. One reason why we apply
different priors to soft mass parameters only is that they play a
dominant r\^{o}le in the determination of the masses of the
superpartners and Higgs bosons. Another important reason is that flat
priors suffer from the ``volume effect'' by putting effectively too
much emphasis on larger values of scanned parameters, while the log
prior is more suitable for exploring smaller values of both $\mhalf$
and $\mzero$ which are anyway more natural in effective low-energy
SUSY models.  Therefore the choice of log priors appears to be
actually more suitable for revealing the structure of the model's
parameter space, similarly as in the CMSSM~\cite{tfhrr1} and the
NUHM~\cite{nuhm1}.

For the nuisance parameters we use flat priors (although this is not
important as they are directly constrained by measurements) and
apply Gaussian likelihoods representing the experimental observations 
(see table~\ref{tab:meas}), as before~\cite{rtr1,rrt2,rrt3,nuhm1}.

We compute our mass spectra and observable quantities using the
publicly available NMSSMTools (version 2.1.1) that includes NMSPEC
with a link to Micromegas; for details see Ref.~\cite{NMSSMTools}. We
list the observables that the current version of NMSPEC, as linked
with statistical subroutines available in SuperBayeS allows us to
include in the likelihood function in Table~\ref{tab:measderived}. The
relic density $\abundchi$ of the lightest neutralino is computed with
the help of Micromegas, which is also linked to NMSPEC. We further use
the same code to compute the cross section for direct detection of
dark matter via its elastic scatterings with targets in underground
detectors but do not include it in the likelihood due to large
astrophysical uncertainties.

The likelihoods for the measured observables are taken as Gaussian
with mean $\mu$, experimental and theoretical errors (see the detailed
explanation in Refs.~\cite{rtr1,rrt2}). In the case where there only an
experimental limit is available, this is given, along with the
theoretical error.  The smearing out of bounds and combination of
experimental and theoretical errors is handled in an identical manner
to Refs.~\cite{rtr1,rrt2}, with the notable exception of the Higgs
mass and LEP limits on sparticle masses, which are calculated as a
step function with values of the cross section times branching ratio
(in the case of the Higgs) or mass that are within two standard
deviations of the experimental limit being accepted. Finally, any
points that fail to provide radiative EWSB, give us tachyons or the
LSP other than the neutralino are rejected.

As our scanning technique we adopt a ``nested sampling''
method~\cite{skilling-nsconv} as implemented in the
MultiNest~\cite{Feroz:2007kg} algorithm, which computes the Bayesian
evidence primarily but produces posterior pdfs in the process.
MultiNest provides an extremely efficient sampler even for likelihood
functions defined over a parameter space of large dimensionality with
a very complex structure. (See, \eg, Refs.~\cite{tfhrr1,nuhm1}.)  This
aspect is very important for the model analyzed here since the
8-dimensional likelihood hyperspace is fragmented and features many
finely tuned regions that are difficult to explore with conventional
fixed grid, random scan or even MCMC methods. For a comparison of
CMSSM posterior maps obtained with a Metropolis-Hastings MCMC
algorithm~\cite{rtr1, rrt2, rrt3} and the MultiNest algorithm see
Ref.~\cite{tfhrr1}.

As we are using nested sampling in this study, the issue of stopping
criteria is handled differently from the MCMC case used in some
earlier paper~\cite{rtr1, rrt2, rrt3}. Our treatments follows
closely that presented in Appendix~A of Ref.~\cite{tfhrr1}. In
nested sampling one is calculating the Bayesian 
evidence, defined by,
 \be
 \mathcal{Z} \equiv  p(d) = \int_{0}^{1}\mathcal{L}(X)dX,
 \ee
where $\mathcal{L}$ is the likelihood and $X$ the prior volume. One
can get the posterior in a nested sampling scan, but the principle
value calculated is the evidence. The stopping criteria takes into
account that in general one is proceeding through shells of increasing
iso-likelihood contours, with the set of ``live points" drawn from
within these contours. One can then define the stopping criterion as
taking the maximum likelihood point in the set of live points, see
Ref.~\cite{skilling-nsconv} ($\mathcal{L}_{\rm max}$) and calculating
the maximum change to the evidence it could make,
$\delta\mathcal{Z}_{i}=\mathcal{L}_{max}X_{i}$. Once this value goes
below a specified value (we take $\delta\mathcal{Z}<0.5$) the run is
terminated.

\begin{table} 
 \centering
\begin{tabular}{|l | l l | l|}
\hline
SM (nuisance) parameter  &   Mean value  & \multicolumn{1}{c|}{Uncertainty} & Ref. \\
 &   $\mu$      & ${\sigma}$ (exper.)  &  \\ \hline
$\mtpole$           &  172.6 GeV    & 1.4 GeV&  \cite{cite:topmass} \\
$m_b (m_b)^{\overline{MS}}$ &4.20 GeV  & 0.07 GeV &  \cite{cite:mb} \\
$\alpha_{\text{s}}(M_Z)^{\overline{MS}}$       &   0.1176   & 0.002 &  \cite{cite:mb}\\
\hline
\end{tabular}
\caption{Experimental mean $\mu$ and standard deviation $\sigma$
 adopted for the likelihood function for SM (nuisance) parameters,
 assumed to be described by a Gaussian distribution.
\label{tab:meas}}
\end{table}

\begin{table}
\centering
\begin{tabular}{|l | l l l | l|}
\hline
Observable &   Mean value & \multicolumn{2}{c|}{Uncertainties} & Ref. \\
 &   $\mu$      & ${\sigma}$ (exper.)  & $\tau$ (theor.) & \\\hline
$\deltagmtwo \times 10^{10}$       &  29.5 & 8.8 &  1 & \cite{cite:g-2}\\
 $\brbsgamma \times 10^{4}$ &
 3.55 & 0.26 & 0.21 & \cite{cite:g-2} \\
$\brbtaunu \times 10^{4}$ &  $1.32$  & $0.49$  & $0.38$
& \cite{cite:CDF} \\
$\abundchi$ &  0.1099 & 0.0062 & $0.1\,\abundchi$& \cite{cite:WMAP} \\\hline
   &  Limit (95\%~\cl)  & \multicolumn{2}{r|}{$\tau$ (theor.)} & Ref. \\ \hline
$\brbsmumu$ &  $ <5.8\times 10^{-8}$ & \multicolumn{2}{r|}{14\%}  & \cite{cite:CDF2}\\
$\mhl$  & As implemented in NMSSMTool. & & & \cite{NMSSMTools} \\ 
sparticle masses  & As implemented in NMSSMTool. & & & \cite{NMSSMTools} \\ \hline 
\end{tabular}
\caption{Summary of the observables used in the analysis. Upper part:
Observables for which a positive measurement has been
made. $\deltagmtwo$ denotes the discrepancy between
the experimental value and the SM prediction of the anomalous magnetic
moment of the muon $\gmtwo$. For central values of the SM input
parameters used here, the SM value of $\brbsgamma$ is
$3.11\times10^{-4}$, while the theoretical error of
$0.21\times10^{-4}$ includes uncertainties other than the parametric
dependence on the SM nuisance parameters, especially on $\mtpole$ and
$\alphas$.  For each quantity we use a
likelihood function with mean $\mu$ and standard deviation $s =
\sqrt{\sigma^2+ \tau^2}$, where $\sigma$ is the experimental
uncertainty and $\tau$ represents our estimate of the theoretical
uncertainty (see Ref.~\cite{rtr1} for details). Lower part: Observables for
which only limits currently 
exist.  The likelihood function is given in
Ref.~\cite{rtr1}, including in particular a smearing out of
experimental errors and limits to include an appropriate theoretical
uncertainty in the observables in $\brbsmumu$. The limit on the light
Higgs mass $\mhl$ is applied in a simplified way, see text for details.
\label{tab:measderived}}
\end{table}

\section{Probability maps of CNMSSM parameters and observables}\label{sec:CNMSSMpars}

In this section we present our numerical results from global scans of
the CNMSSM parameter space.  We begin with the CNMSSM parameters and
next show probability maps for several observables, including, in
turn, the Higgs bosons, some superpartners and other collider
signatures, and finally dark matter cross sections.

 \begin{figure}[tbh!]
  \begin{center}
  \begin{tabular}{c c}
     \includegraphics[width=0.35\textwidth]{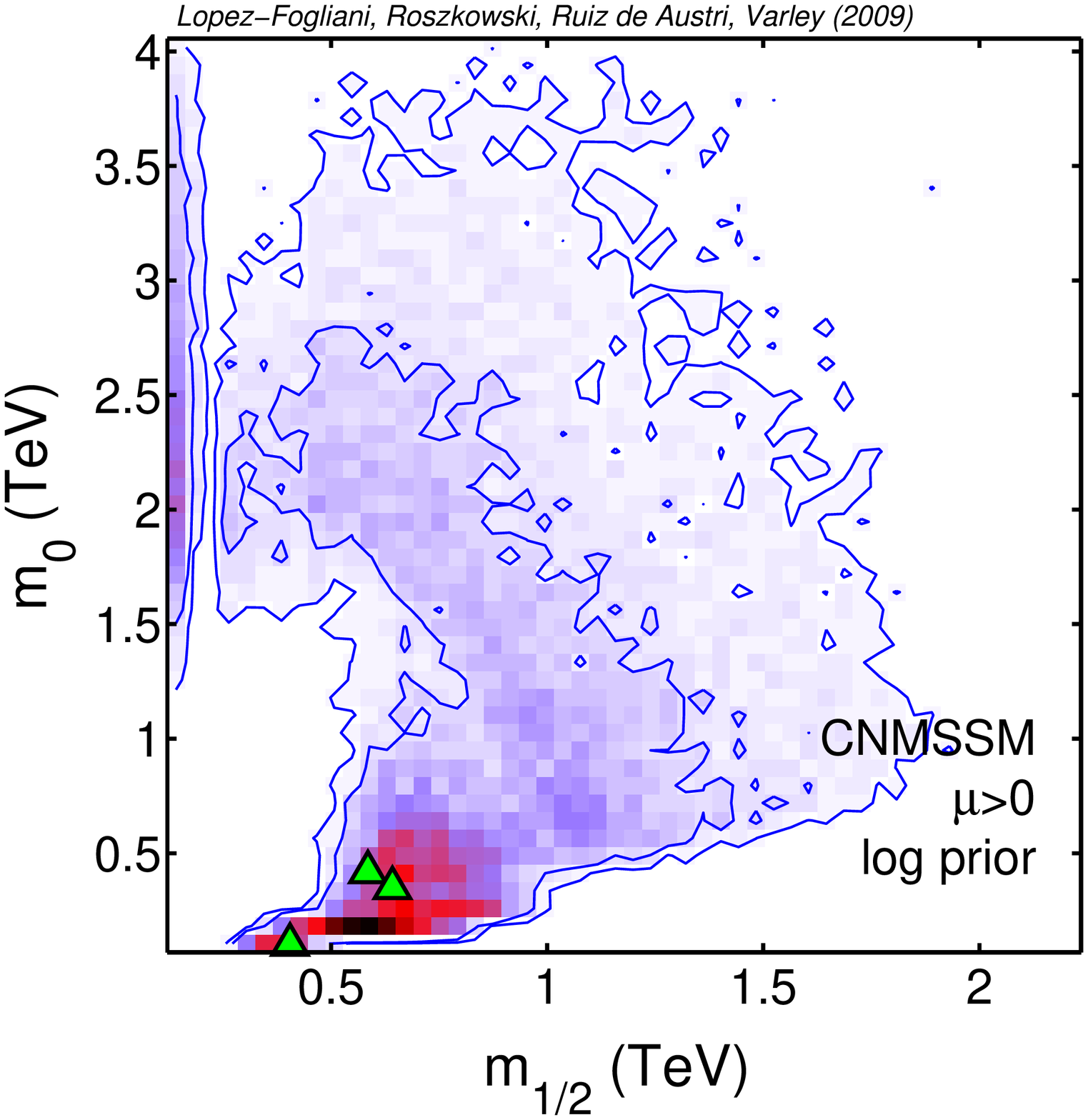}
   & \includegraphics[width=0.35\textwidth]{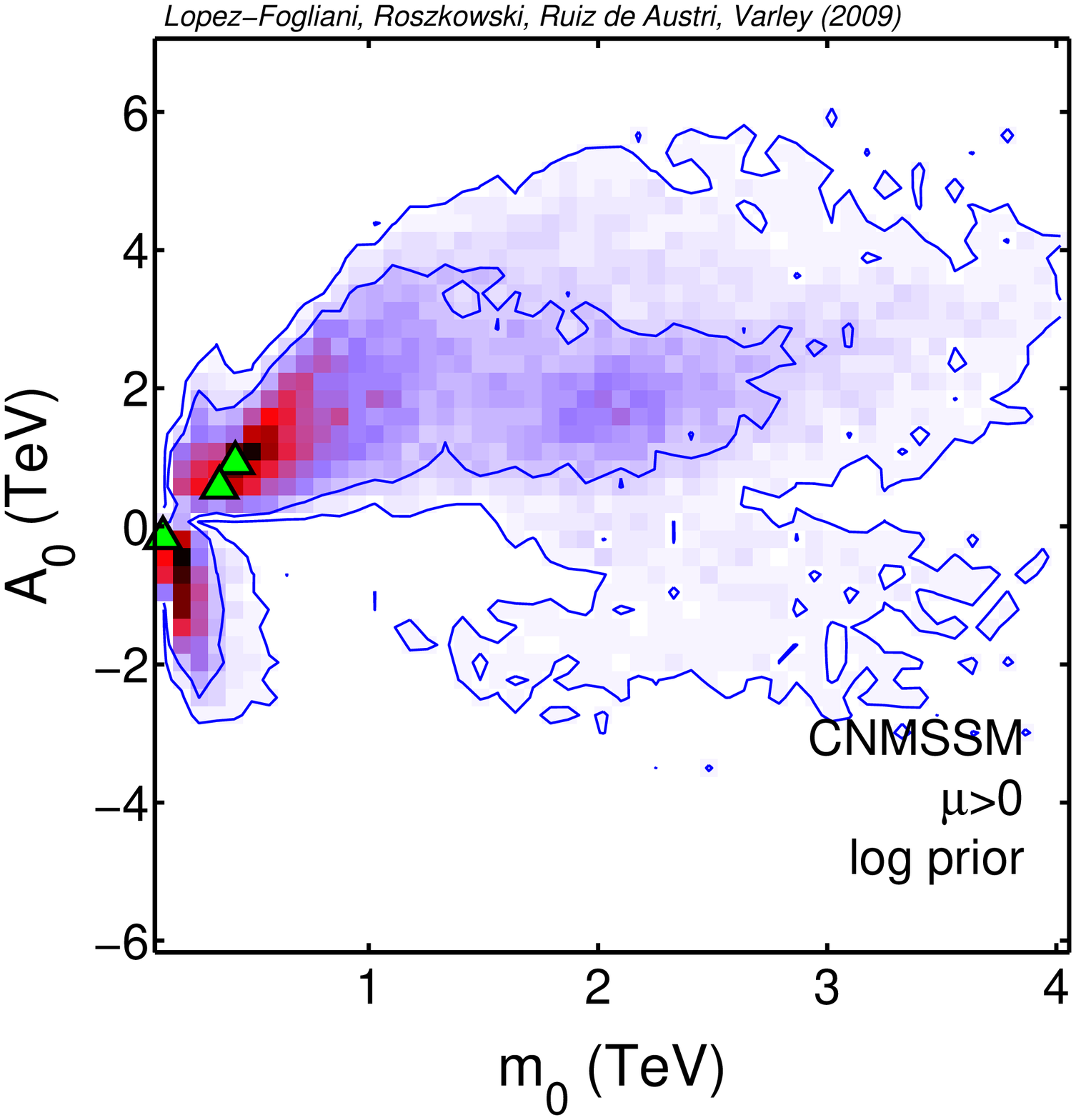}\\
     \includegraphics[width=0.35\textwidth]{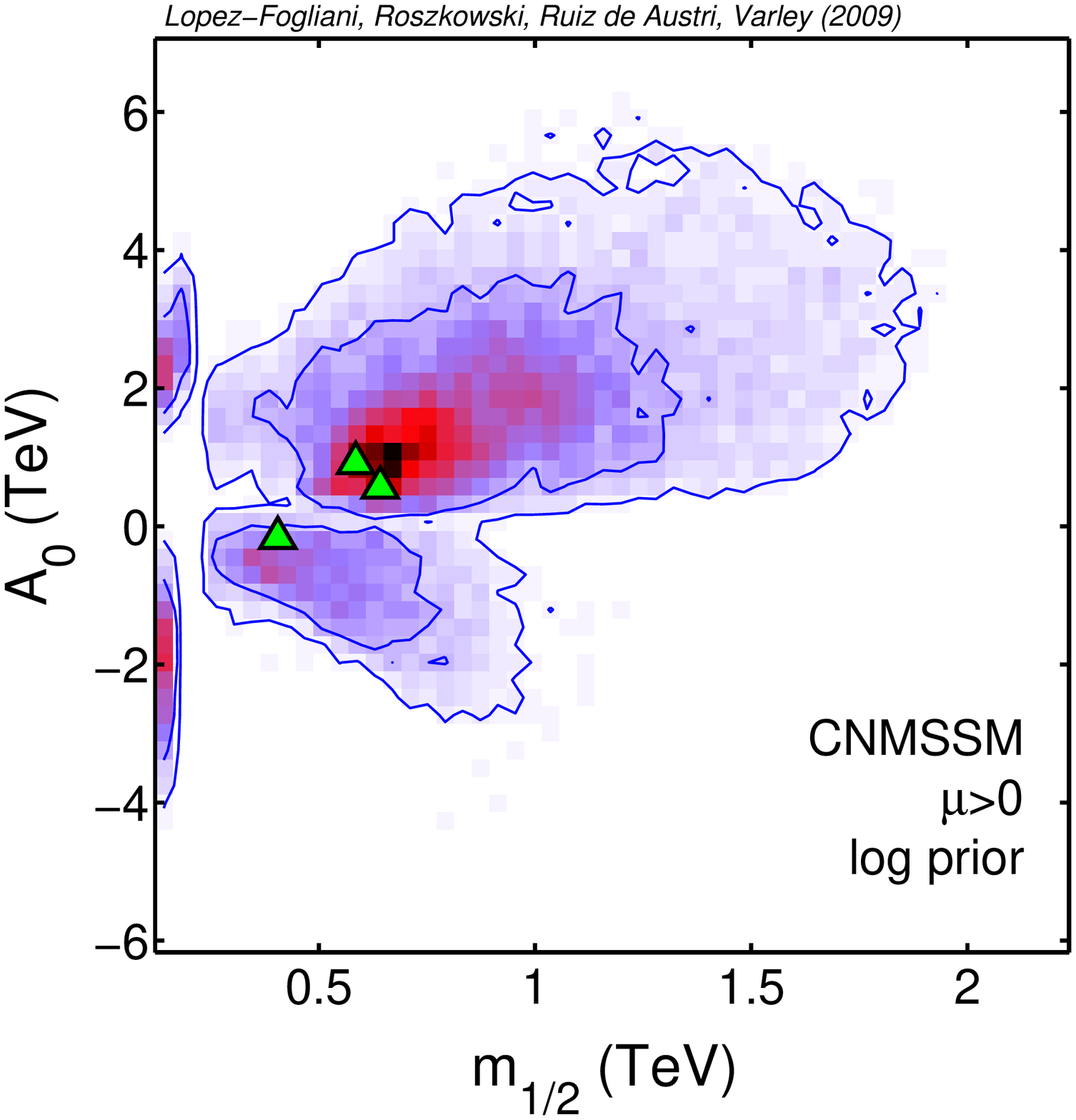}
   & \includegraphics[width=0.35\textwidth]{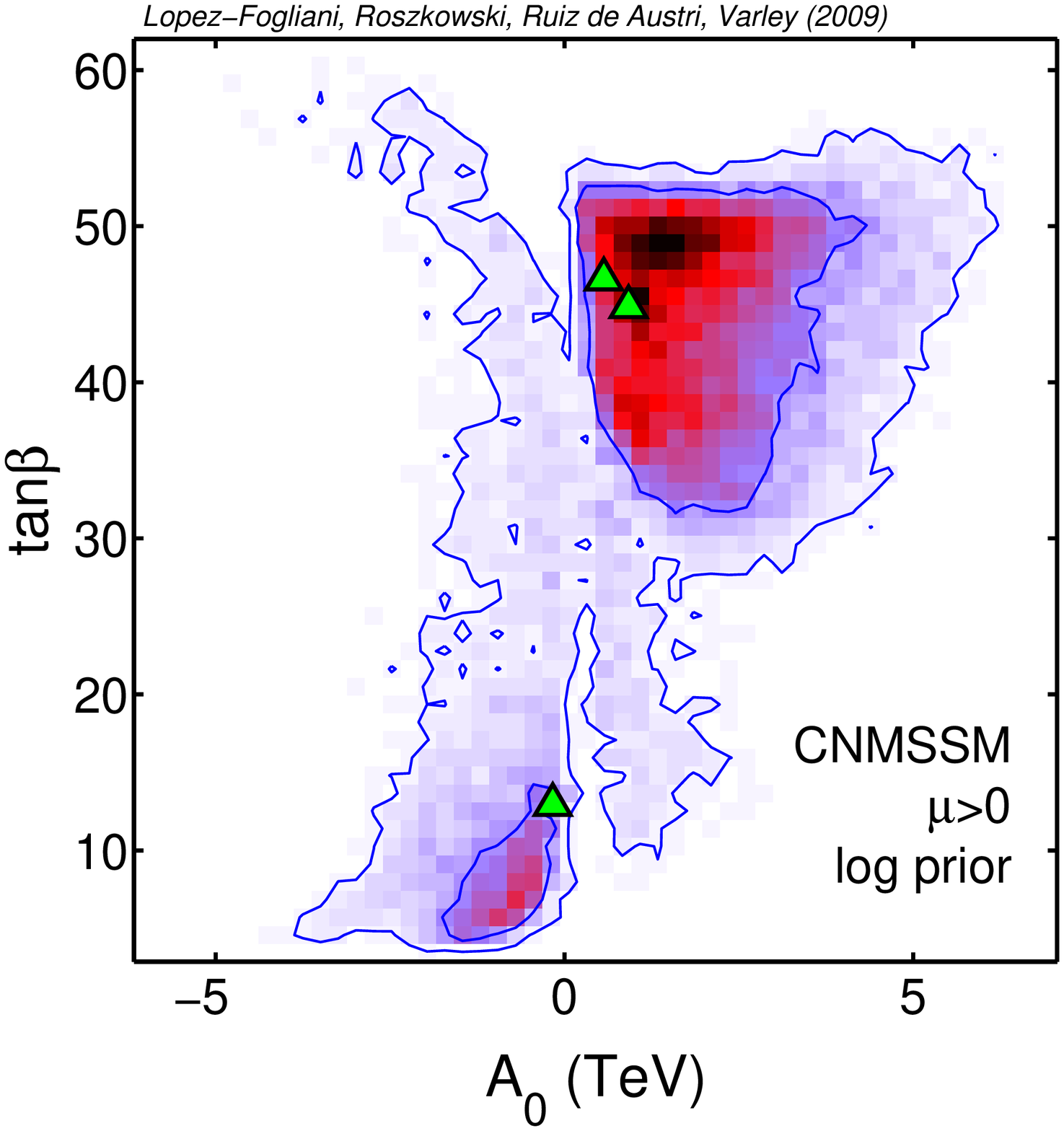}\\
     \includegraphics[width=0.35\textwidth]{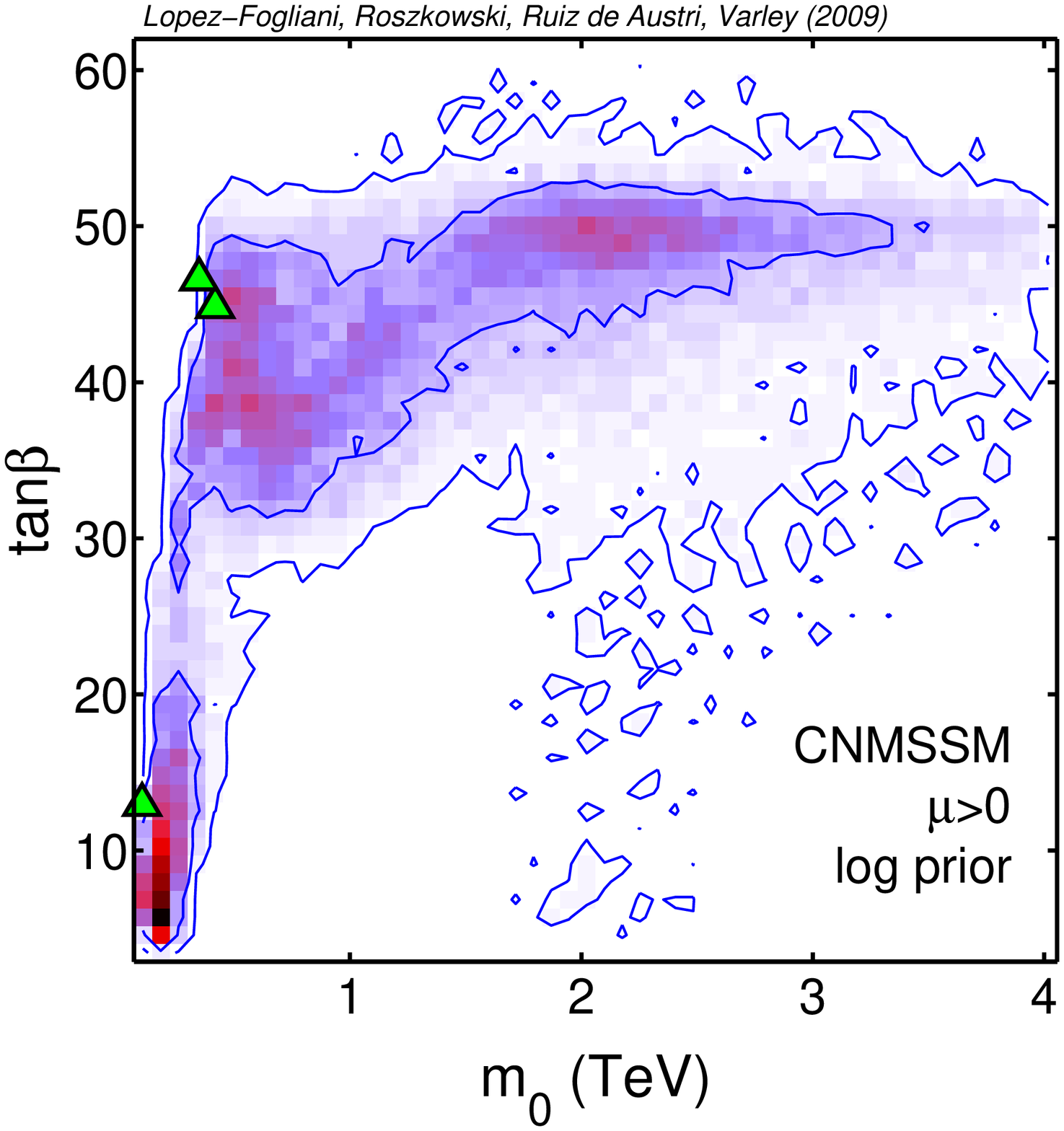}
   & \includegraphics[width=0.35\textwidth]{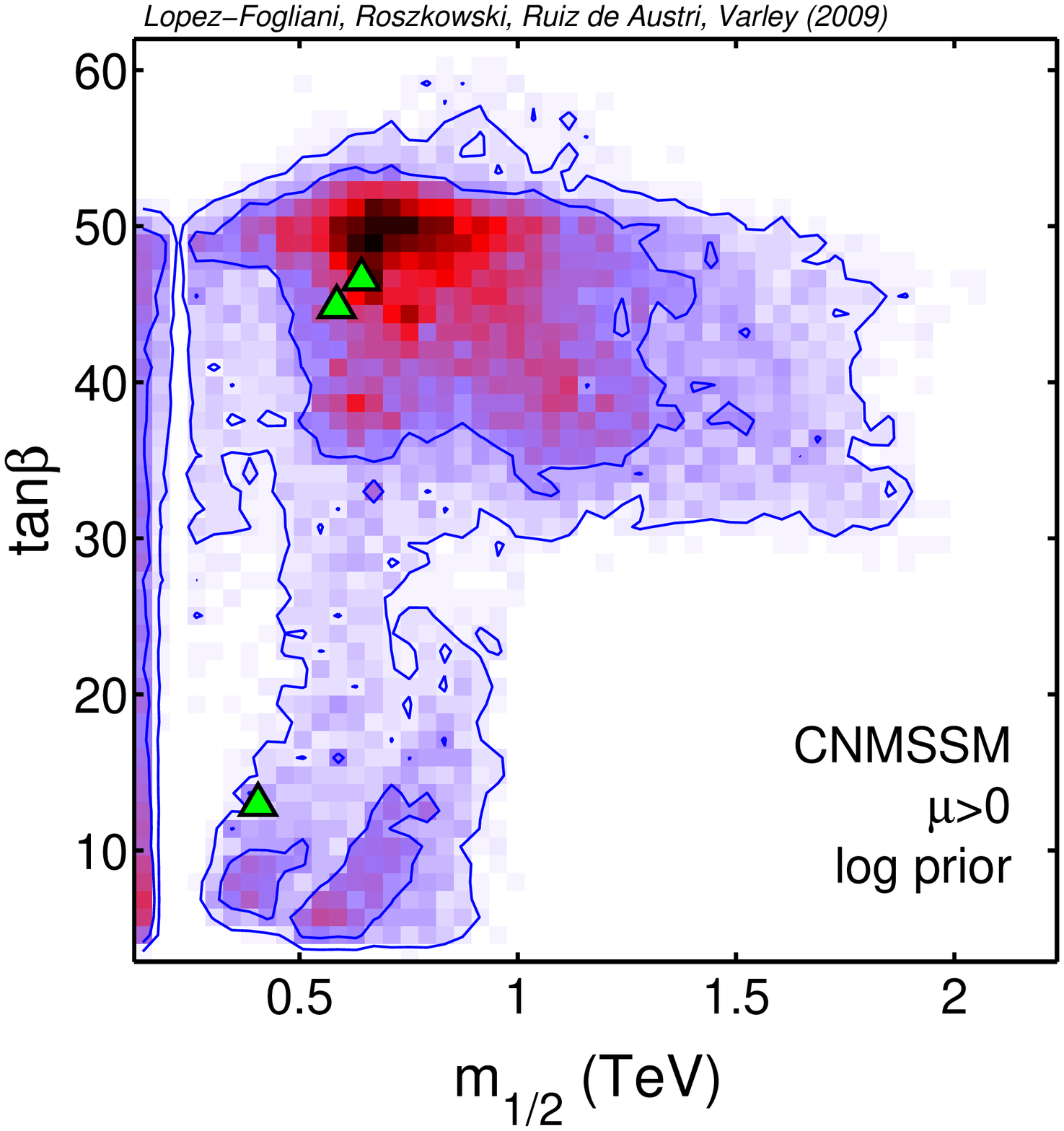}\\
   \end{tabular}
  \includegraphics[width=0.3\textwidth]{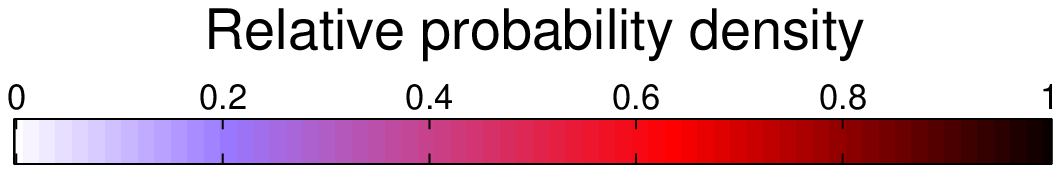}
  \end{center}
  \caption{\label{fig:CNMSSMps2dpdf_scan1} The 2D relative probability density
    functions in the planes spanned by the CNMSSM parameters $\mhalf$,
    $\mzero$, $\tanb$ and $\azero$ for the log prior.  The pdfs are normalized to unity
    at their peak. The inner (outer) blue solid contours delimit regions
    encompassing 68\% and 95\% of the total probability,
    respectively. All other basis parameters, both CNMSSM and SM ones,
   in each plane have been marginalized over (i.e., integrated
    out). Blue dots denote some best fit points. }
  \end{figure}


To begin with, in Fig.~\ref{fig:CNMSSMps2dpdf_scan1} we plot joint 2D
relative probability density functions (pdfs) for some combinations of
the CNMSSM parameters in our default case as given in
Table~\ref{table:prior} and taking the log prior, as explained above.
In this,  and figures below showing 2D pdfs, the inner (outer) contours
delineate the 68\% (95\%) total probability regions and the color code
is given in the bar at the bottom.

First, we can see that higher probability regions for all the
parameters but $\mzero$ are confined well within the assumed priors
and show clear high probability peaks. Focusing on the left panel in
the plane spanned by $\mhalf$ and $\mzero$, we can see some prominent
features: a rather strong preference for the stau coannihilation
region of $\mhalf\gsim\mzero\simeq0.5\tev$, although the 68\% total
probability region extends to larger $\mzero$ because of the
pseudoscalar funnel effect contribution to $\abundchi$, and to much
larger values of the parameter of the focus point (FP)
region~\cite{Chan:1997bi,focuspoint-fmm}. 
Green triangles indicate some of the best fit points.  The wedge
of $\mhalf\gg\mzero$ is disallowed because of charged LSP (normally
the stau). 

Examining the other panels of Fig.~\ref{fig:CNMSSMps2dpdf_scan1} one
can see a rather strong preference for large $\tanb$ (as in the
CMSSM), although small values below some 15 are also
favored. $\azero$ appears to have two distinct branches of opposite
sign, with $\azero=0$, although not excluded, proving to be hard to
find solutions for, with some resemblance to the CMSSM.  In some sense
the latter may even be as favored as large positive values, as
indicated by one of the best fit points.

Clearly, these are familiar features of the CMSSM, as one can see by
comparing the high probability regions of the CNMSSM in
Fig.~\ref{fig:CNMSSMps2dpdf_scan1} with analogous figures for the
common set of parameters shared with the CMSSM, as shown in Fig.~13 of
Ref.~\cite{tfhrr1} (obtained with the NS scan), or with Fig.~1 of
Ref.~\cite{rrt3} (obtained with the MCMC scan).

\begin{figure}[tbh!]
  \begin{center}
  \begin{tabular}{c c}
     \includegraphics[width=0.35\textwidth]{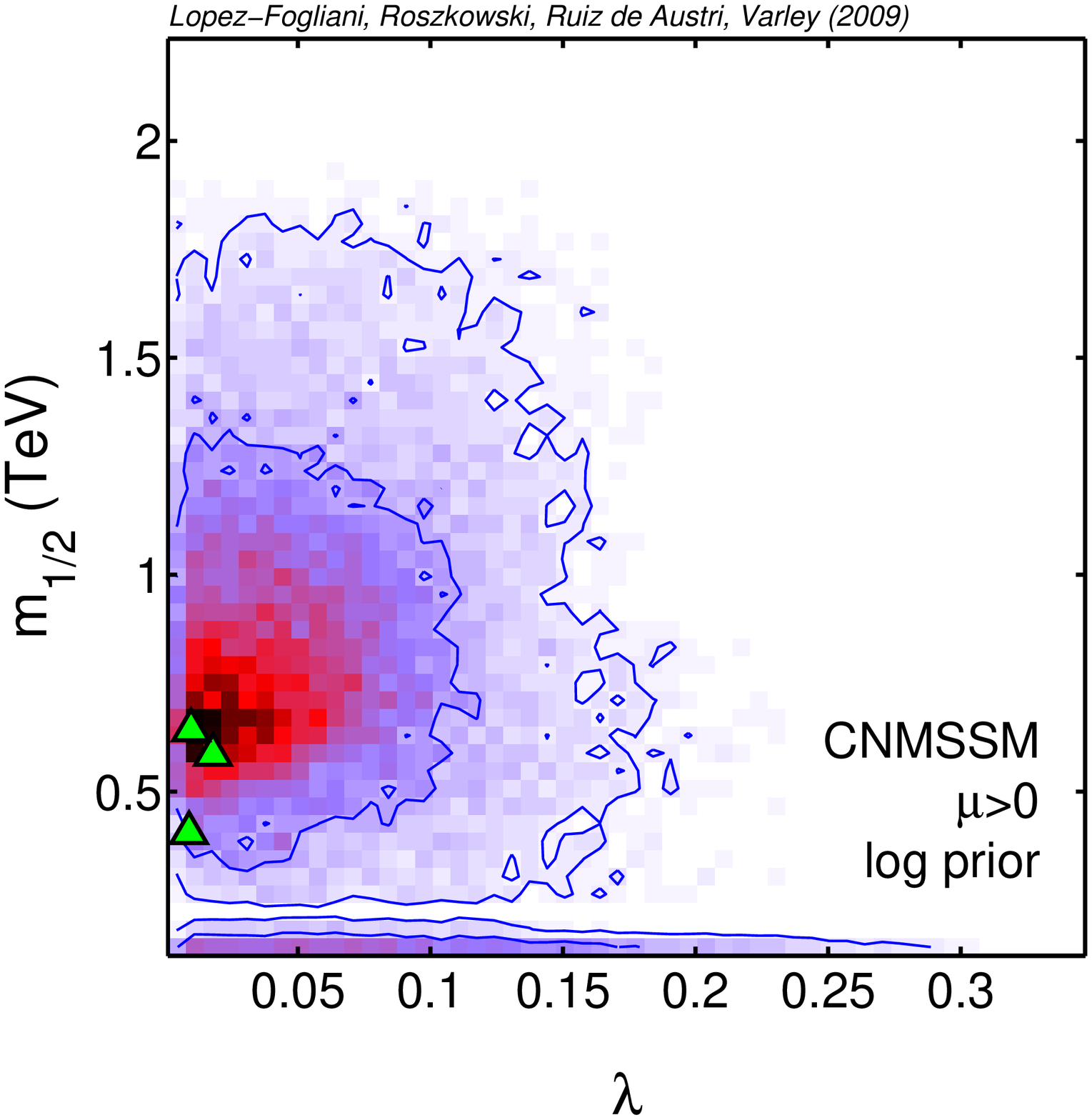}
    & \includegraphics[width=0.35\textwidth]{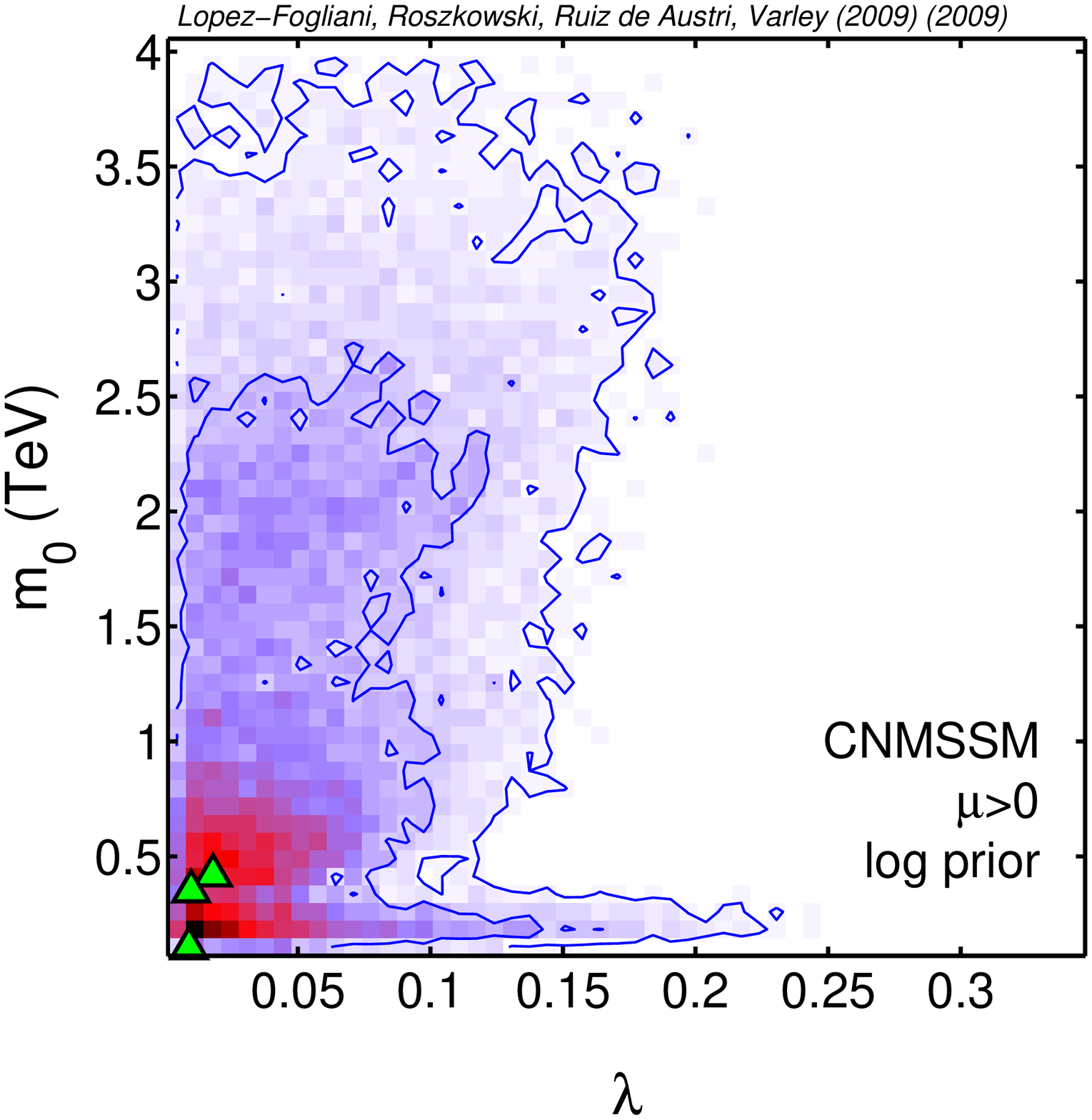}\\
     \includegraphics[width=0.35\textwidth]{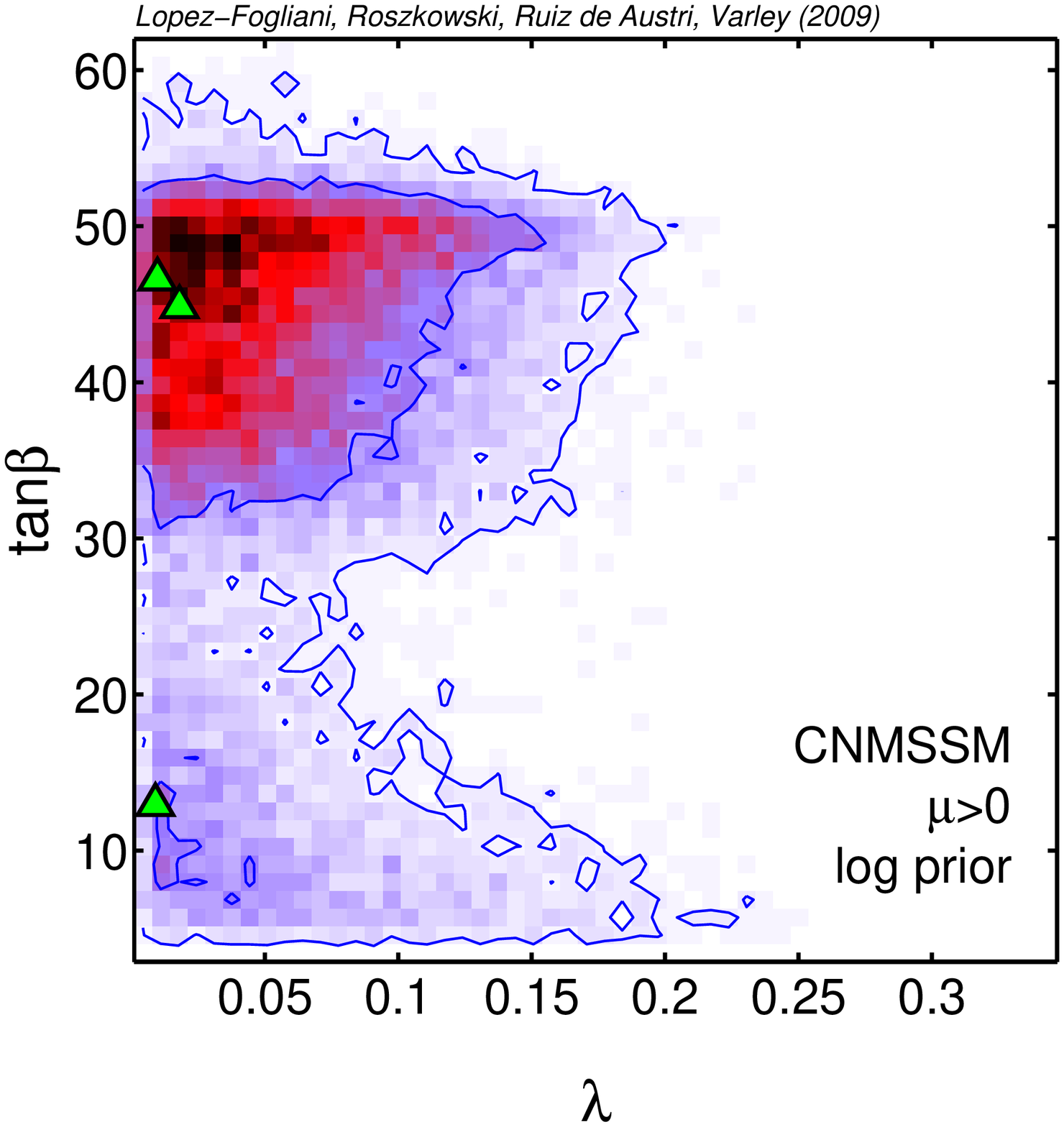}
     & \includegraphics[width=0.35\textwidth]{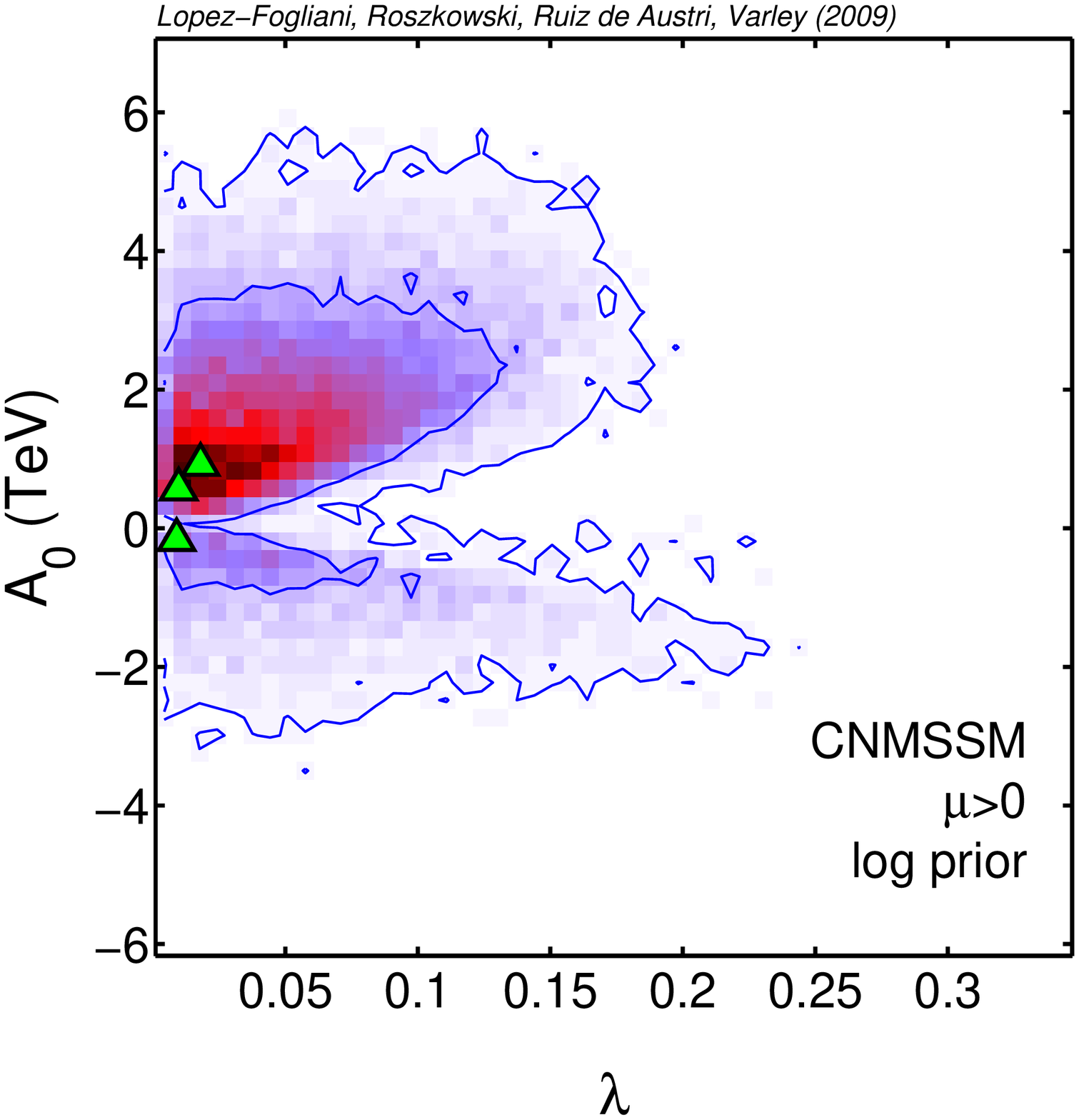}\\
   \end{tabular}
  \includegraphics[width=0.3\textwidth]{figures/colorbar.ps}
  \end{center}
\caption{\label{fig:CNMSSM_lambda2dpdf_scan1} The 2D relative
    probability density functions in the planes spanned by $\lambda$
    and the other CNMSSM parameters for the log prior.  The pdfs are
    normalized to unity at their peak. The inner (outer) blue solid
    contours delimit regions encompassing 68\% and 95\% of the total
    probability, respectively. All other basis parameters, both CNMSSM
    and SM ones, in each plane have been marginalized over. Blue dots
    denote some best fit points.}
  \end{figure}

\begin{figure}[tbh!]
\begin{center}
 \begin{tabular}{c}
    \includegraphics[width=0.35\textwidth]{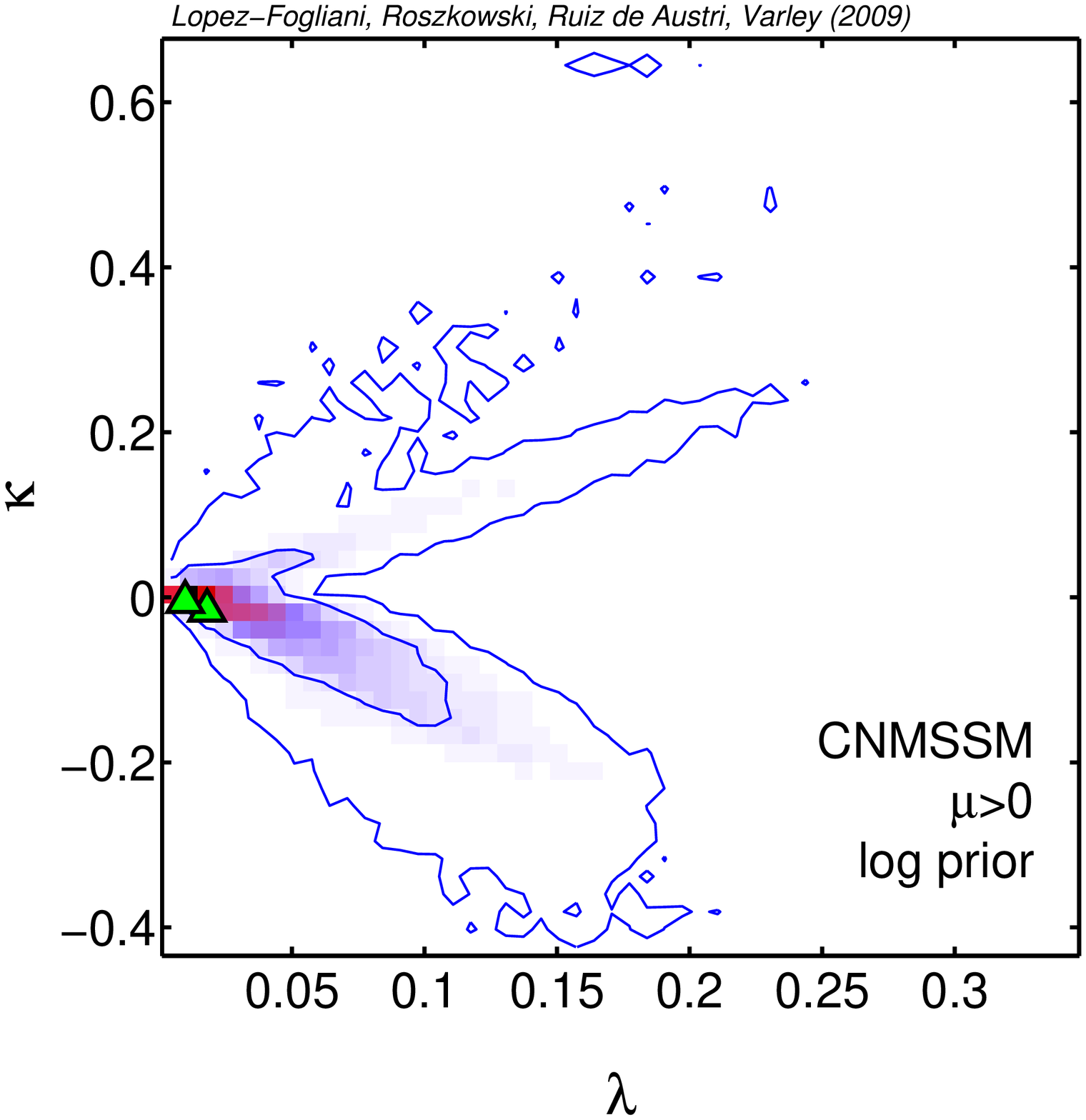}\\
  \includegraphics[width=0.3\textwidth]{figures/colorbar.ps}
  \end{tabular}
\end{center}
\caption{\label{fig:lk} 
The 2D
relative pdfs in the plane of $(\lambda,\kappa)$ for the log prior.
}
\end{figure}

The similarity of the high probability regions of $\mhalf$ and
$\mzero$ in both models suggests the parameters of the CNMSSM tend to
favor the decoupling limit, $\lambda \to 0$.  In
Fig.~\ref{fig:CNMSSM_lambda2dpdf_scan1} we show 2D pdfs of $\lambda$
with the CMSSM-like parameters. One can see that in
general $\lambda$ prefers to be small, which leads to a statistical
preference for a very CMSSM-like behavior.  Large values of $\lambda$,
bigger than around 0.6 are disfavored due to a Landau pole in the
running of $\lambda$. At ``intermediate'' values,
$0.1\lsim\lambda\lsim0.6$, the constraints become weaker but there
remain problems with tachyons, seen most clearly in the
$(\lambda,\kappa)$ plane, shown in Fig.~\ref{fig:lk}, which also shows
how both parameters are rather closely correlated, $\kappa \propto
\lambda $, and favor small values, towards the decoupling limit.  The
region with $\kappa\gg \lambda$ is disfavored by the presence of
tachyonic CP-odd scalars, and similarly with CP-even scalars for
$\kappa\ll \lambda$.
The
preference for low $\lambda$ could be due to the fact that there are
fewer tachyonic directions in the potential close to the decoupling
limit $\lambda \to 0$ (for this to happen $\lambda \lsim 0.1$ is
sufficient). For a more detailed discussion, see Ref.~\cite{Cerdeno:2004xw}.

In conclusion, the parameters of the CNMSSM seem to favor the CMSSM
limit since at small values of $\lambda \lsim 0.1$ it is simply much easier to find physical
solutions. This has been confirmed with exploratory runs with only the
requirement of correct EWSB and a neutralino LSP enforced. 
Indeed, a flat (and nor a log) prior in $\lambda$ was
chosen so as not to  emphasize low values of the parameter
and instead to ``force'' the scan away from the decoupling limit. Even with
the flat prior, however, the preference for small $\lambda$ remains
strong. For $\lambda$ below $0.1$ there are more solutions because for
instance tachyons are less of a problem (due to less mixing with
singlets) and also we are further away from the Landau pole
region.

 \begin{figure}[tbh!]
  \begin{center}
  \begin{tabular}{c c}
     \includegraphics[width=0.35\textwidth]{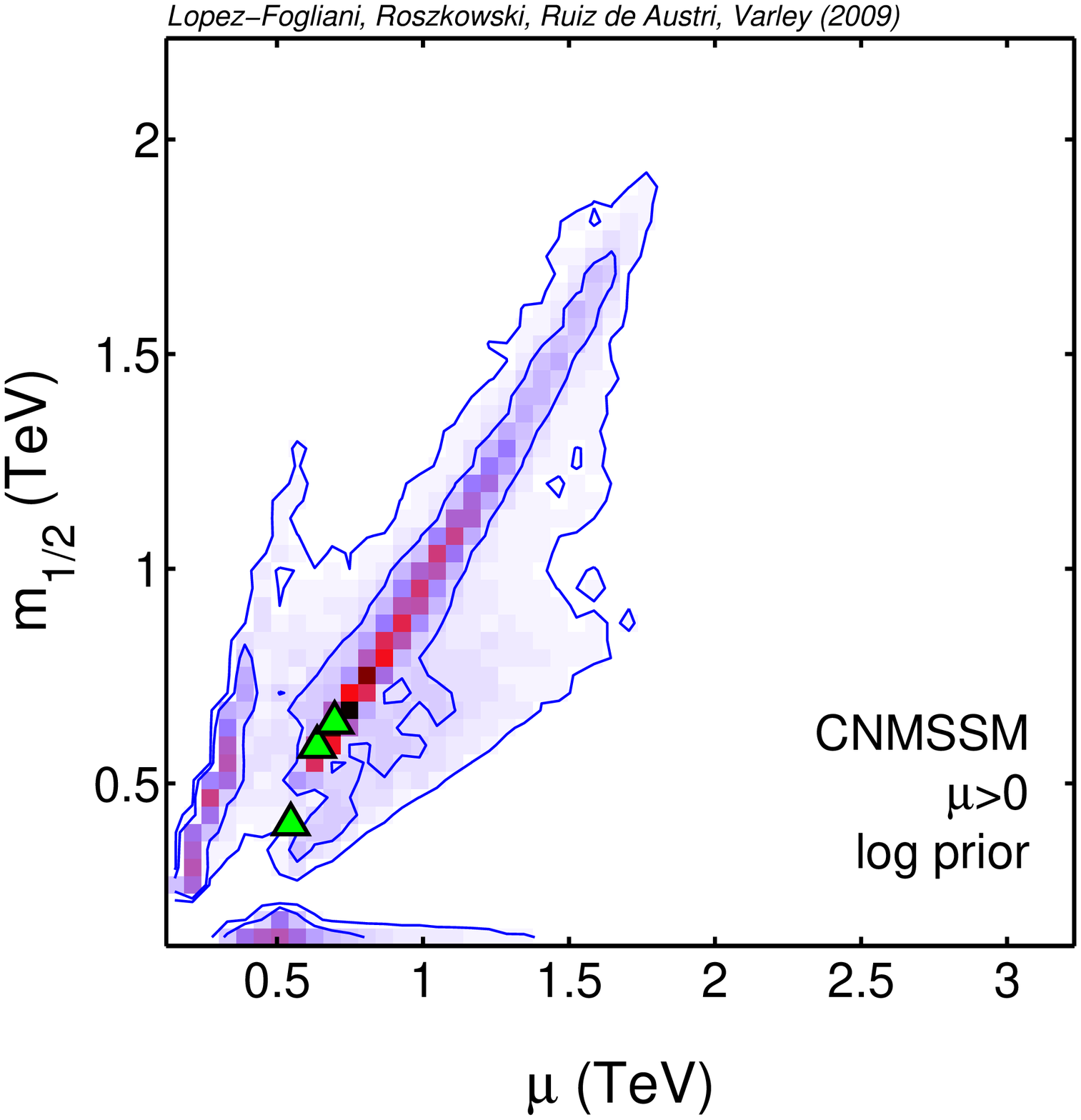}
    & \includegraphics[width=0.35\textwidth]{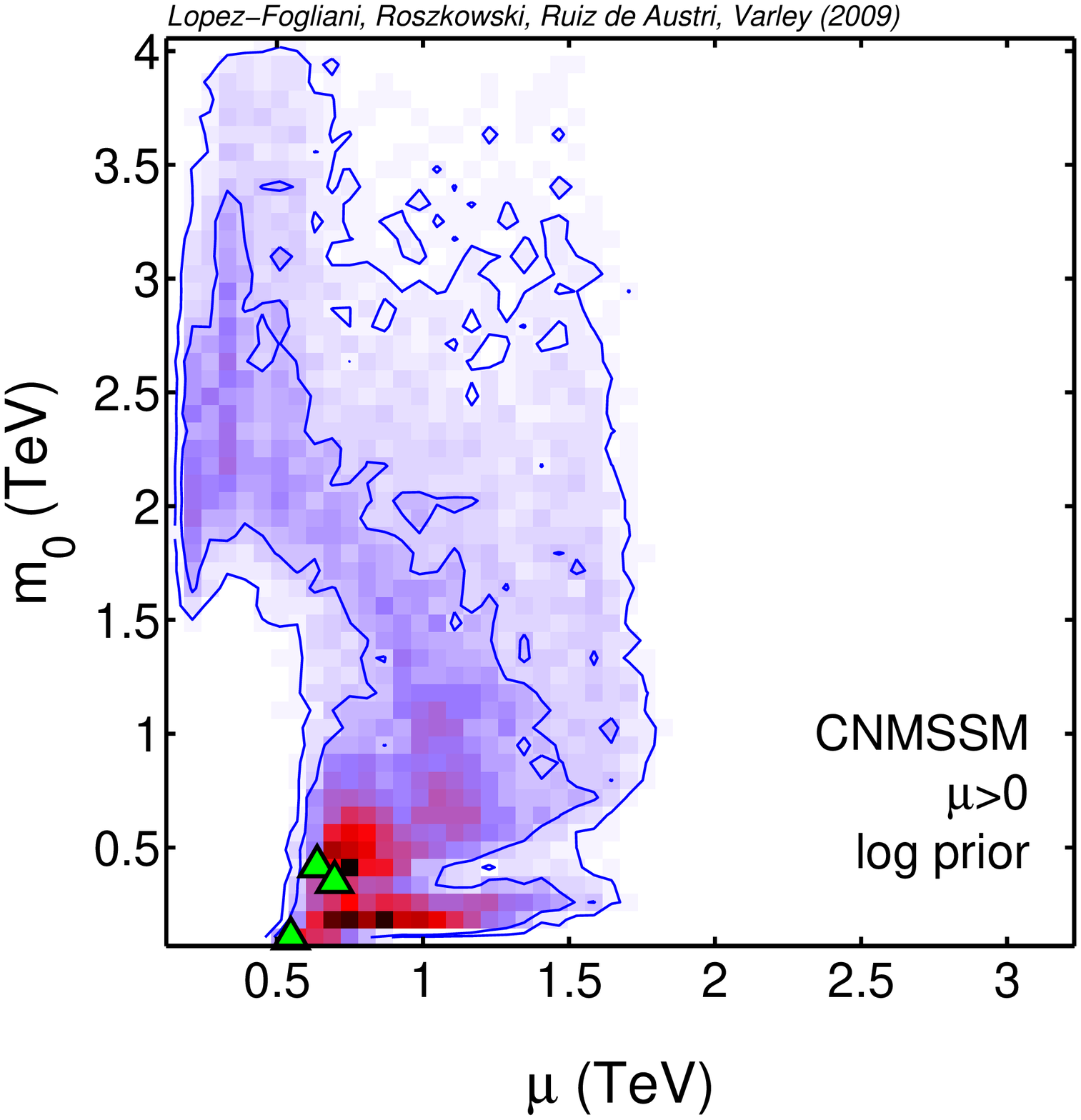}\\
      \includegraphics[width=0.35\textwidth]{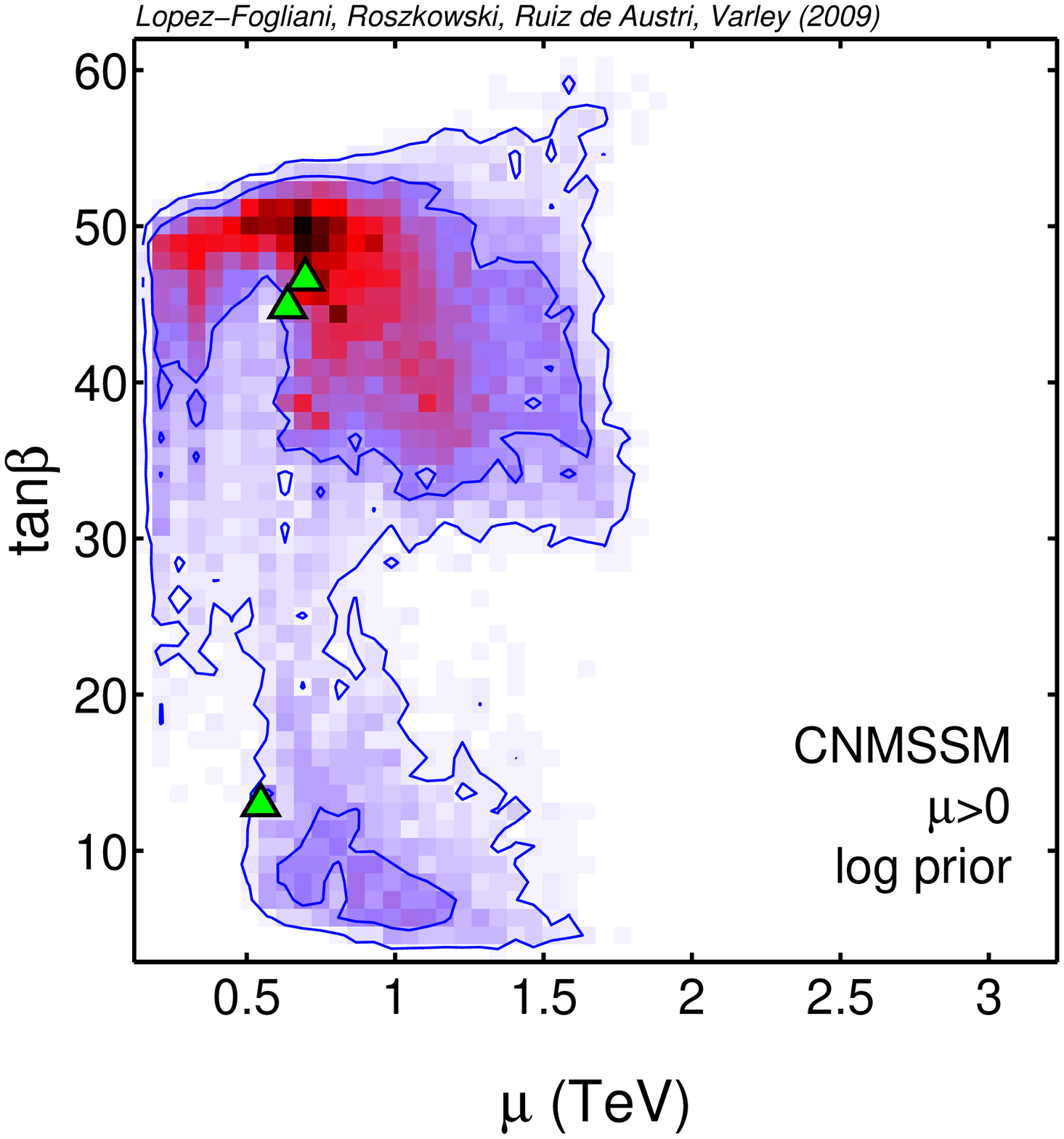}
     & \includegraphics[width=0.35\textwidth]{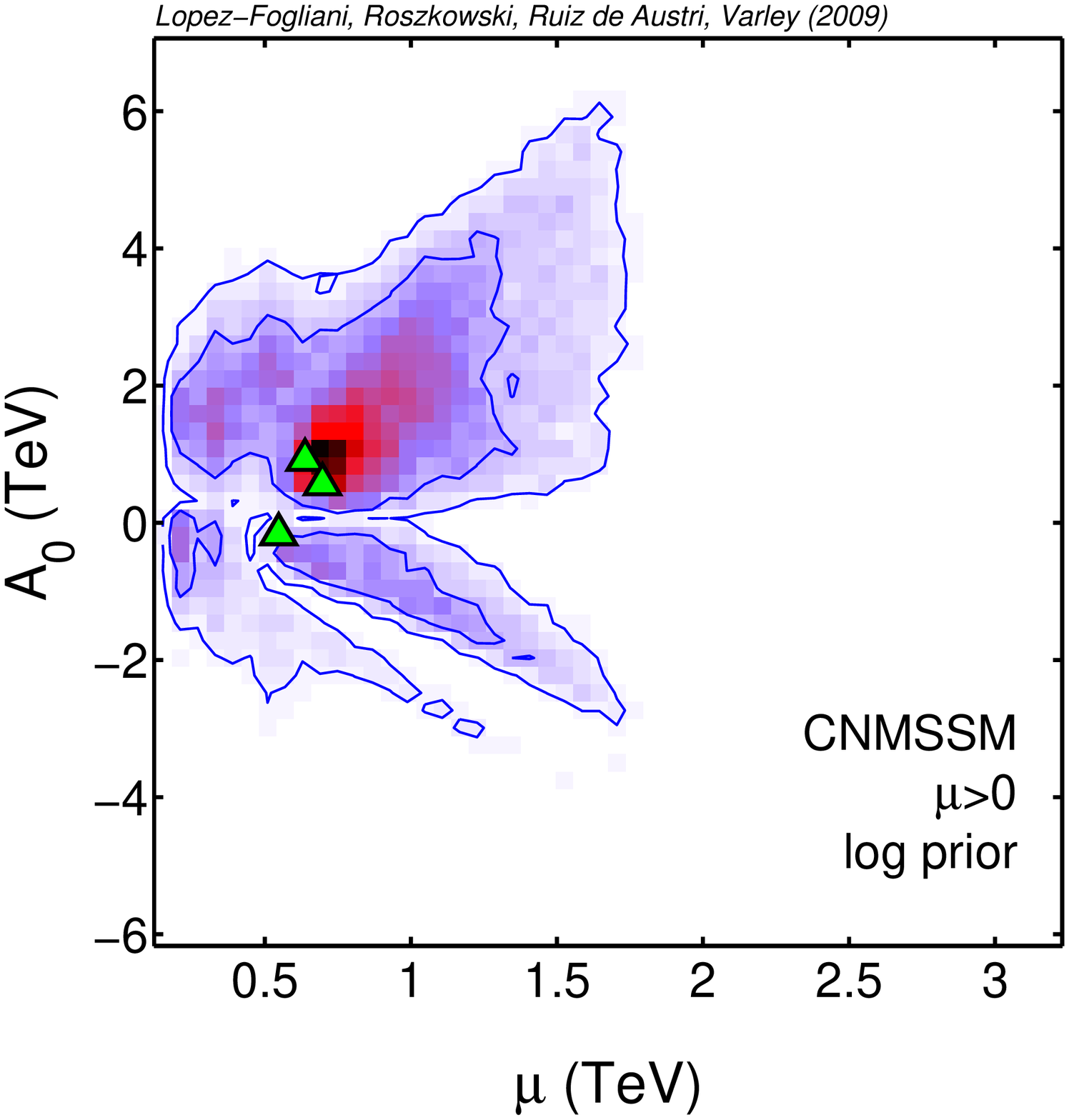}\\ 
   \end{tabular}
  \includegraphics[width=0.3\textwidth]{figures/colorbar.ps}
  \end{center}
\caption{\label{fig:CNMSSM_mu2dpdf_scan1} The 2D relative probability
    density functions in the planes spanned by $\mu$ and the CNMSSM parameters
    that are the same for the CMSSM for the log
    prior.  The pdfs are normalized to unity at their peak. The inner
    (outer) blue solid contours delimit regions encompassing 68\% and
    95\% of the total probability, respectively. All other basis
    parameters, both CNMSSM and SM ones, in each plane have been
    marginalized over. Blue dots denote some
    best fit points.}
  \end{figure}

 \begin{figure}[tbh!]
  \begin{center}
  \begin{tabular}{c c}
     \includegraphics[width=0.35\textwidth]{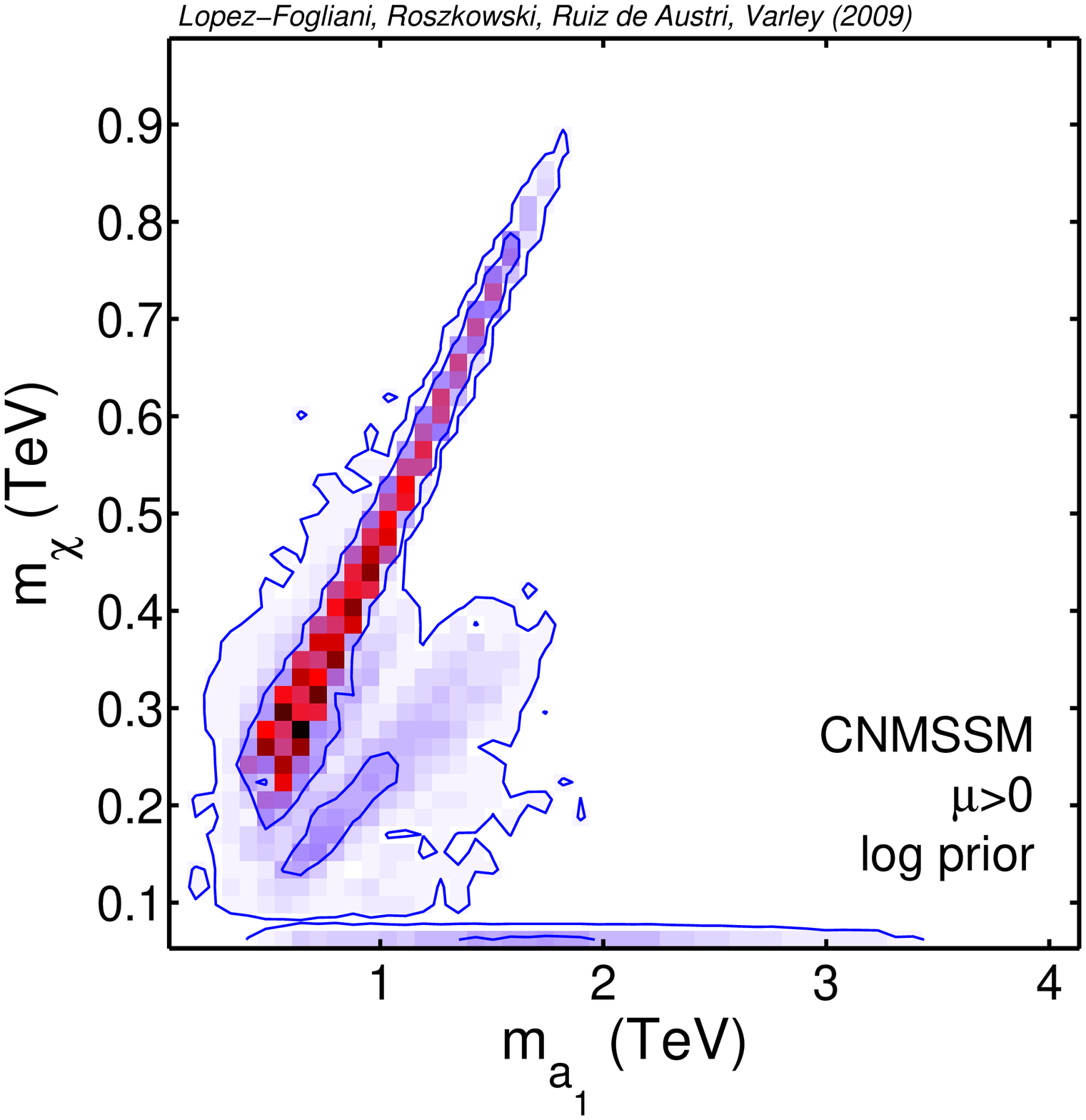}
    & \includegraphics[width=0.35\textwidth]{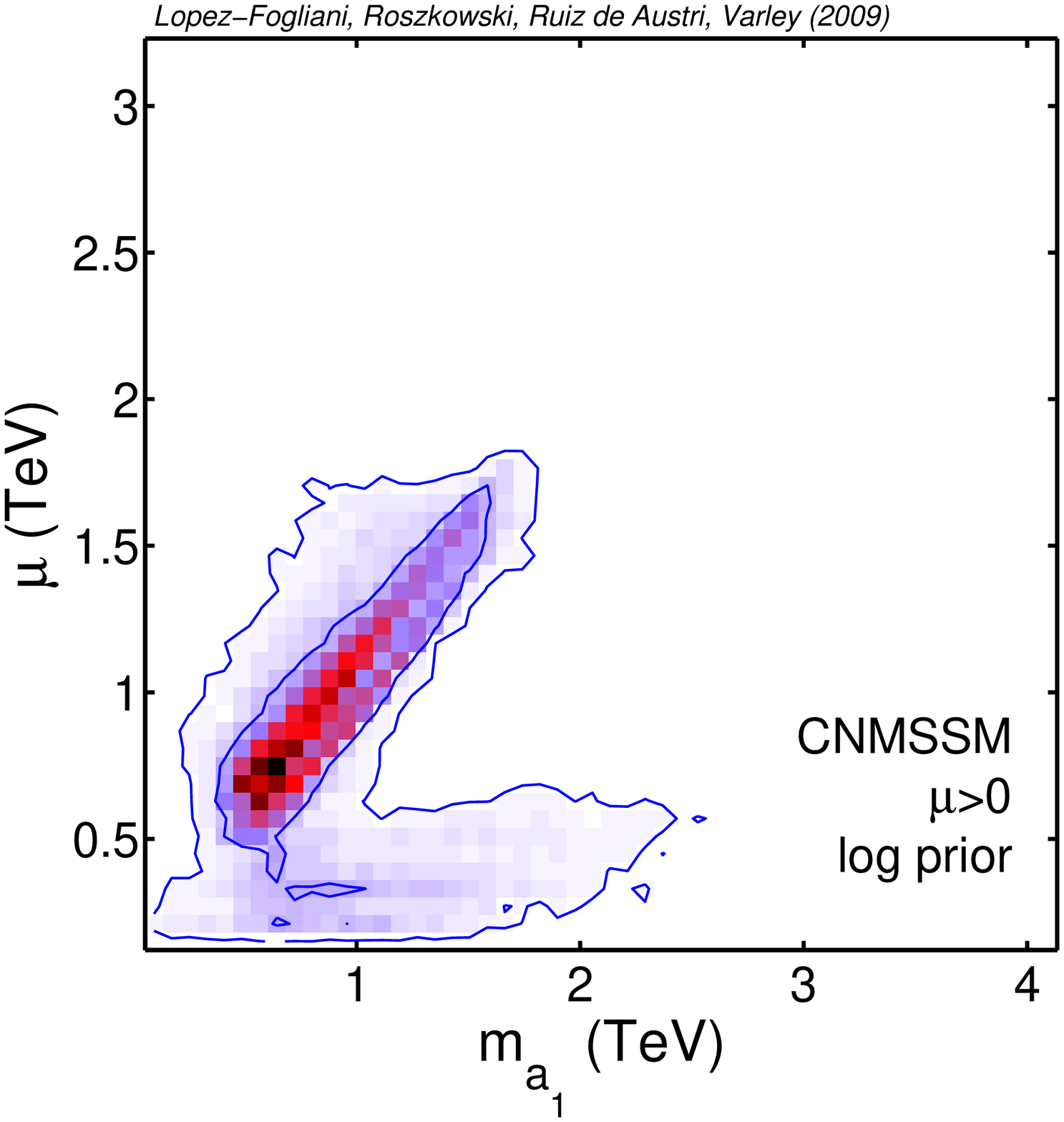}\\
   \end{tabular}
  \includegraphics[width=0.3\textwidth]{figures/colorbar.ps}
  \end{center}
\caption{\label{fig:CNMSSM_mumaone_scan1} The 2D relative probability
    density functions in the plane of $(\mchi,m_{a_1})$ (left panel)
    and $(\mu,m_{a_1})$ (right panel) for the log
    prior.  The pdfs are normalized to unity at their peak. The inner
    (outer) blue solid contours delimit regions encompassing 68\% and
    95\% of the total probability, respectively. All other basis
    parameters, both CNMSSM and SM ones, in each plane have been
    marginalized over. }
  \end{figure}

\begin{figure}[tbh!]
\begin{center}
\begin{tabular}{c c c c}
\includegraphics[width=0.245\textwidth]{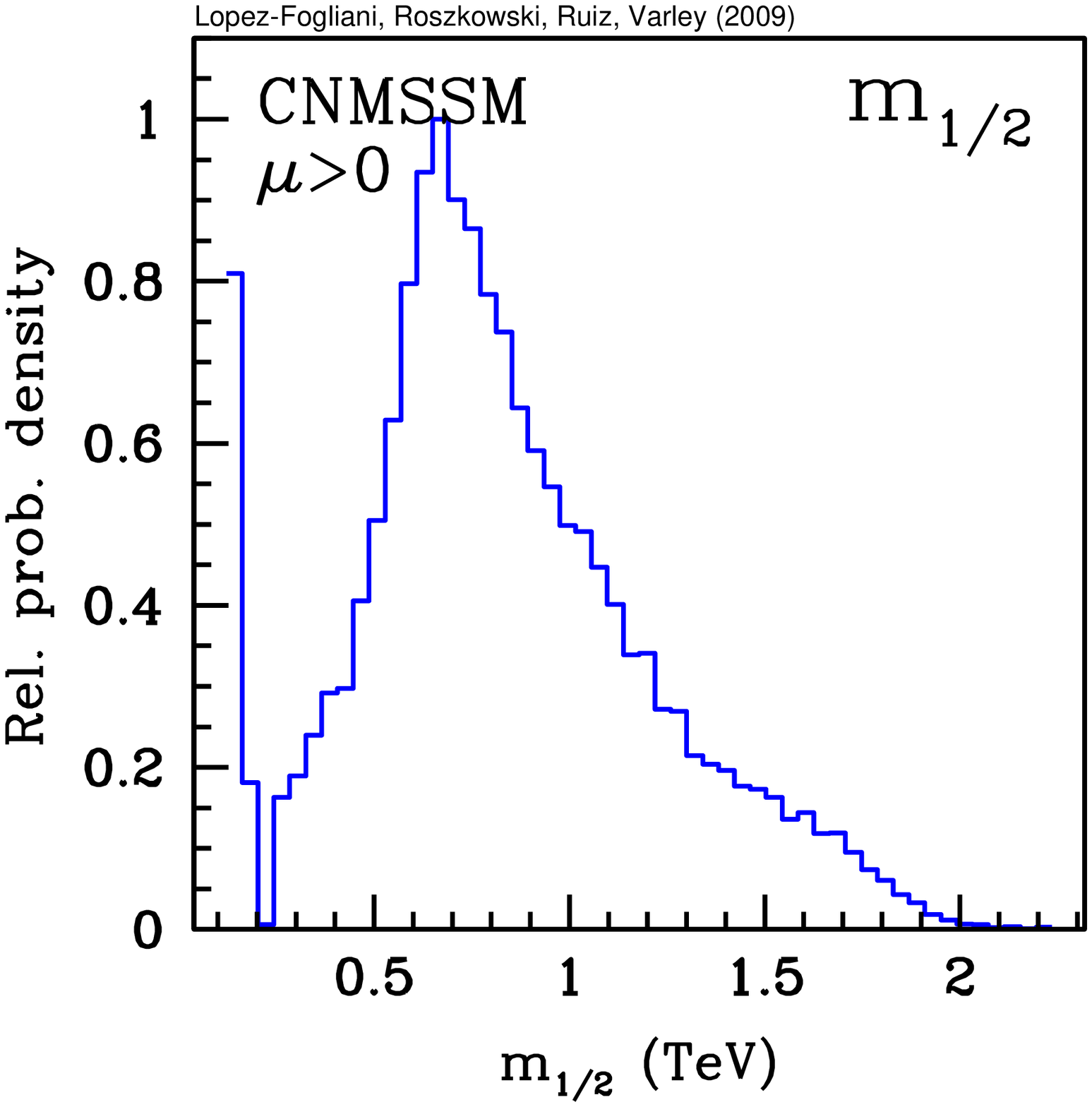}&
\includegraphics[width=0.245\textwidth]{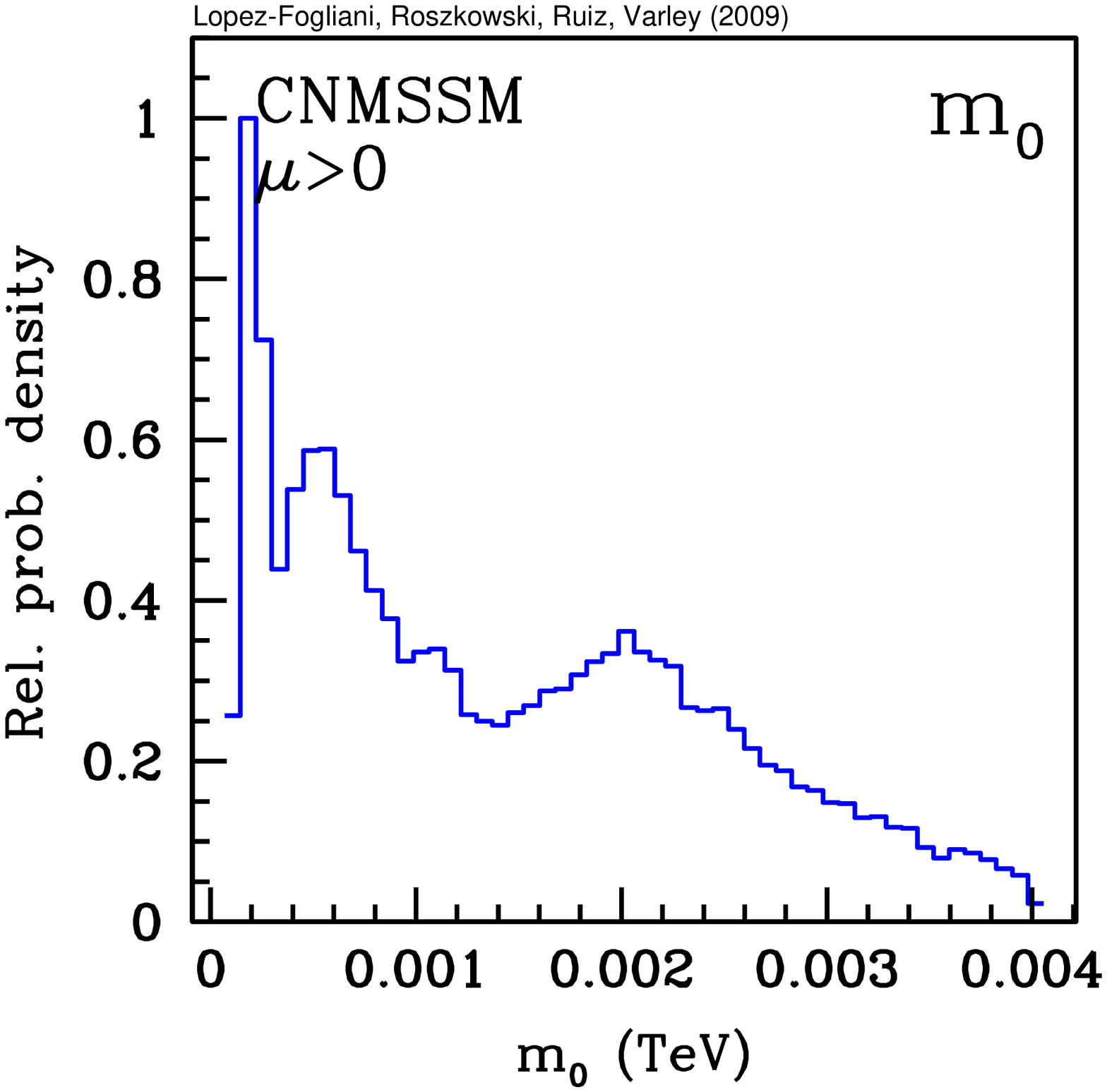}&
\includegraphics[width=0.245\textwidth]{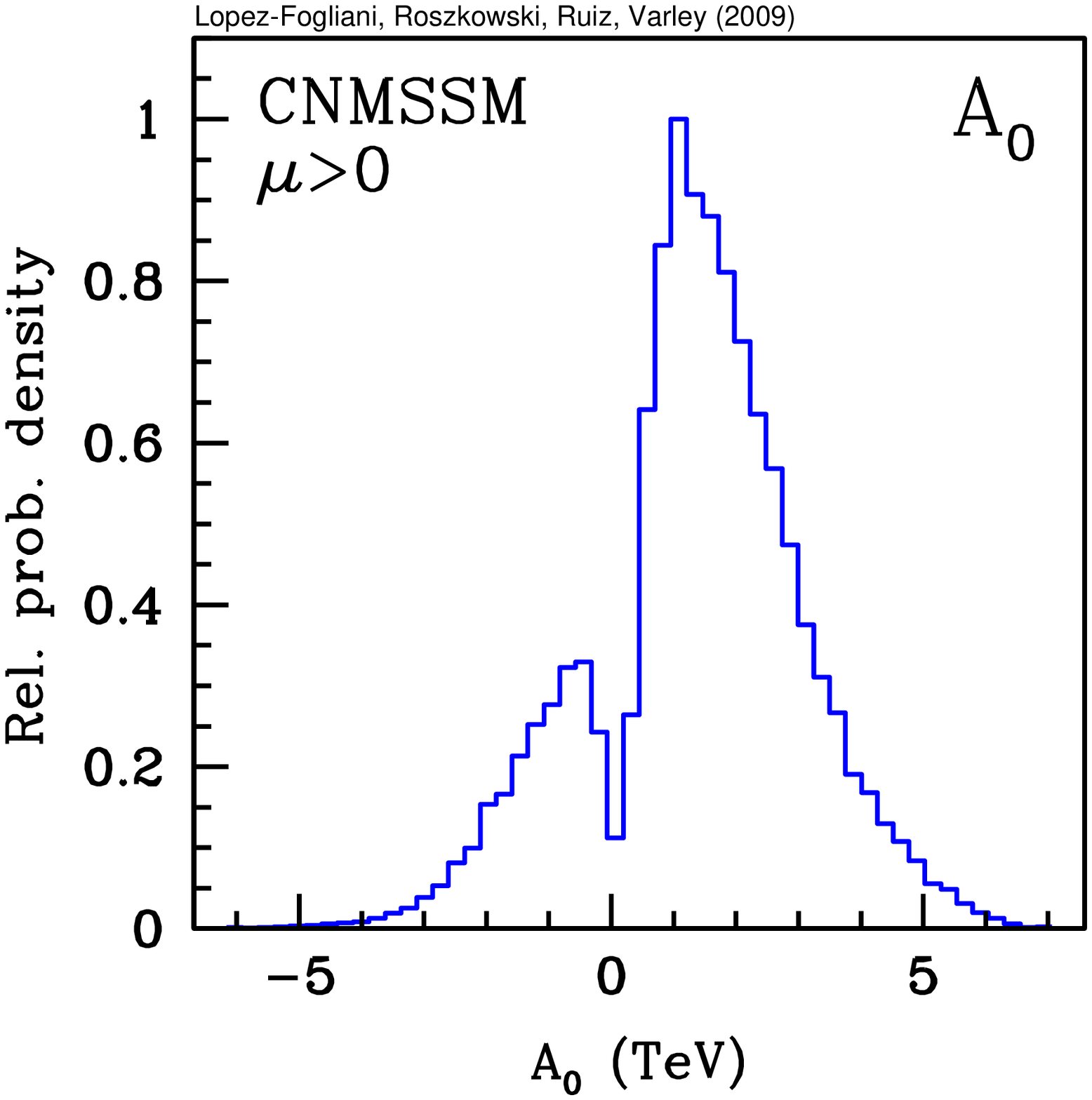}&
\includegraphics[width=0.245\textwidth]{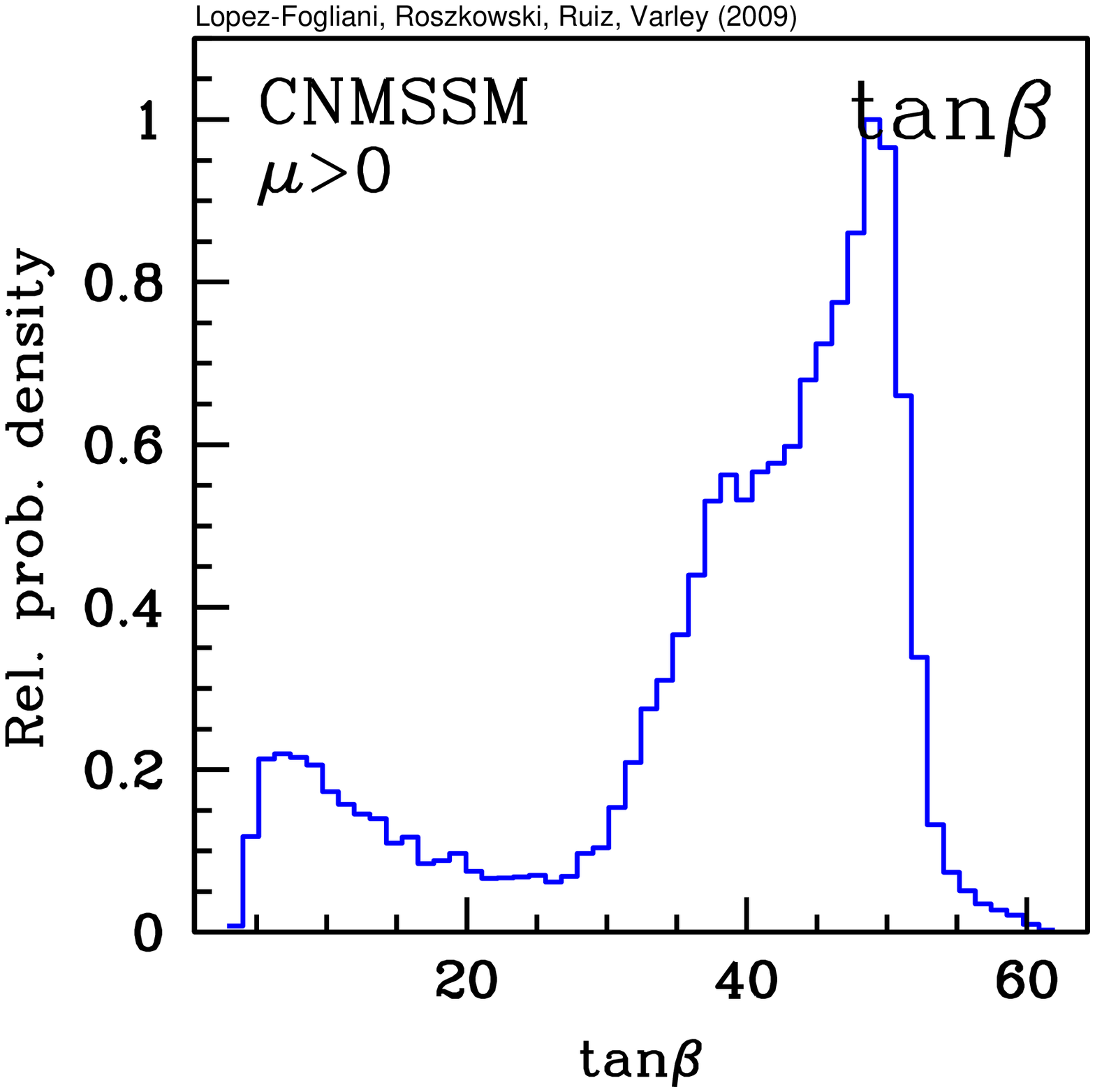}\\
\includegraphics[width=0.245\textwidth]{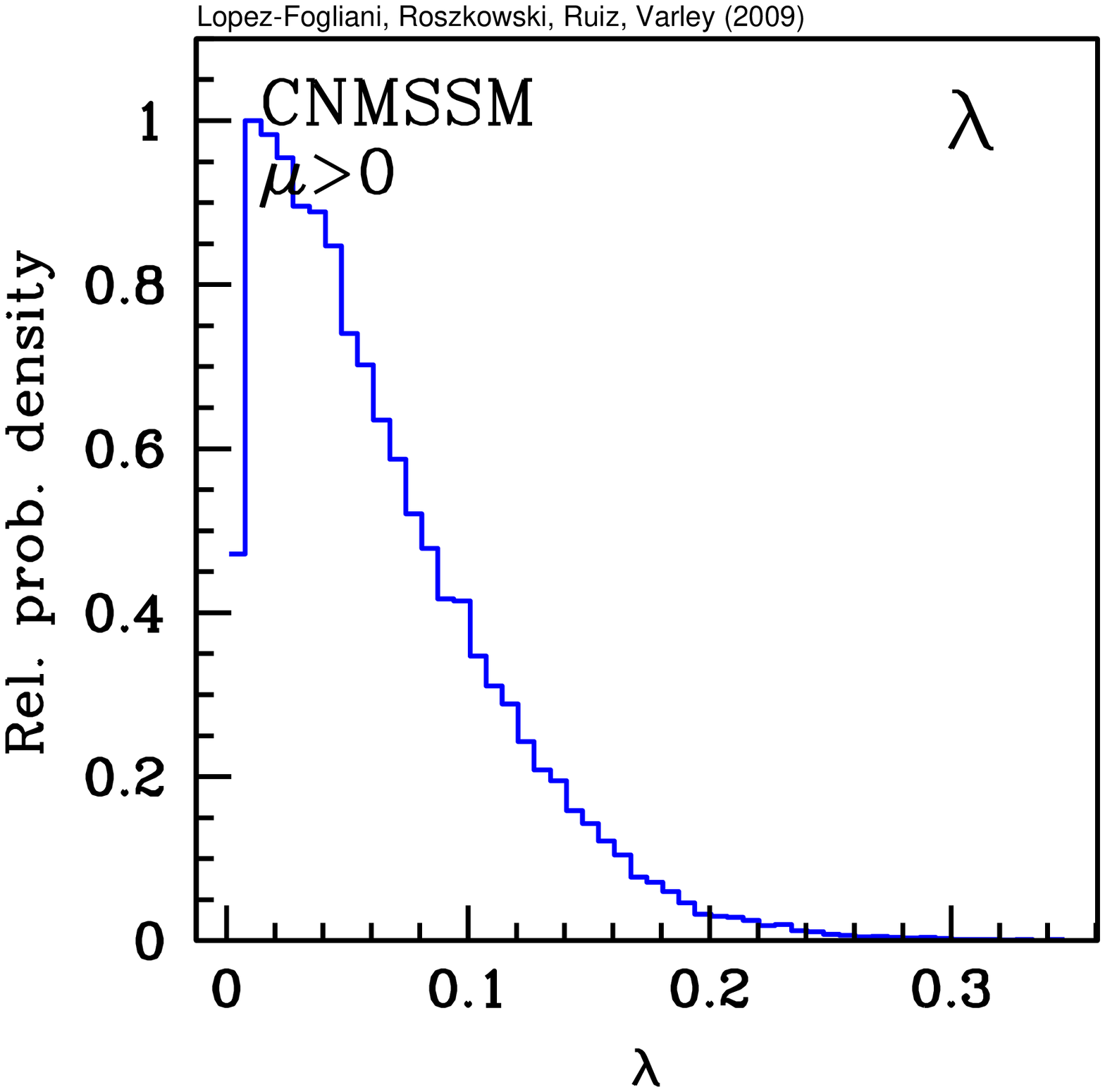}&
\includegraphics[width=0.245\textwidth]{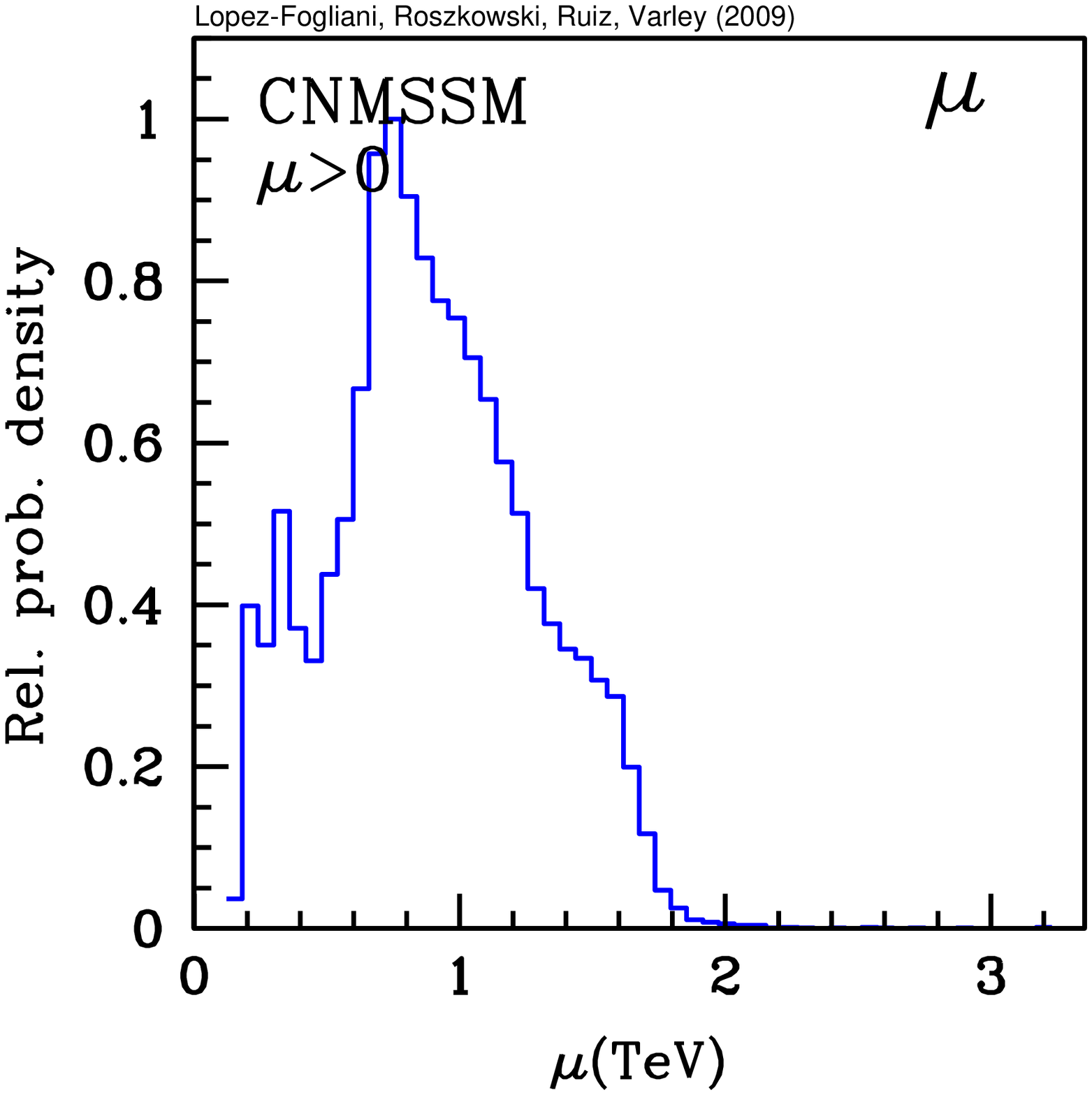}&
\includegraphics[width=0.245\textwidth]{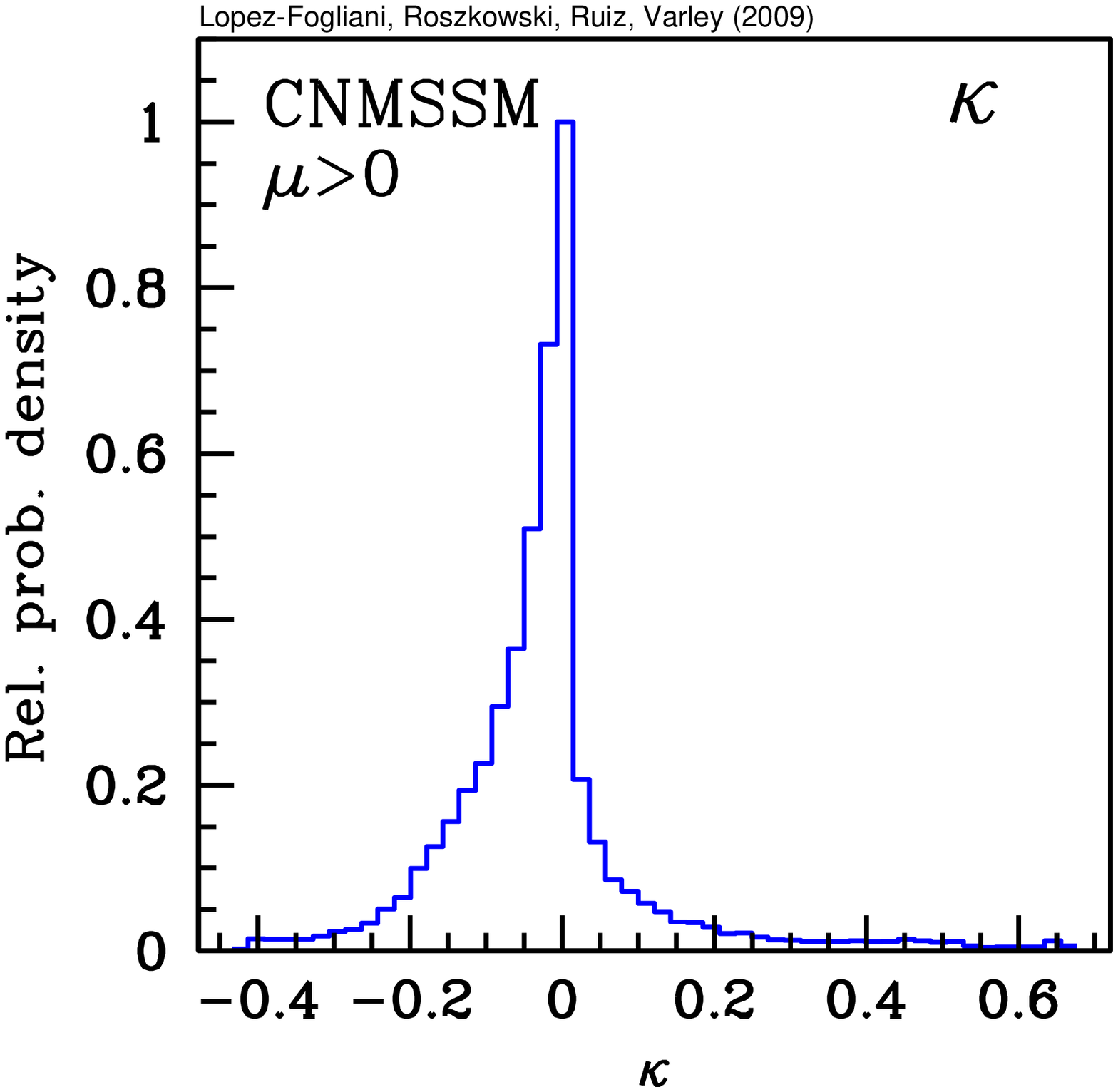}&
\includegraphics[width=0.245\textwidth]{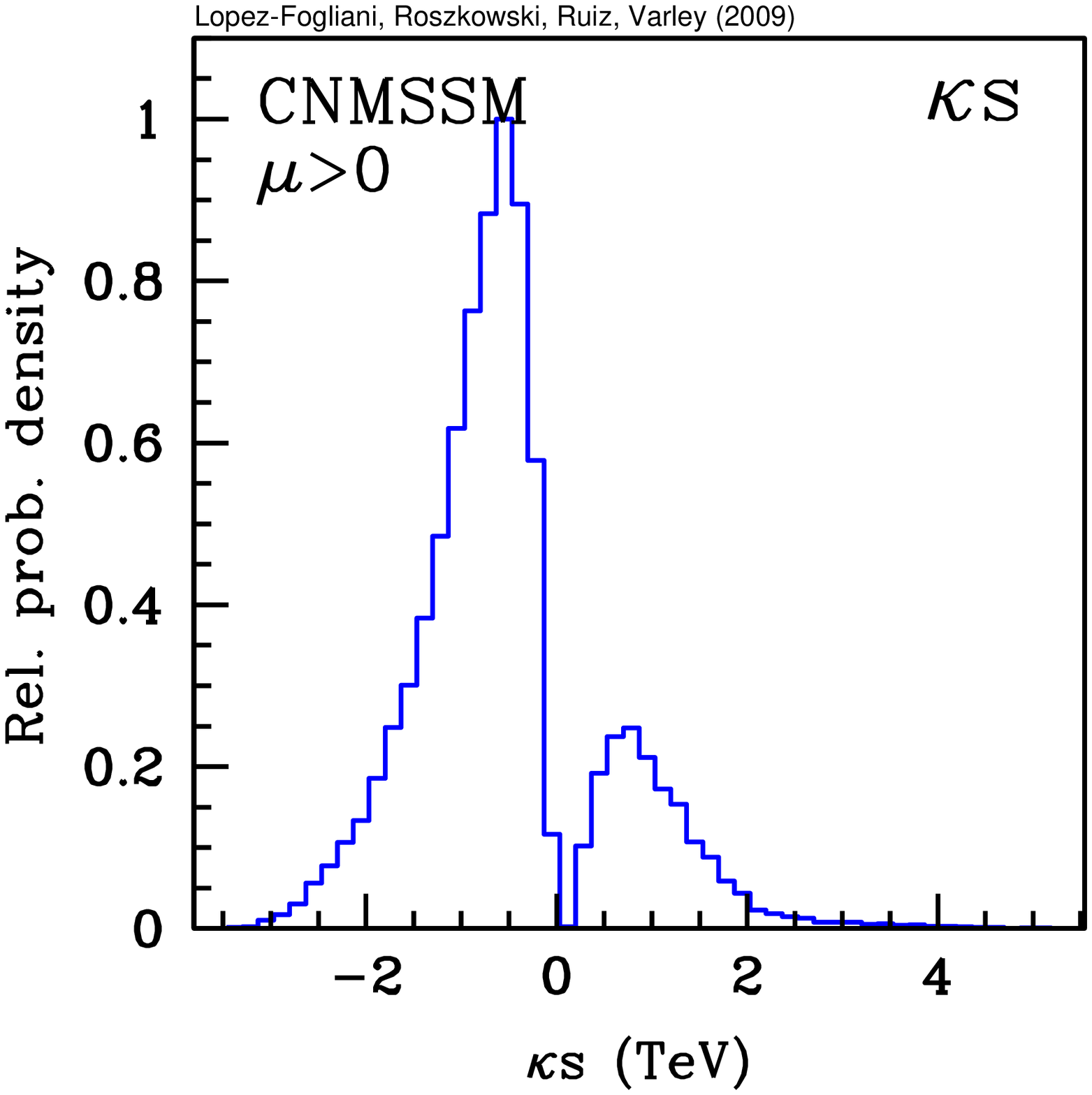}\\
\end{tabular}
\end{center}
\caption{\label{fig:CNMSSMps1dpdf_scan1} 
The 1D relative
probability densities for all the CNMSSM parameters, plus $\kappa$,
$\mu$, $\lambda s$ and $\kappa s$. }
\end{figure}

It is also instructive to show 2D probability maps of the effective $\mu$
parameter \vs\ some of the CNMSSM parameters. This is presented in
Fig.~\ref{fig:CNMSSM_mu2dpdf_scan1}. We can see that the consistency
of the model and the applied set of constraints favor $\mu$ below some
$2\tev$, the range comparable to $\msusy$, as expected. In other
words, in the CNMSSM the $\mu$ problem is solved without any need for
additional fine tuning of parameters. We can also see an interesting
correlation with $\mhalf$ but not with the other
C(N)MSSM parameters. This is caused primarily by the CP-odd Higgs $a_1$
funnel effect and the fact that its mass is correlated with
$\mu$. These features are presented in Fig.~\ref{fig:CNMSSM_mumaone_scan1}.
Finally, in Fig.~\ref{fig:CNMSSMps1dpdf_scan1} we present 1D pdfs of several
key parameters which show more clearly their high probability ranges.

\begin{figure}[tbh!]
 \begin{center}
  \begin{tabular}{c c}
  \includegraphics[width=0.35\textwidth]{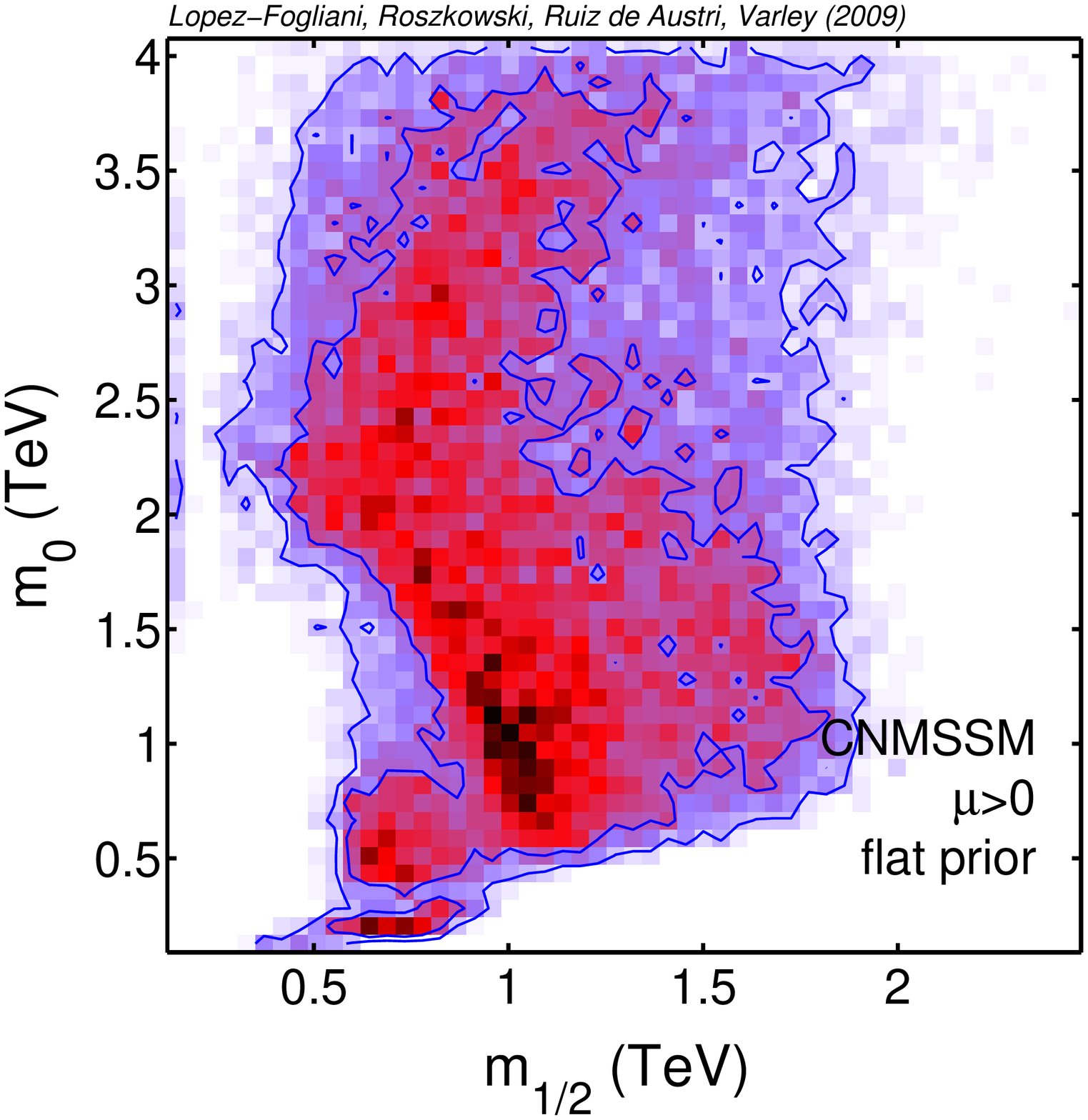}
   & \includegraphics[width=0.35\textwidth]{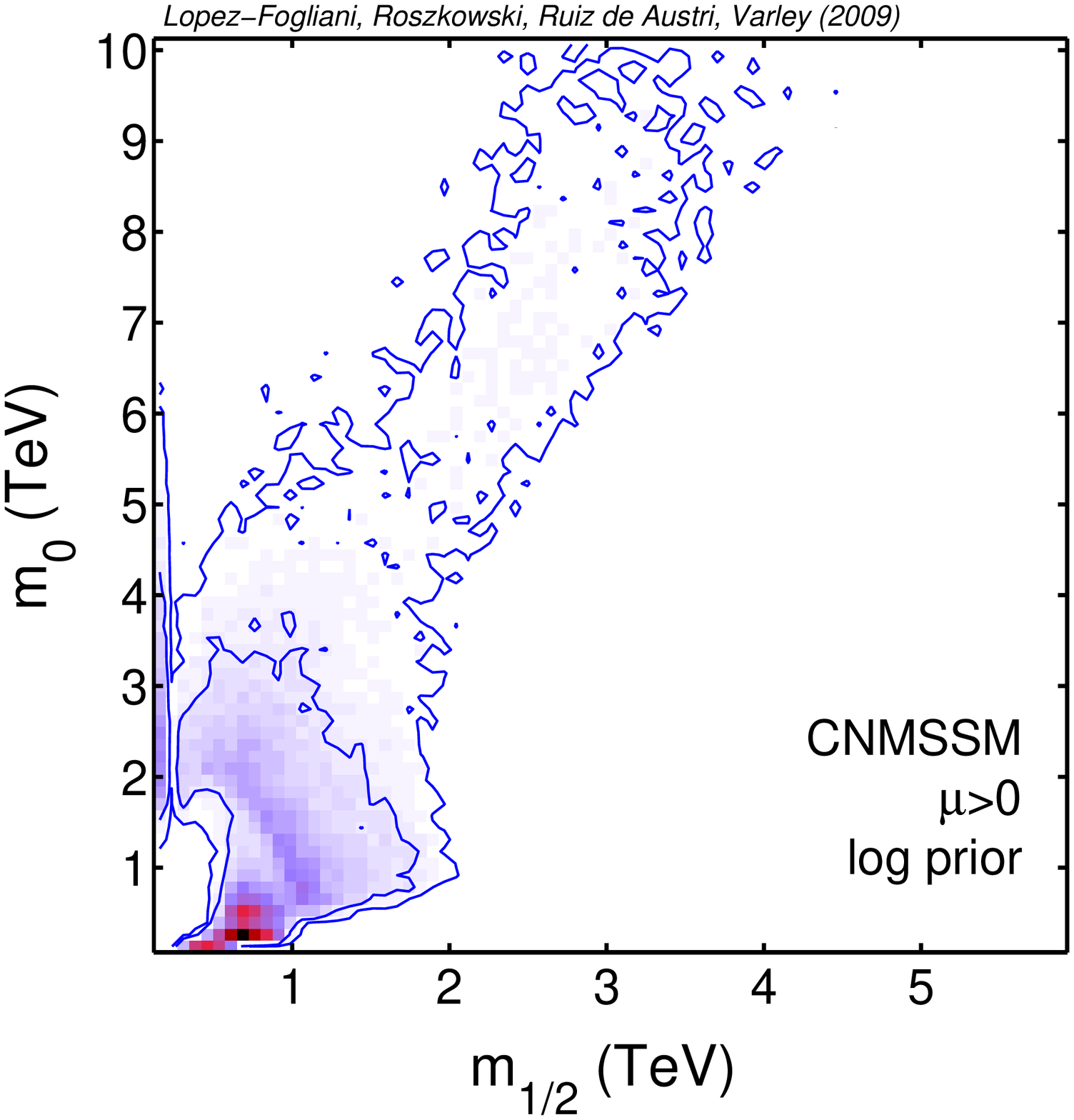}\\
   \end{tabular}
  \end{center}
 \caption{ \label{fig:CNMSSMps2dpdf_comp} The same as in
   fig.~\protect\ref{fig:CNMSSMps2dpdf_scan1} but for a flat prior
   (left panel) and a log prior with a greatly extended range,
   $50\gev<\mhalf,\mzero<10\tev$ (right panel).  }
 \end{figure}

Prior dependence is often an issue in Bayesian statistics and needs to
be addressed. Even the CMSSM tends to be under-constrained which leads
to a fairly strong prior dependence~\cite{tfhrr1}, and in the NUHM,
with two extra parameters, the situation becomes
worse~\cite{nuhm1}. In the left panel of
Fig.~\ref{fig:CNMSSMps2dpdf_comp} we show a 2D pdf in the
($\mhalf,\mzero$) plane assuming the flat prior in all the CNMSSM
parameters. By comparing with the analogous panel of
Fig.~\ref{fig:CNMSSMps2dpdf_scan1} we indeed see a substantial shift
in the high probability region to larger values, as typical for the
flat prior due to the volume effect.

A related issues is that of the assumed range of input parameters. We
have already seen in the left panel of
Fig.~\ref{fig:CNMSSMps2dpdf_comp} that $\mzero$ was not well confined
to the assumed range below $4\tev$. In order to examine this, in the
right panel of Fig.~\ref{fig:CNMSSMps2dpdf_comp} we show a 2D pdf in
the ($\mhalf,\mzero$) plane with greatly extended ranges of both
parameters ($50\gev<\mhalf,\mzero<10\tev$) and taking the log
prior. Clearly, there is basically no cap on the 95\% total
probability range in both parameters, although of course such large
values of soft mass parameters can hardly be considered as well
motivated in effective low-energy SUSY models.

\begin{figure}[tbh!]
\begin{center}
\begin{tabular}{c c}
 \includegraphics[width=0.35\textwidth]{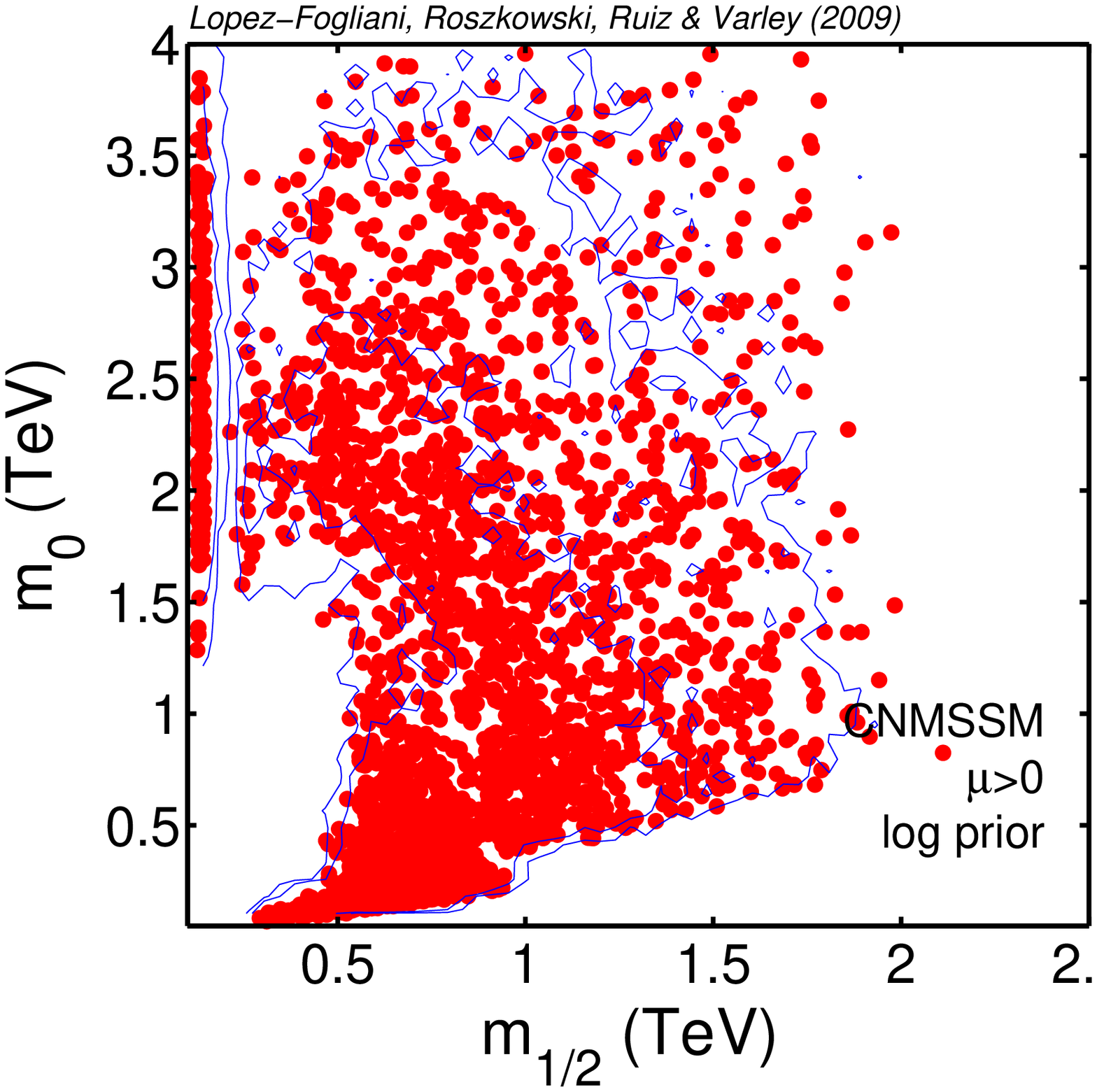}
    & \includegraphics[width=0.35\textwidth]{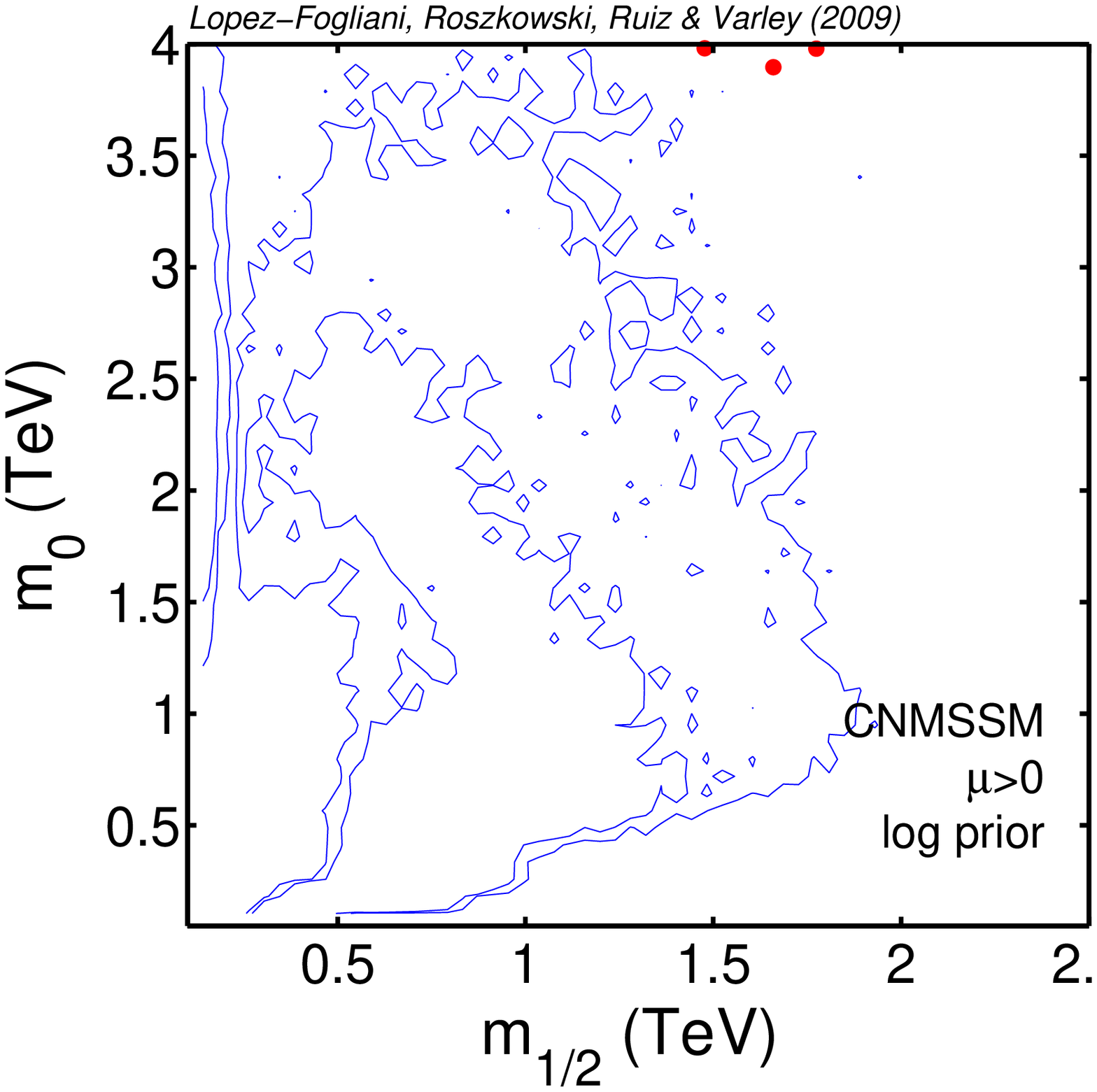}\\
     \includegraphics[width=0.35\textwidth]{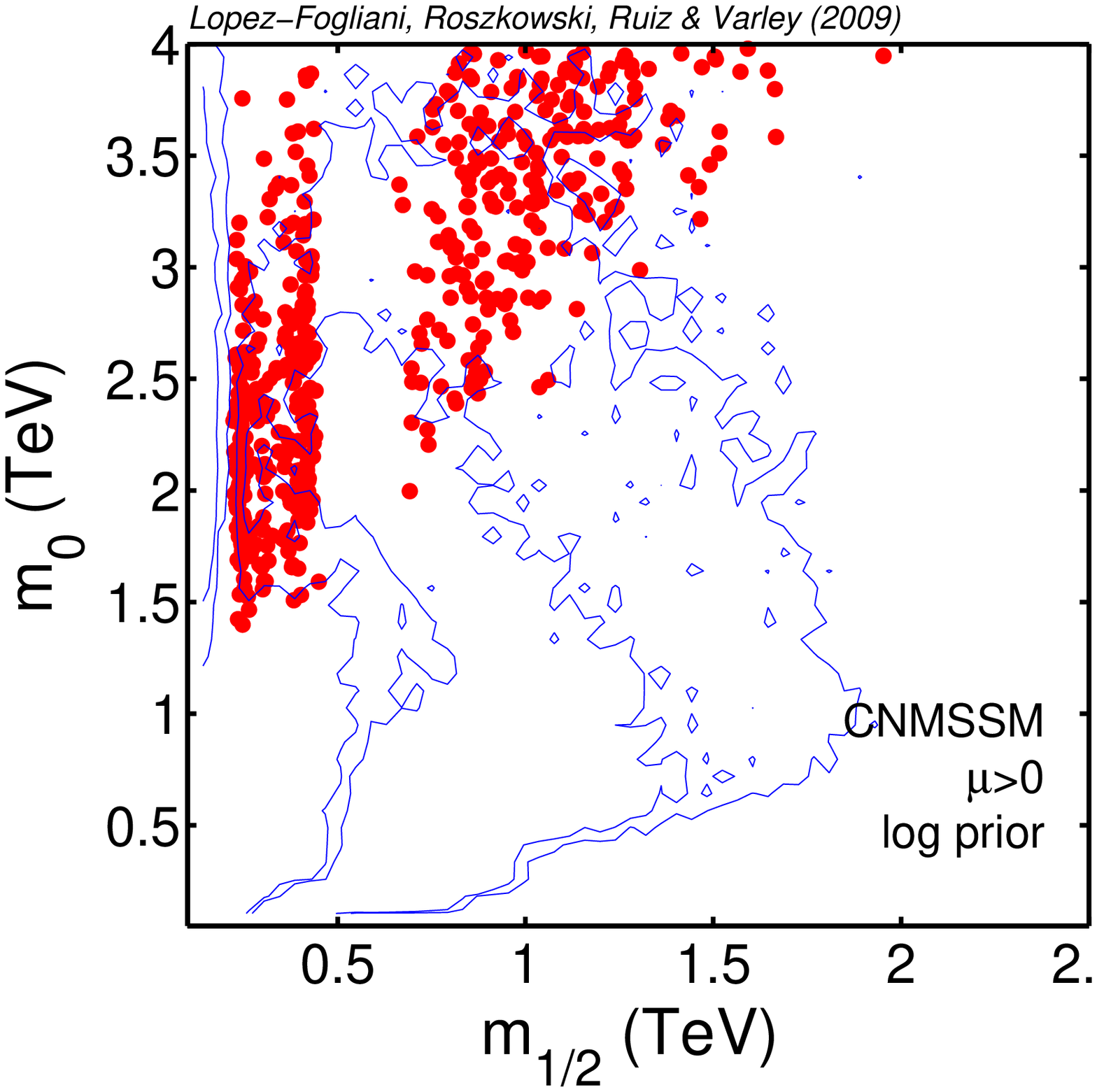} 
    & \includegraphics[width=0.35\textwidth]{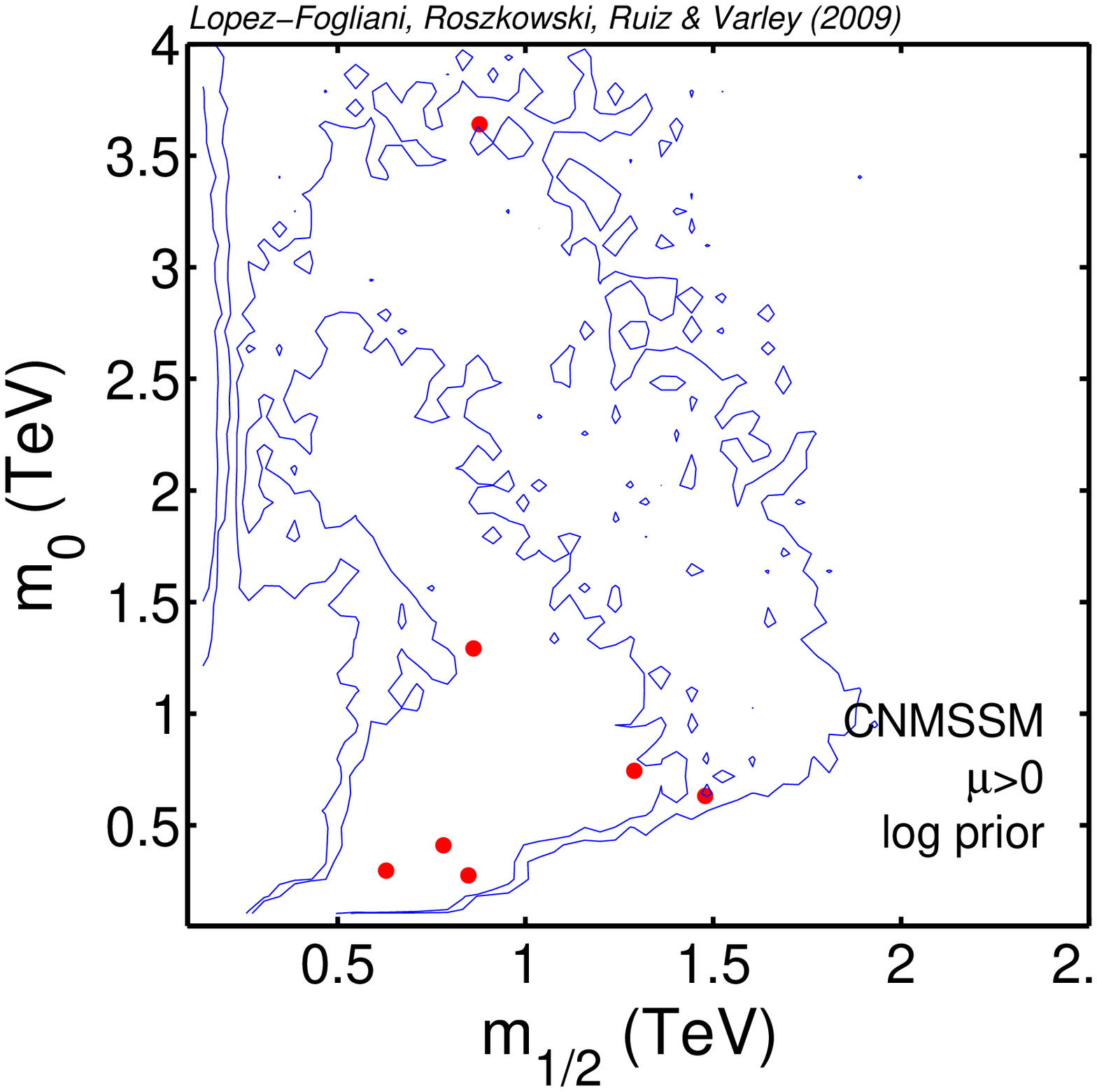}\\
\end{tabular}
\end{center}
\caption{\label{fig:CNMSSM3D_gaugino_scan1} In the plane of
  $(\mhalf,\mzero)$ for the log prior we show values of the gaugino
  fraction $Z_g=Z^2_{11}+Z^2_{12}$ (upper left panel), doublet higgsino
  fraction $Z_h=Z^2_{13}+Z^2_{14}$ (upper right panel), mixed region
  (lower left panel), as well as the singlino fraction
  $Z_s=Z^2_{15}$ (lower right panel). 
Also shown are the 69\% (95\%)
  total probability regions from the upper left panel of
  Fig.~\protect{\ref{fig:CNMSSMps2dpdf_scan1}}. }
\end{figure}

\begin{figure}[tbh!]
 \begin{center}
  \includegraphics[width=0.35\textwidth]{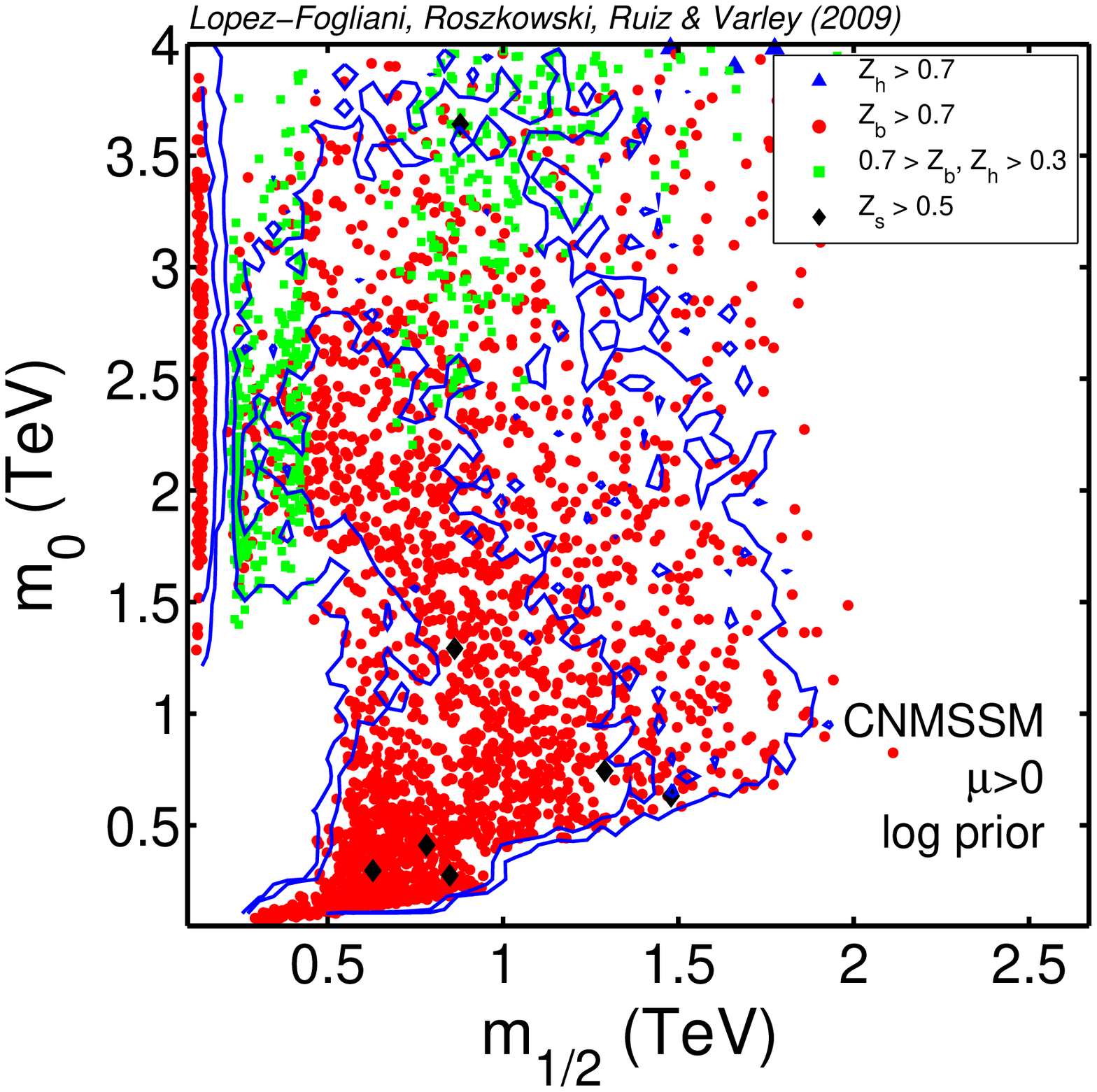}
 \includegraphics[width=0.35\textwidth]{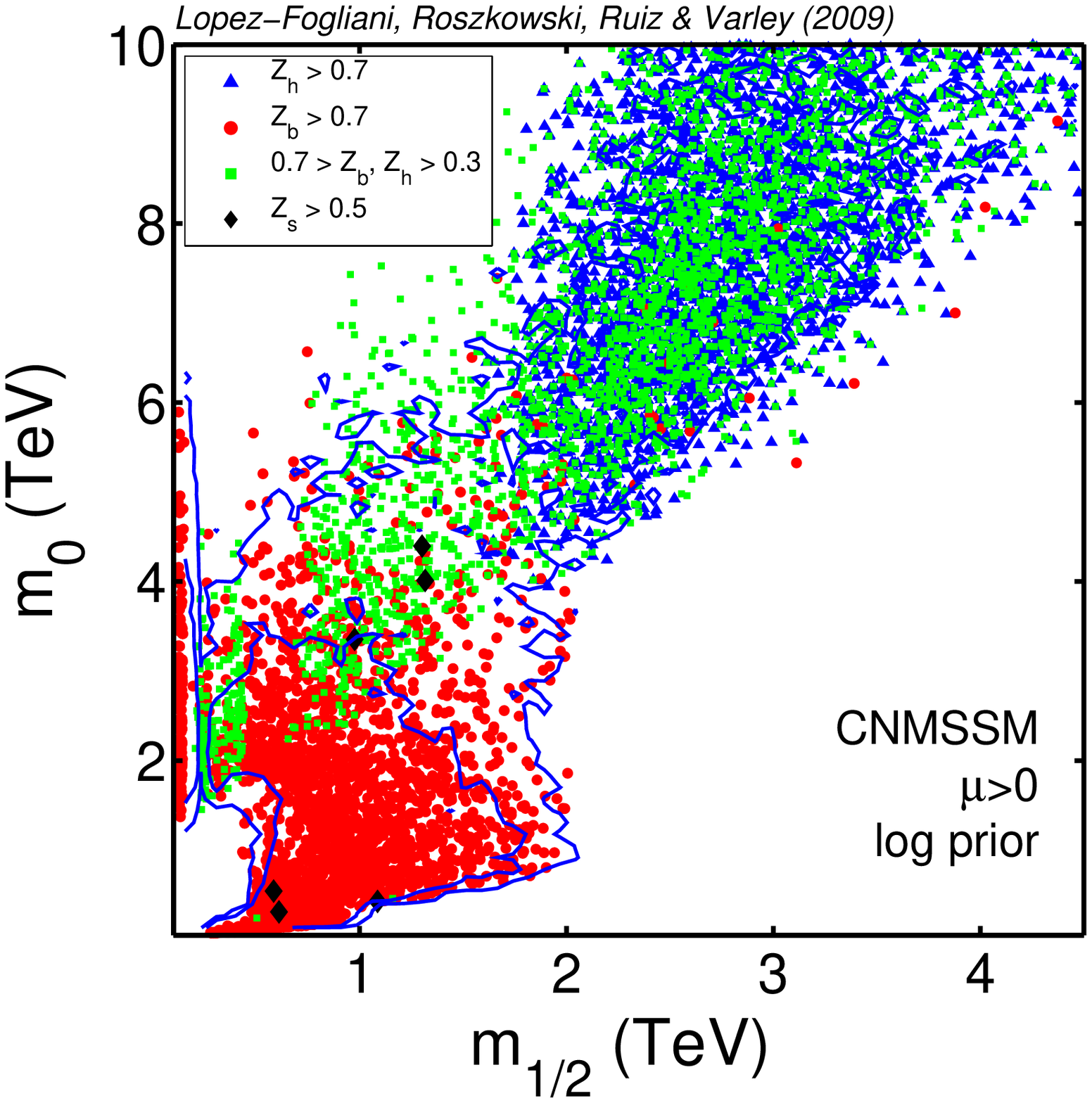}
  \end{center}
 \caption{ \label{fig:CNMSSMps2dpdf_comp2} A comparison of the bino ($Z_b>0.7$),
 doublet higgsino ($Z_h>0.7$), mixed ($0.3<Z_g,Z_h<0.7$ and $Z_s<0.5$) and singlino
 dominated ($Z_s>0.5$)  cases of the LSP in the plane spanned by $\mhalf$ and $\mzero$,
with our default case on the left hand side and with
 the log prior and much extended range of $10\tev$ on the right. Also
 shown are the 69\% (95\%)   total probability regions.} 
 \end{figure}

We have already emphasized that the high probability regions of the
crucial parameters $\mhalf$ and $\mzero$ in the CNMSSM are quite
similar to the well studied CMSSM case. One of the key features of the
CMSSM is that the neutralino LSP is mostly a bino, except for the FP
region of large $\mzero$, where a larger admixture of the higgsino is
present. In the CNMSSM, with an additional singlet field whose mass is
controlled by $\kappa s$, the picture could in principle be very
different from the CMSSM.  We examine this in
Fig.~\ref{fig:CNMSSM3D_gaugino_scan1} where we separately show the
regions where the LSP is mostly gaugino ($Z_g=Z^2_{11}+Z^2_{12}>0.7$)
(left panel), doublet higgsino ($Z_h=Z^2_{13}+Z^2_{14}>0.7$) (middle
left panel), the mixed region ($0.3<Z_g,Z_h<0.7$ and
$Z_s=Z^2_{15}<0.5$) (middle right panel), as well as mostly singlino
($Z_s>0.5$) (right panel). (The wino component is always negligible
and below we will show only the bino fraction $Z_b=Z^2_{11}$.) We can
see that in an overwhelming fraction of cases the LSP still remains
predominantly bino-like. This is in agreement with the panel showing
$\kappa s$ in Fig.~\ref{fig:CNMSSMps1dpdf_scan1} where small values of
the product tend to be strongly disfavored.  We expose those non-CMSSM
like cases in Fig.~\ref{fig:CNMSSMps2dpdf_comp2}. Clearly, points
corresponding to singlino LSP cases exist but are rare.

\begin{figure}[tbh!]
\begin{center}
\begin{tabular}{c c}
\includegraphics[width=0.35\textwidth]{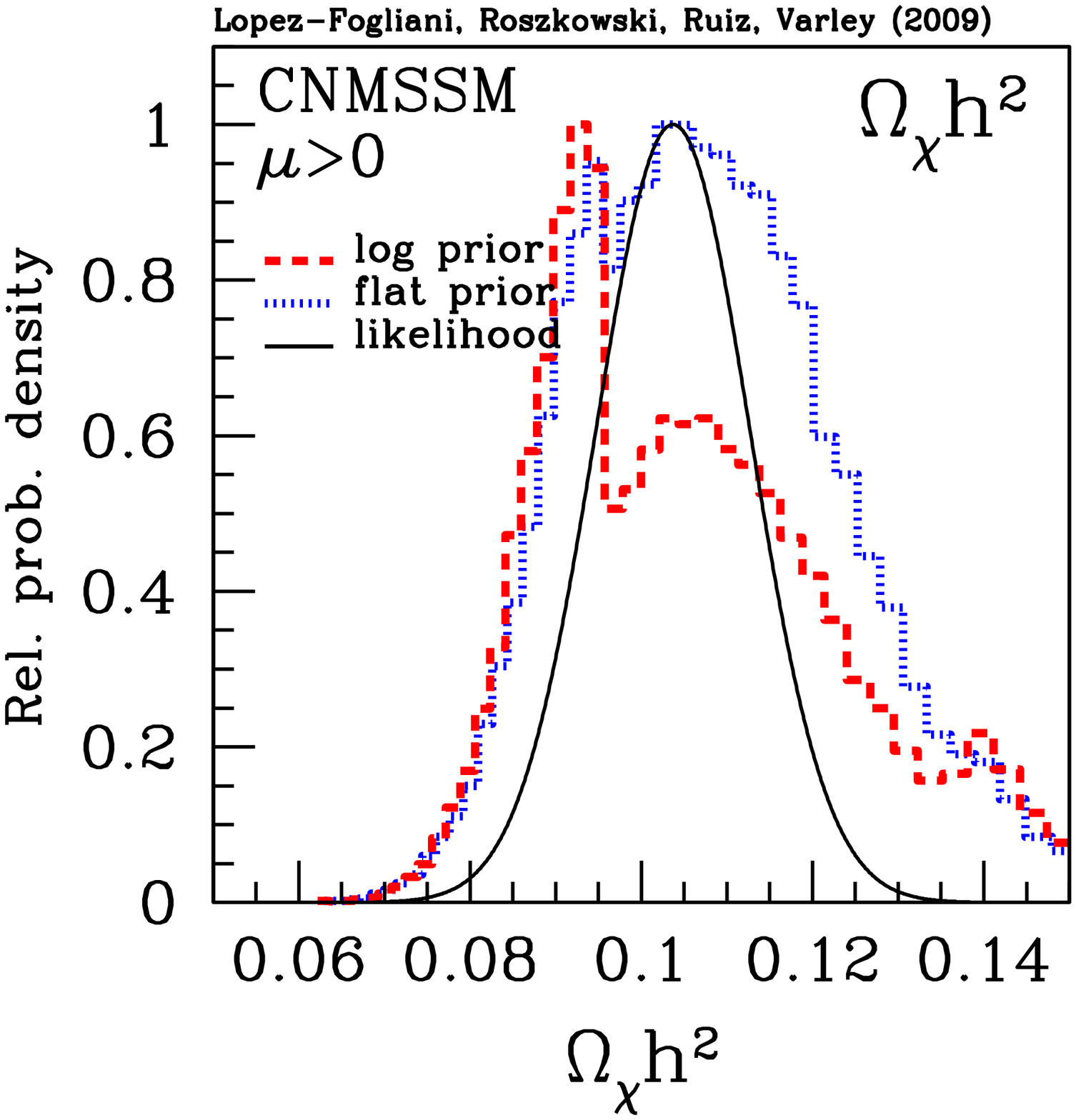} &
\includegraphics[width=0.35\textwidth]{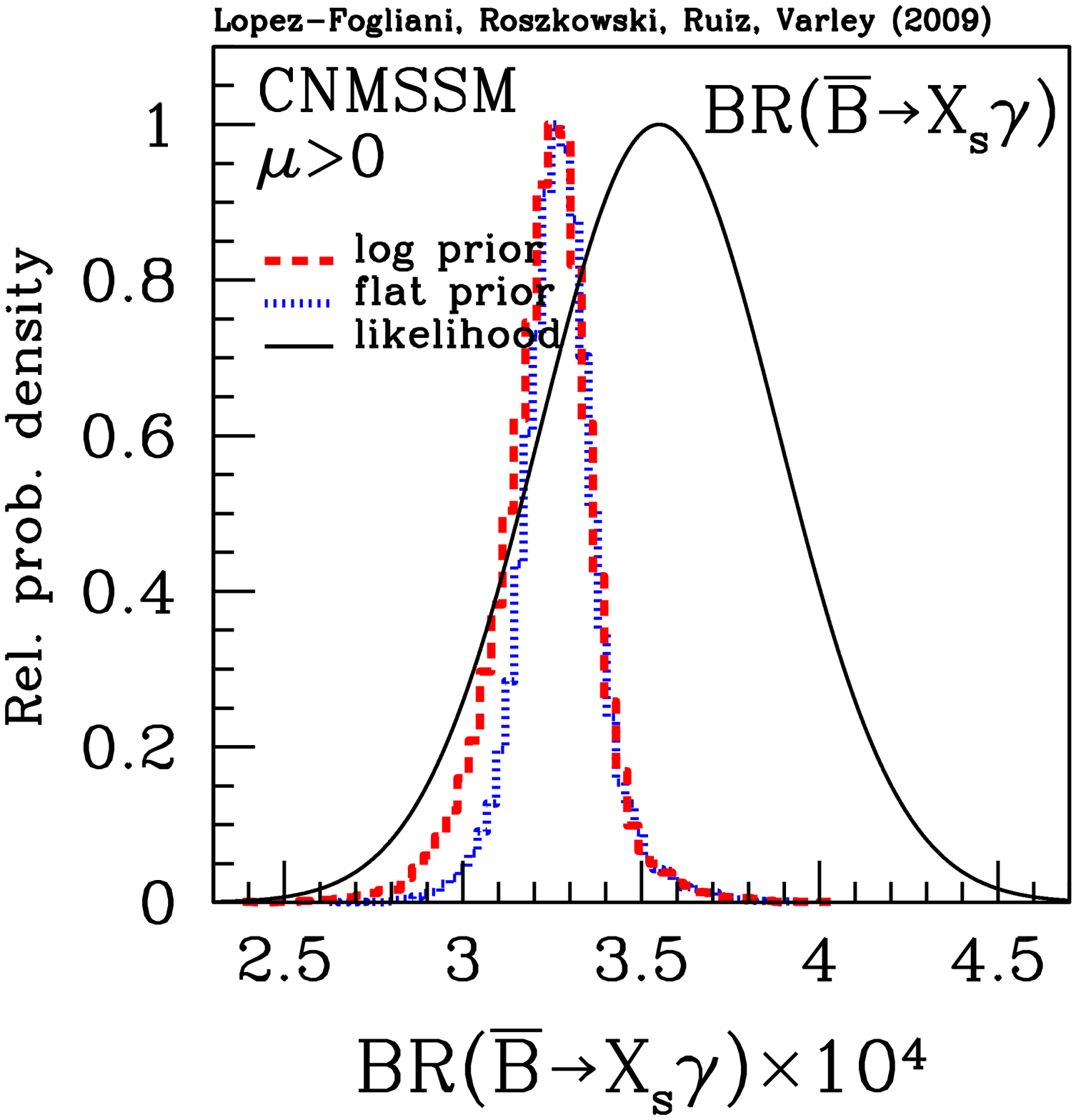}\\
\includegraphics[width=0.35\textwidth]{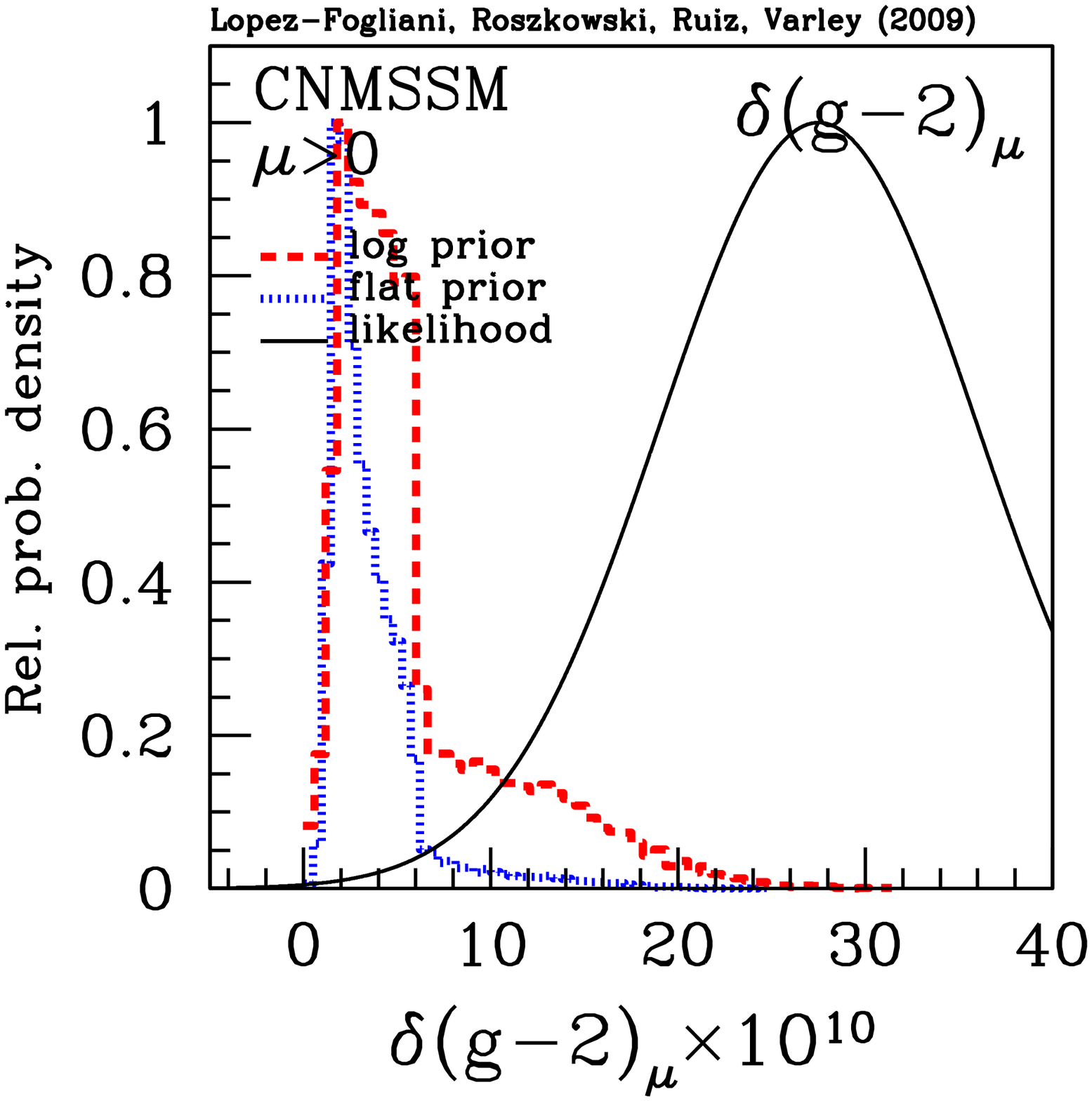} &
\includegraphics[width=0.35\textwidth]{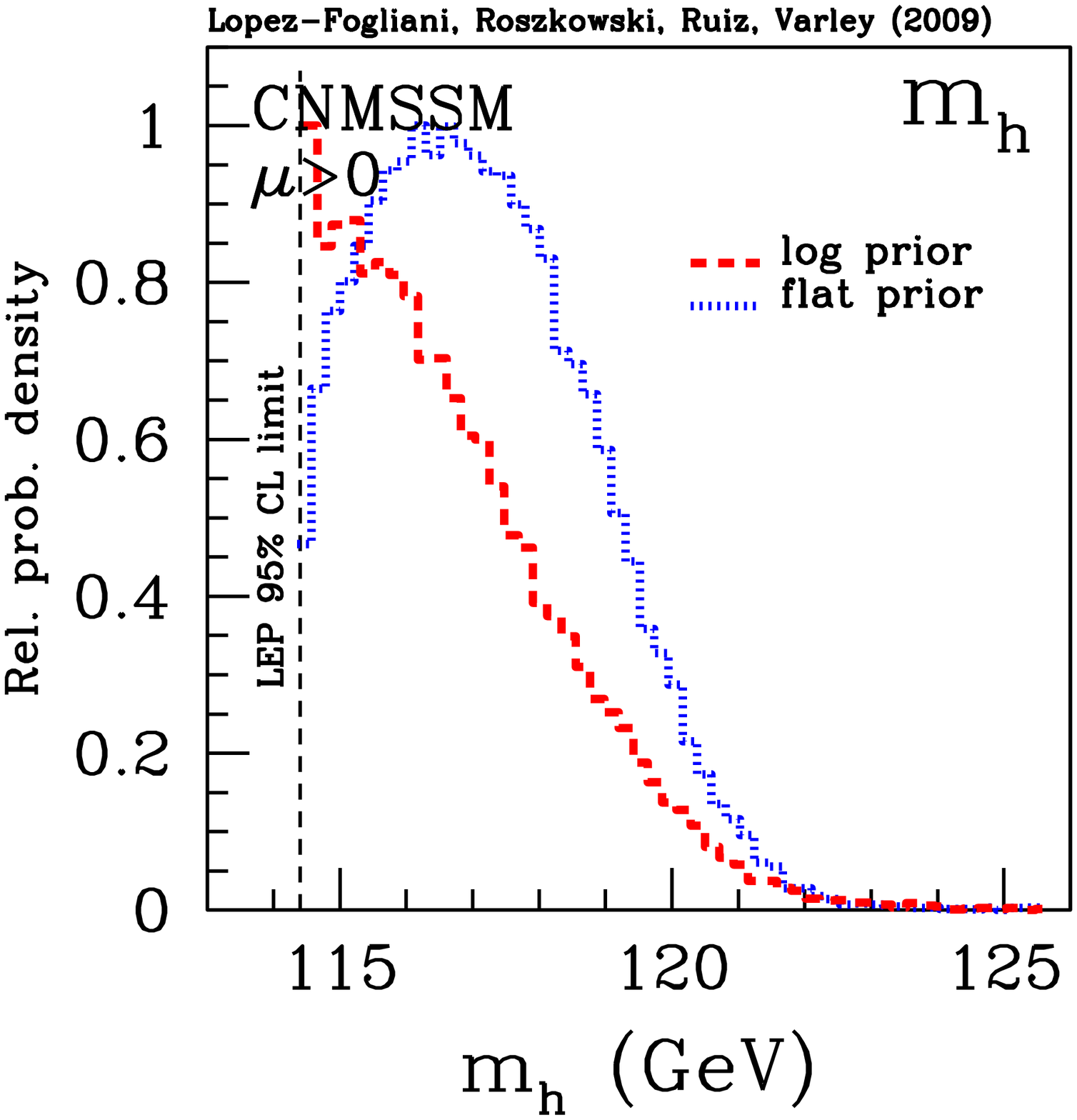}
\end{tabular}
\end{center}
\caption{\label{fig:NMSSM_oh2+bsg+gm2+mhl_1d} The 1D relative
probability densities for $\abundchi$ (upper left panel), $\brbsgamma$
(upper right panel), $\deltagmtwo$ (lower left panel) and the light Higgs
mass $\mhl$ (lower right panel).  In each panel we show the posterior
for the flat prior (dotted blue) and the log prior (long-dashed red)
and the likelihood function (solid black). }
\end{figure}

Similarly to the CMSSM, the key observables shaping the high
probability regions of the model are: $\abundchi$, $\brbsgamma$,
$\deltagmtwo$ and the light Higgs mass $\mhl$. Their 1D pdfs are
presented in Fig.~\ref{fig:NMSSM_oh2+bsg+gm2+mhl_1d}. We can see that,
apart from $\deltagmtwo$, they reproduce the likelihood rather well,
especially for our default log prior (long-dashed red curve), although
even for the flat prior the fit is not bad, except for $\deltagmtwo$, again with much
resemblance to the CMSSM~\cite{tfhrr1}.

The sharp cutoff in $\mhl$ at the LEP limit results from the
approximation that we have adopted, as mentioned above. Given the
complexities of the Higgs sector and a much larger list of possible
decay channels, we have taken an approximate approach of setting the
likelihood to one or zero for points that are accepted or excluded by
LEP data using the NMSPEC code. A more correct would would have
allowed for some ``tail'' at lower masses, as in the
CMSSM~\cite{rrt2}, and would have expanded allowed ranges of $\mhalf$
and $\mzero$ towards smaller values but this would have a limited
effect on this analysis in which we are mostly interested in
presenting mainly global features of the CNMSSM. Nevertheless, it is
worth stressing that in an exploratory run with the LEP limit smeared out
we have found only a few points for which 
the singlet component is barely enough to escape the limit of
$114.4\gev$. In this sense we expect that a full analysis would have
basically reproduced the case of the CMSSM~\cite{rrt2}.

\begin{figure}[tbh!]
\begin{center}
\begin{tabular}{c c c}
  \includegraphics[width=0.33\textwidth]{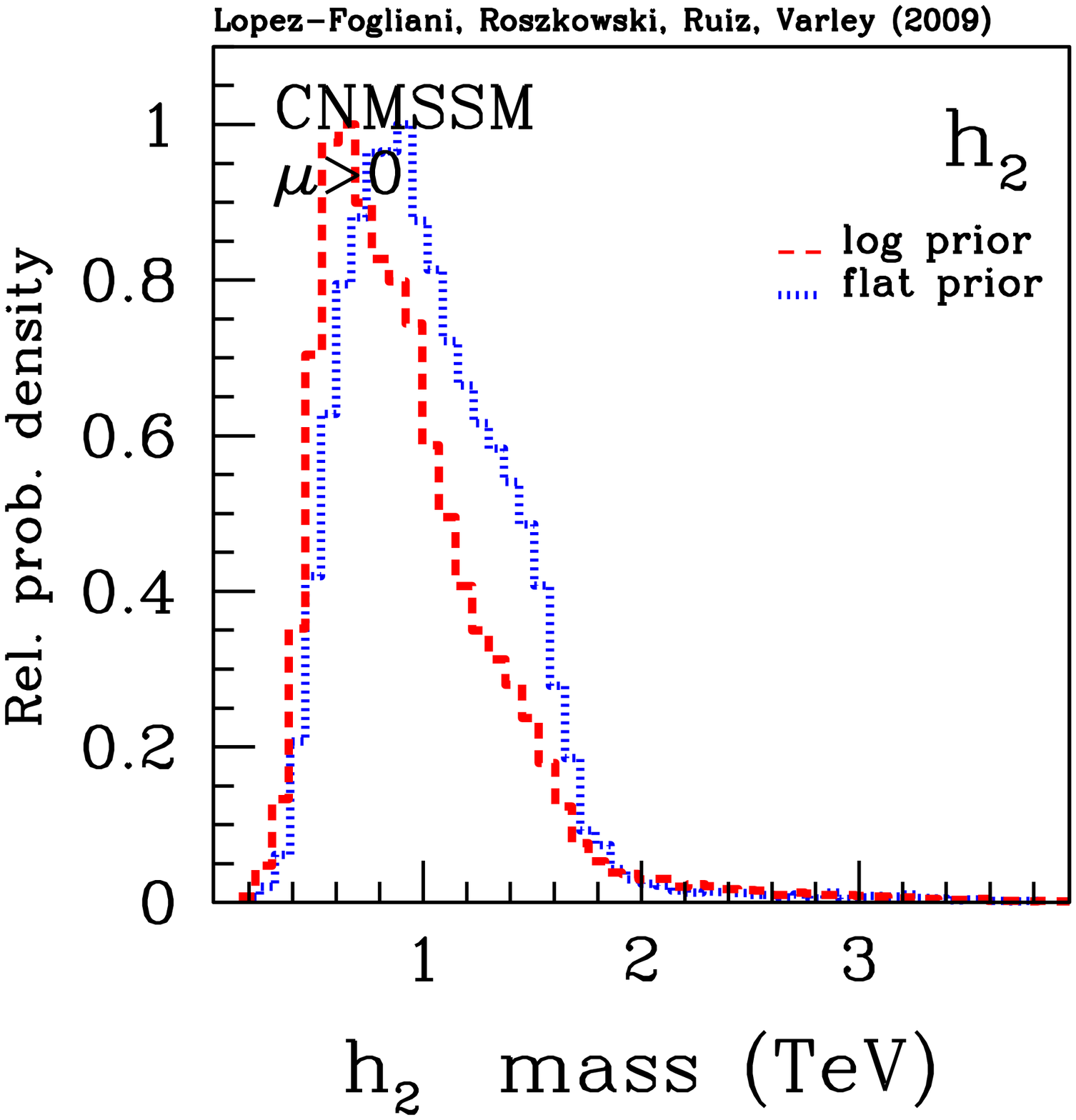}&
      \includegraphics[width=0.33\textwidth]{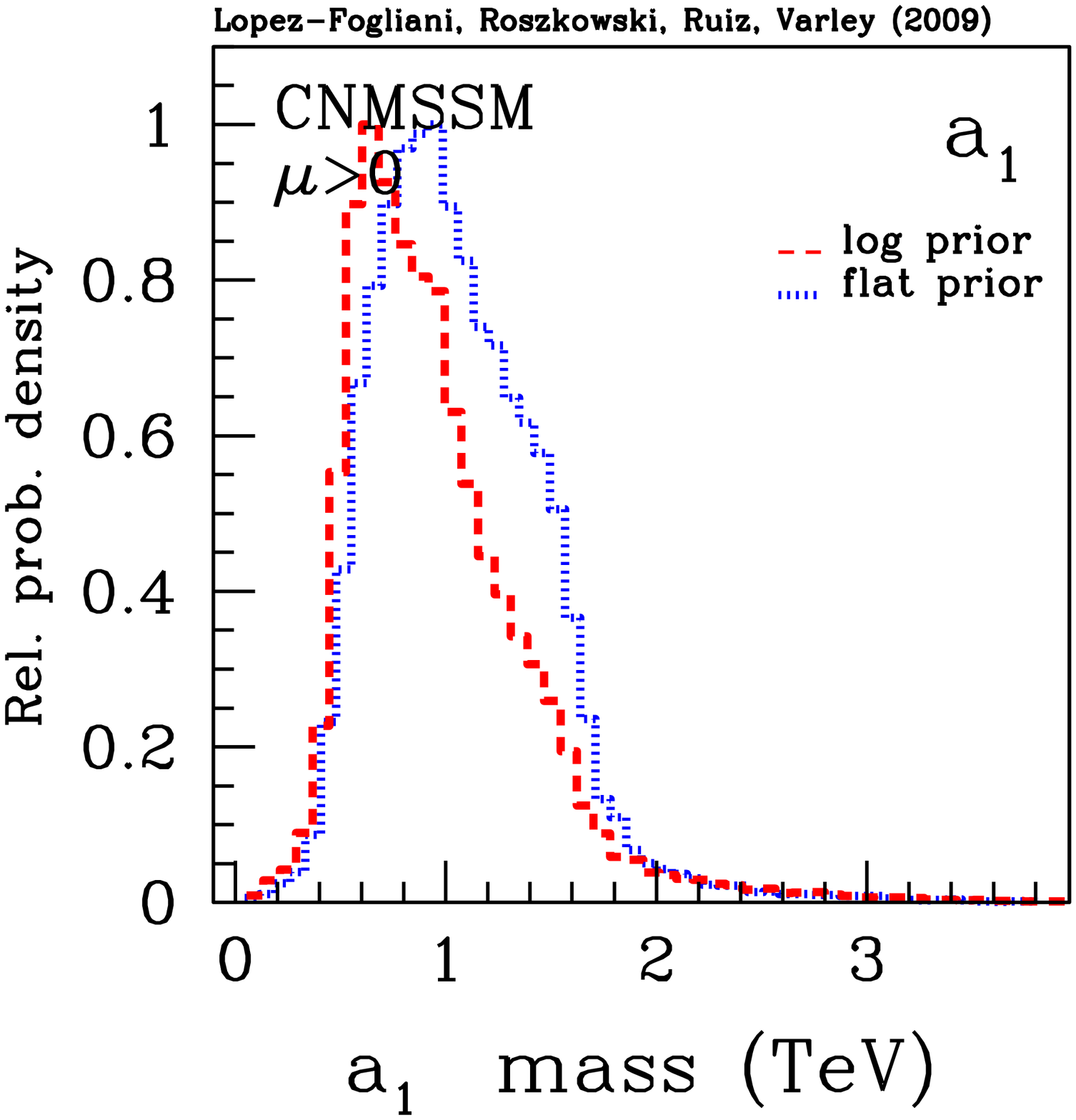}
    & \includegraphics[width=0.33\textwidth]{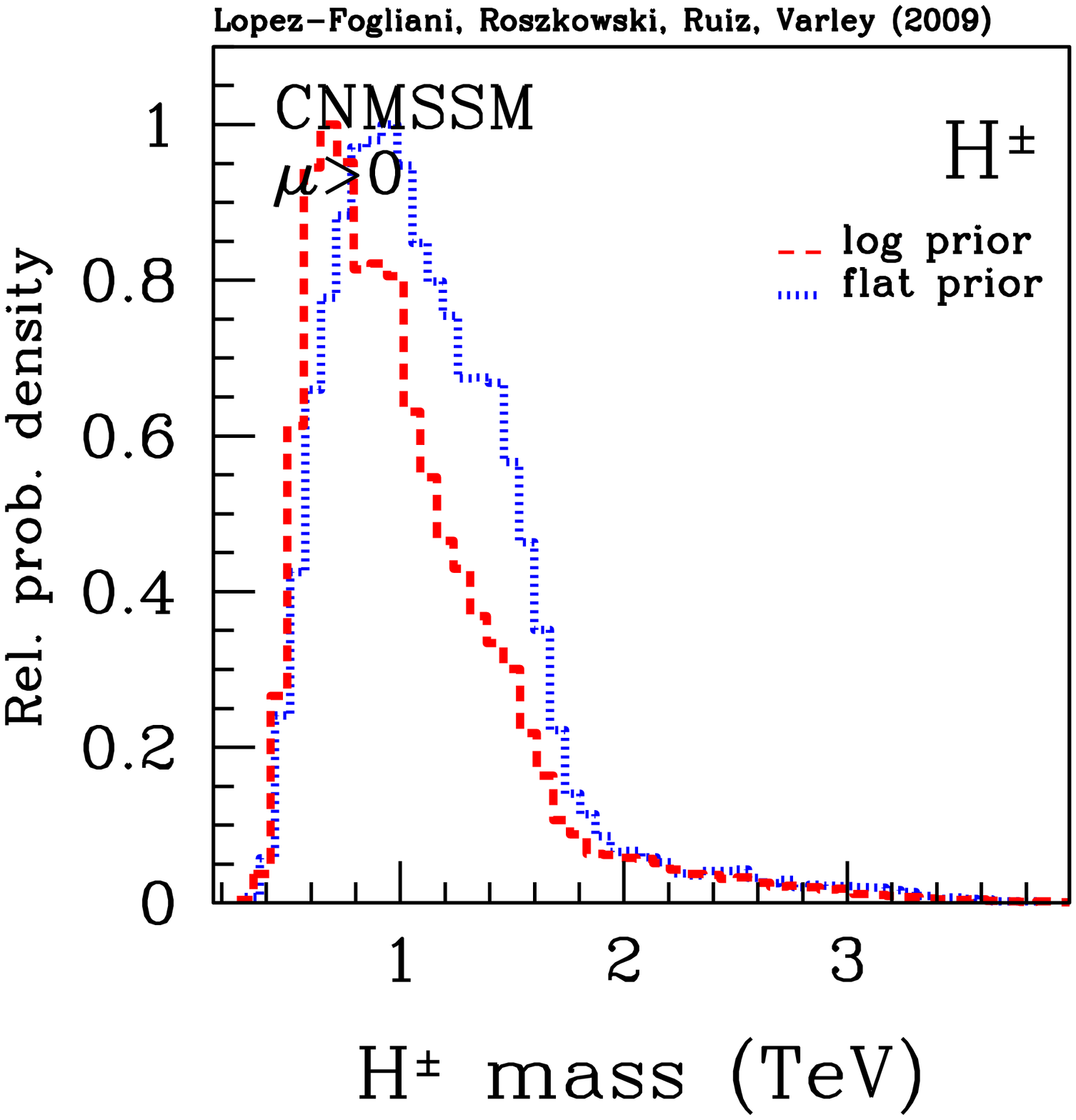}
    \end{tabular}
\end{center}
\caption{\label{fig:CNMSSM_mhiggs_1d_scan1} The relative 1D pdfs of some of the
  Higgs masses: the second to lightest scalar $h_2$ (left panel), the
  lightest pseudoscalar $a_1$ (middle panel) and the charged Higgs
  $H^\pm$ (right panel). In each panel we show
the posterior for the flat prior (dotted blue) and the log prior
(long-dashed red). 
}
\end{figure}

The Higgs sector of the NMSSM contains three CP-even bosons $h_1\equiv
\hl$, $h_2$ and $h_3$, two CP-odd ones $a_1$ and $a_2$, as well as a
pair of the charged Higgs
$H^\pm$. Fig.~\ref{fig:CNMSSM_mhiggs_1d_scan1} shows the relative 1D
pdfs of $h_2$ (left panel), $a_1$ (middle panel) and $H^\pm$ (right
panel) for both our default log prior (long-dashed red) and the flat
prior (dotted blue) for comparison. We can see that the prior
dependence is not very strong, with the flat prior favoring larger
values, as usual. Also, by comparing with Fig.~5 of
Ref.~\cite{rrt2}, we can see that the pdfs are quite similar to the analogous
states in the CMSSM.

\begin{figure}[tbh!]
\begin{center}
\begin{tabular}{c c c}
     \includegraphics[width=0.33\textwidth]{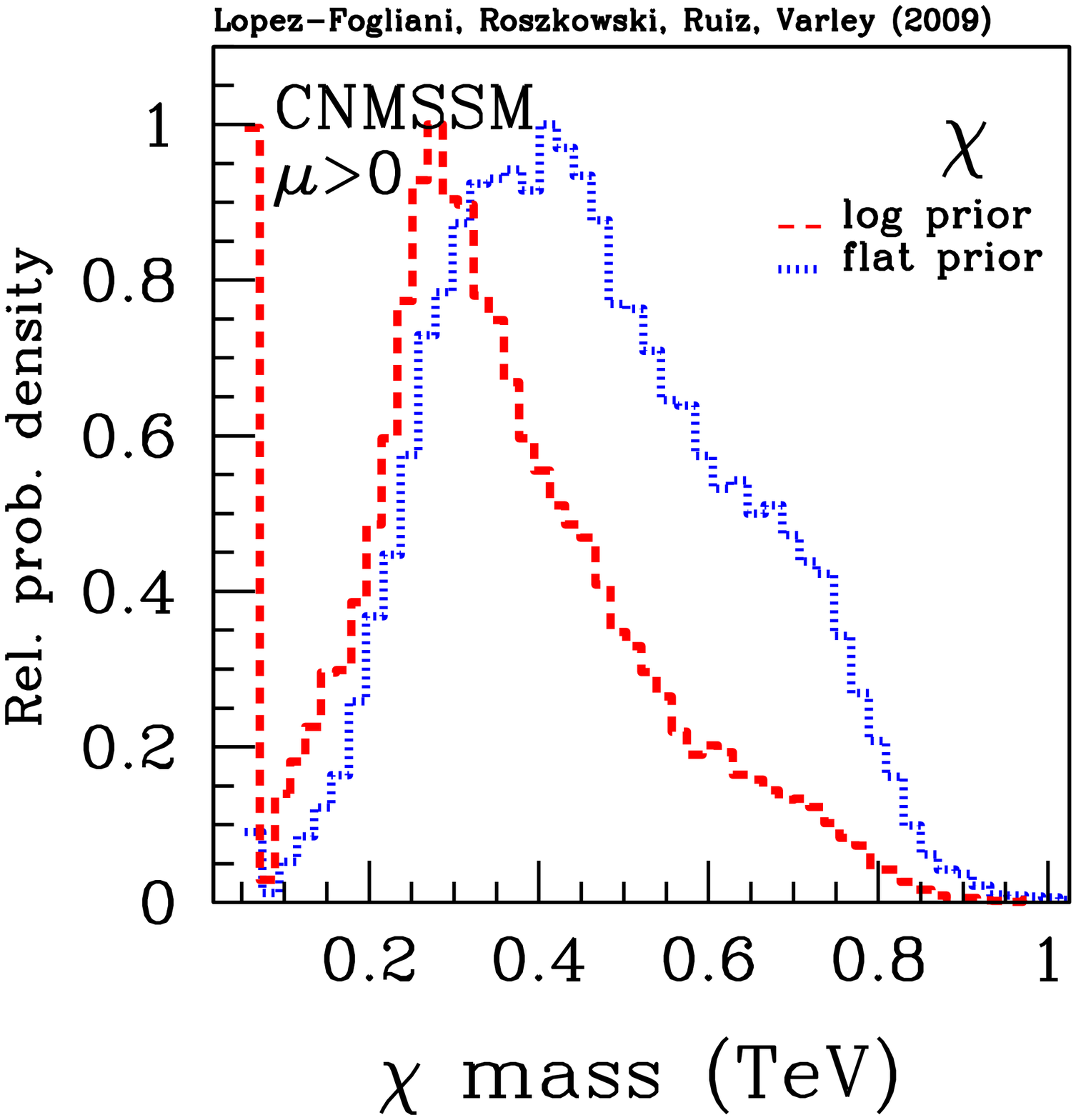}
   & \includegraphics[width=0.33\textwidth]{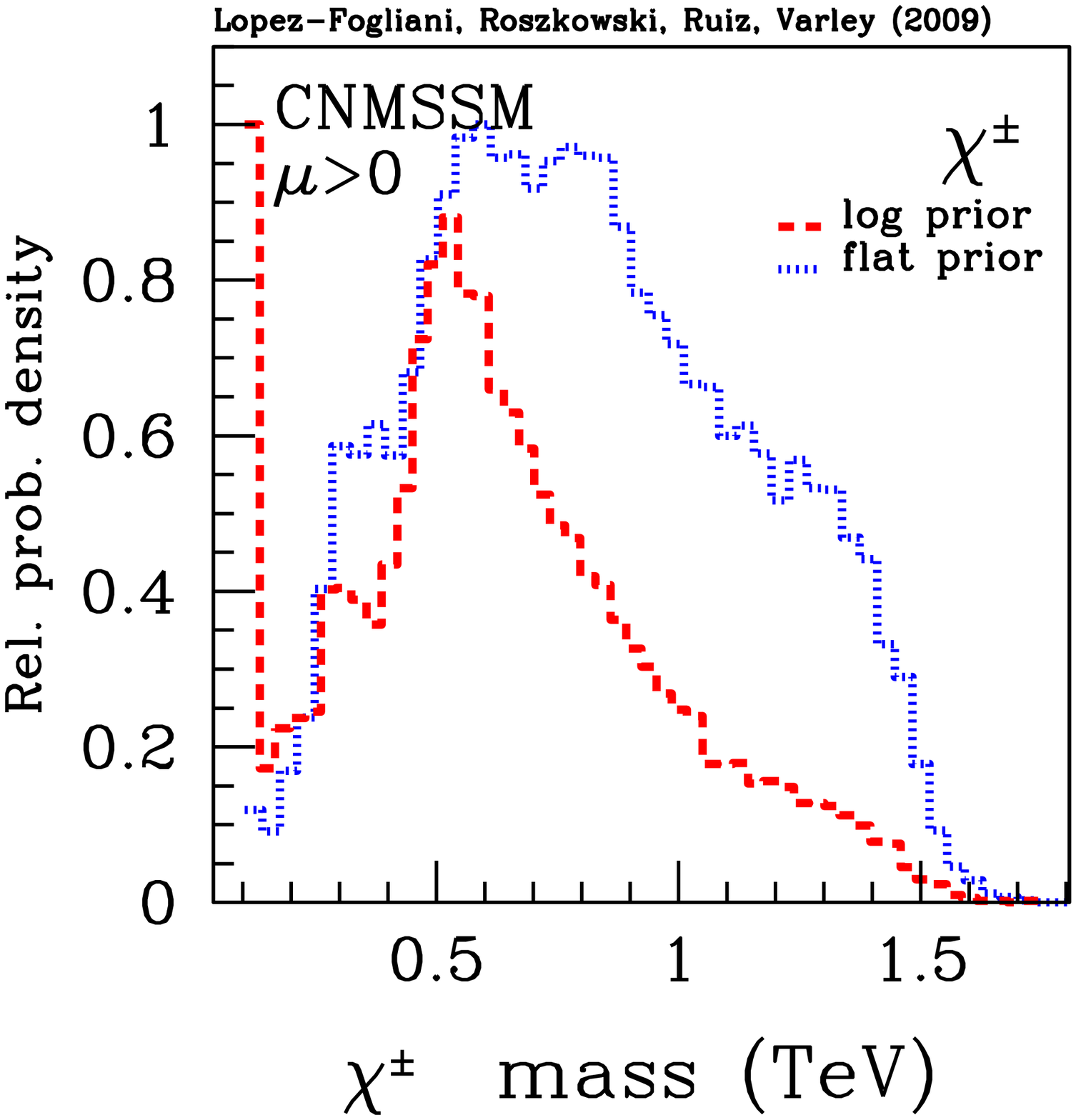}
   & \includegraphics[width=0.33\textwidth]{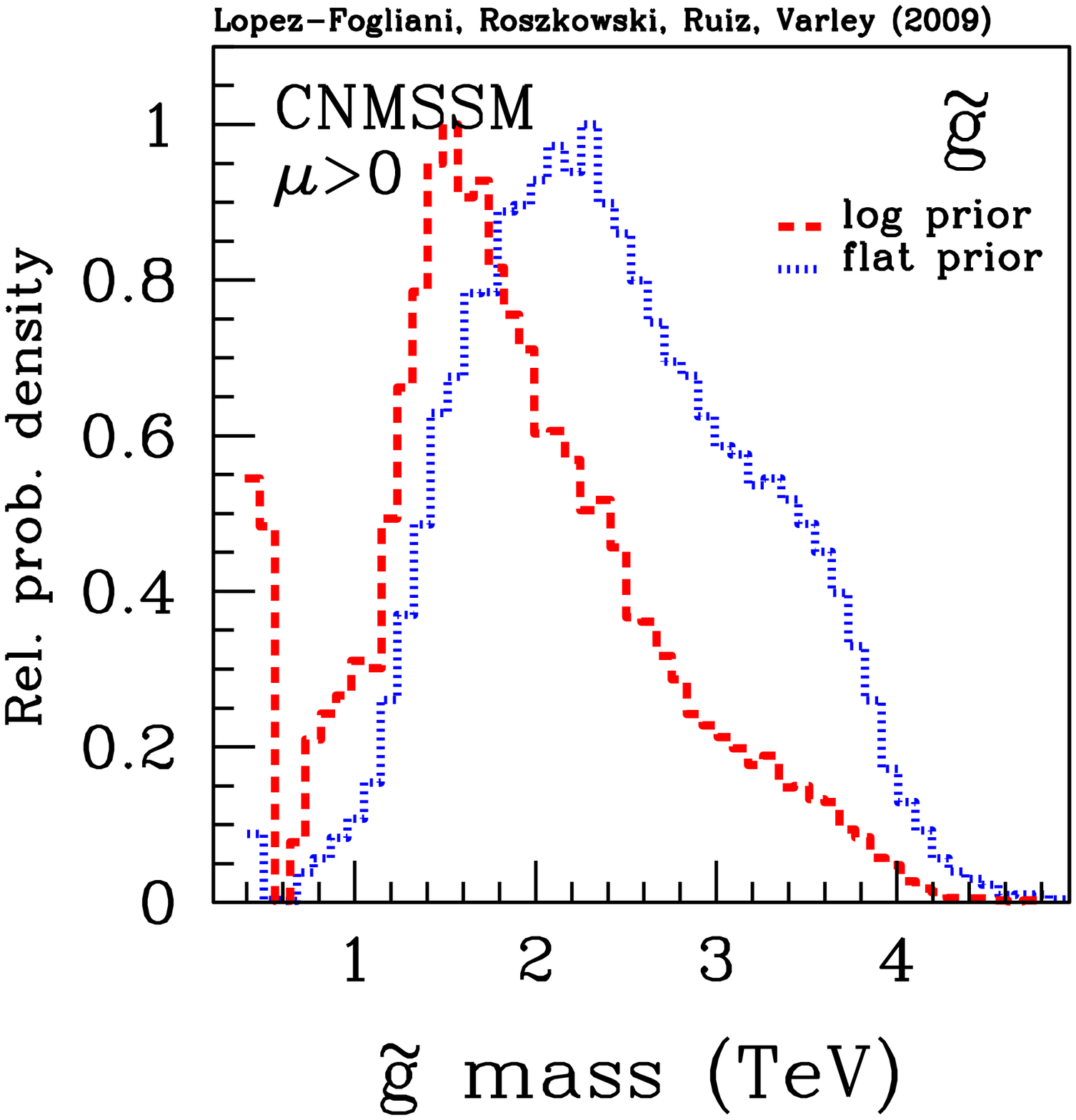}\\
      \includegraphics[width=0.33\textwidth]{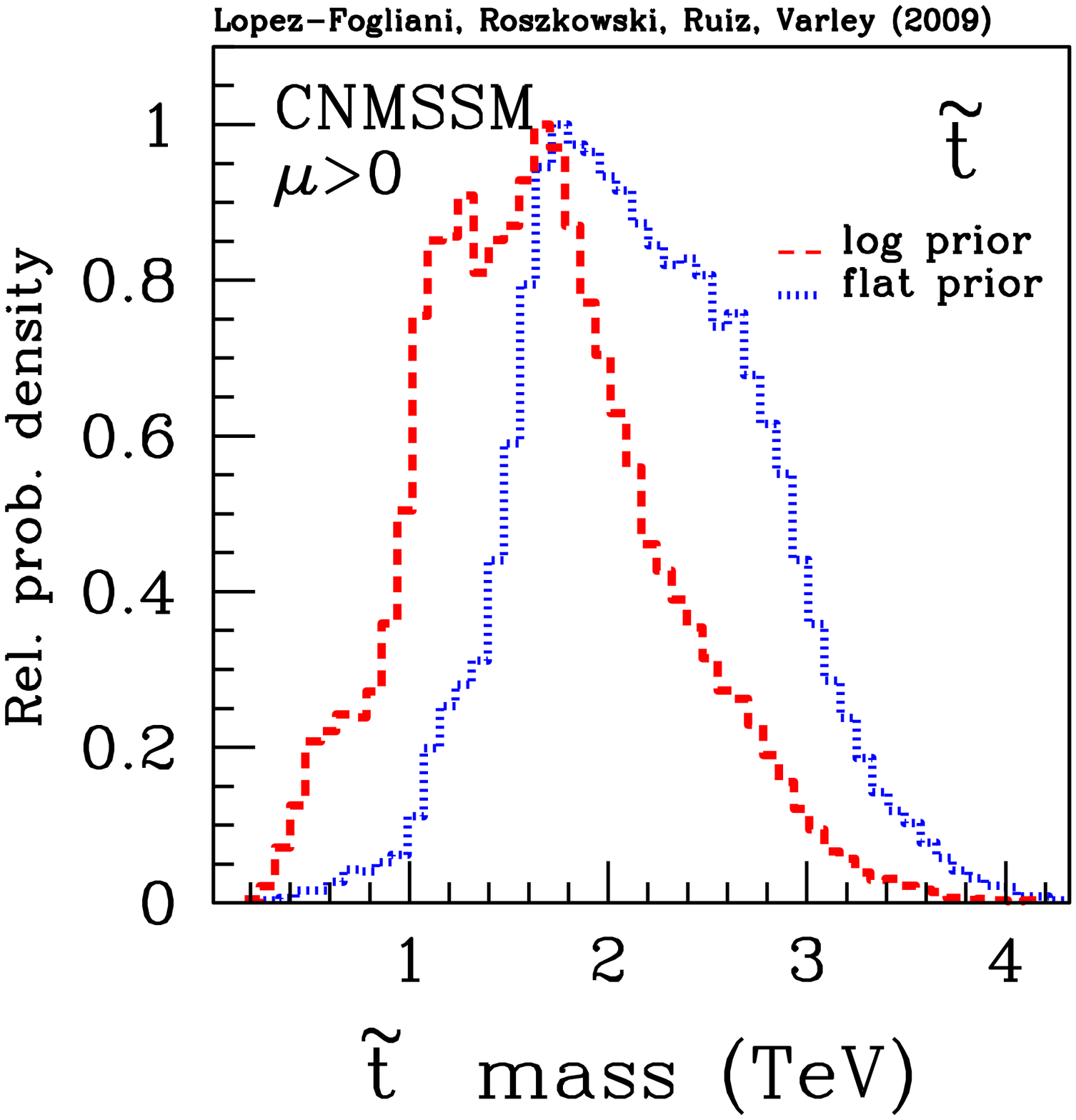}
   & \includegraphics[width=0.33\textwidth]{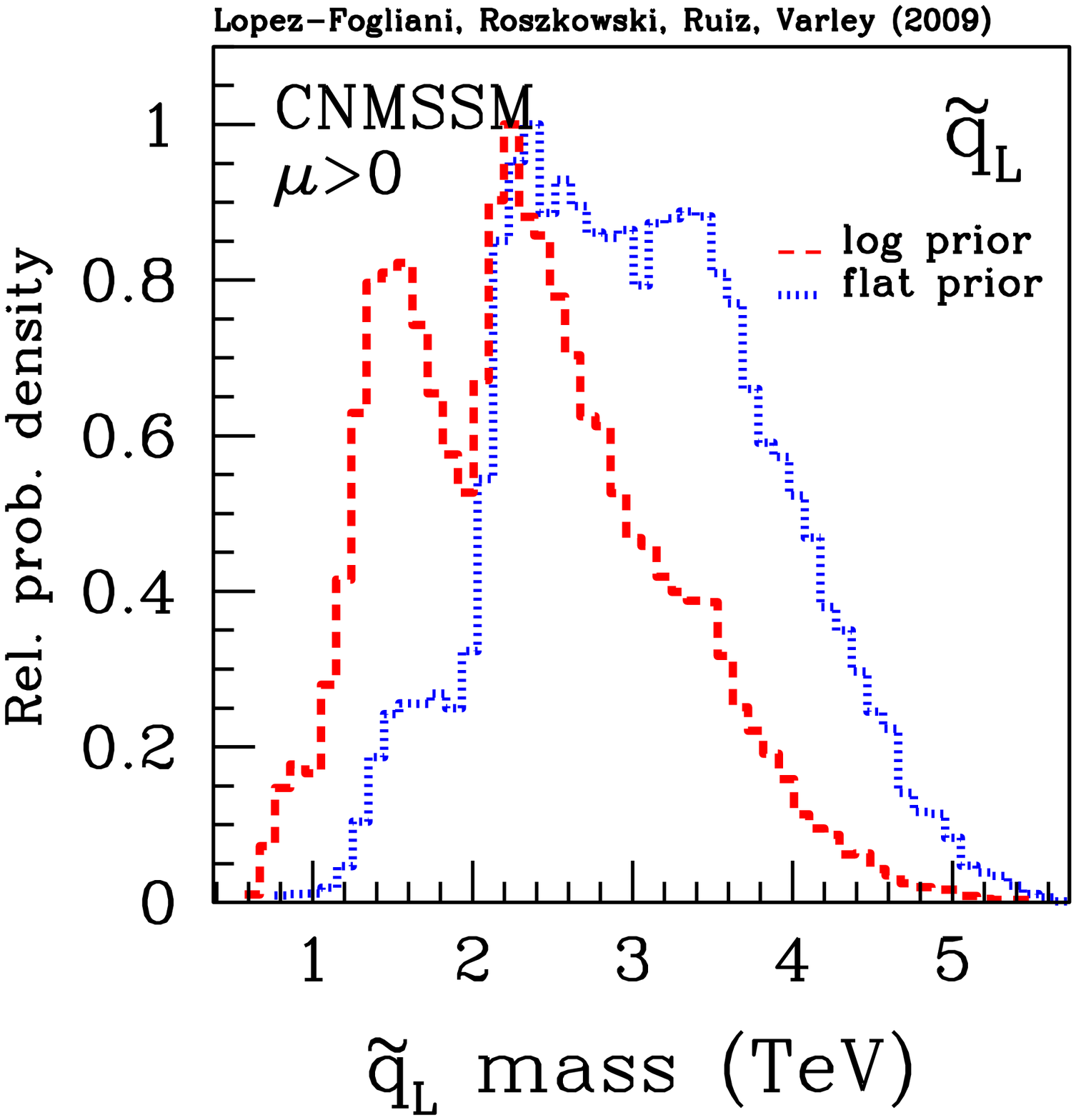}
   & \includegraphics[width=0.33\textwidth]{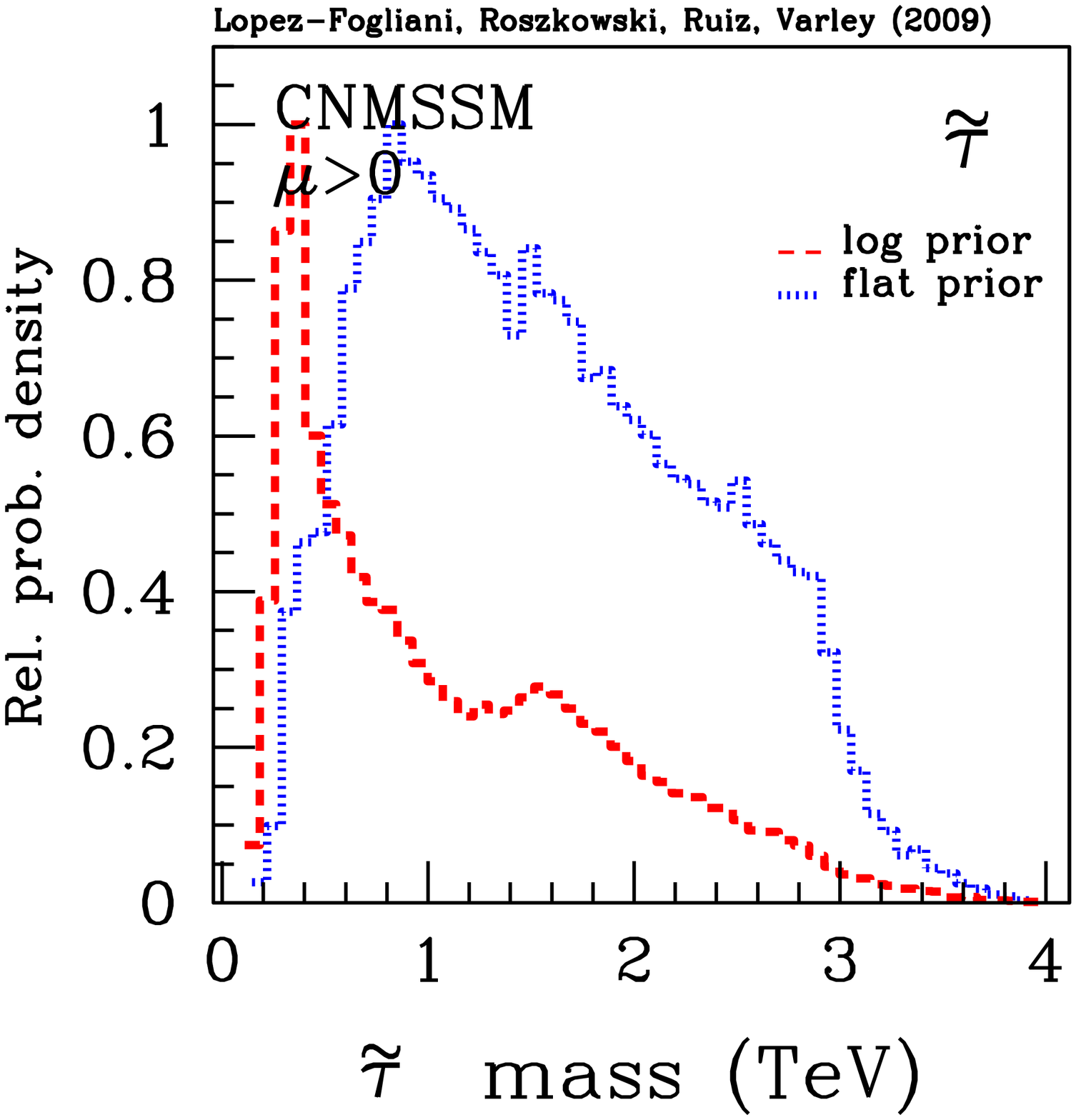}\\
\end{tabular}
\end{center}
\caption{\label{fig:NMSSM_mchi_1d} The 1D relative probability
densities for the mass of the lightest neutralino $\mchi$ (upper left
panel), the lightest chargino $\mcharone$ (upper middle panel), the
gluino $\mgluino$ (upper right panel), the lighter stop $\stopone$
(lower left panel), left squark $\tilde{q}_L$ (lower middle panel) and
the lighter stau $\stauone$ (lower right panel). In each panel we show
the posterior for the flat prior (dotted blue) and the log prior
(long-dashed red).  }
\end{figure}

In Fig.~\ref{fig:NMSSM_mchi_1d} we present the relative 1D pdfs for
several superpartners, in a fashion similar to the previous
Figure. Again, we can see that the log prior gives somewhat lower
ranges of masses, especially for the scalars, which primarily depend
on $\mzero$ and that the distributions are rather similar to the
corresponding ones in the CMSSM; compare, eg, Fig.~17 of
Ref.~\cite{rtr1}.  It is clear that there will be a rather mixed
chance of detecting those states at the LHC. For example, with the
gluino to be probed up to some $2.7\tev$, nearly
the whole range will be tested with the log prior, but much less so
with the flat prior. The scalars, on the other hand, will be much more
challenging for both priors.

In Table~\ref{table:best_fit} we list the best fit values for a number
of observables for our default log prior and also, for comparison, for
the flat prior. We stress, however, that the log prior appears more
appropriate for exploring unified low-energy SUSY models, as already
emphasized above.

  \begin{table}
   \centering
  \begin{tabular}{|l | c| c |}
  \hline
  Parameter & Best fit (log) & Best fit (flat) \\
  \hline
  \hline
  $\mhalf$ &  101\gev&  478\gev           \\
 $\mzero$ &  404\gev&  632\gev           \\
 $\azero$ &  -165\gev& 1.20\tev          \\
 $\tanb$ &  12.9 & 42.4  \\
   $\lambda$ &  0.009 & 0.0252 \\
 \hline
  $\mu$  & 547\gev  & 672\gev  \\
  $m_{a_1}$ & 274\gev  & 476\gev  \\
  \hline
  $\abundchi$ & 0.093 & 0.094    \\
  $\brbsgamma$ &  $3.10\times 10^{-4}$&  $3.27\times 10^{-4}$  \\
  $\brbsmumu$ &  $2.8\times 10^{-9}$&$1.6\times 10^{-8}$    \\
  $\brbtaunu$ &  $1.28\times 10^{-4}$& $0.93 \times 10^{-4}$   \\
  $\deltagmtwo$ & $16.9\times 10^{-10}$&  $14.8\times 10^{-10}$ \\
  $\mhl$ &  114.4\gev & 114.3\gev        \\
  $\mchi$ &164\gev & 263\gev  \\
 $m_{\chi^{\pm}_1}$ & 309\gev & 491\gev          \\
 $\mgluino$ & 950\gev & 1.45\tev  \\
  \hline
  $\chi2$&   9.6965&   9.4635    \\
  \hline
  \end{tabular}
  \caption{\label{table:best_fit} A table showing the values of various
   parameters for the best fitting point in both the log and flat prior
   case.}
  \end{table}

Finally, we move to discussing the model's predictions for the
detection of the lightest neutralino assumed to be the DM in the
Universe in direct detection searches via its elastic scatterings with
targets in underground detectors. We follow the same procedure and
formalism as previously in~\cite{rtr1,rrt3,tfhrr1}.  The underlying
formalism can be found in several sources, \eg, in
Refs.~\cite{dn93scatt:ref,susy-dm-reviews,efo00,knrr1}. 

\begin{figure}[tbh!]
\begin{center}
\begin{tabular}{c c}
   \includegraphics[width=0.35\textwidth]{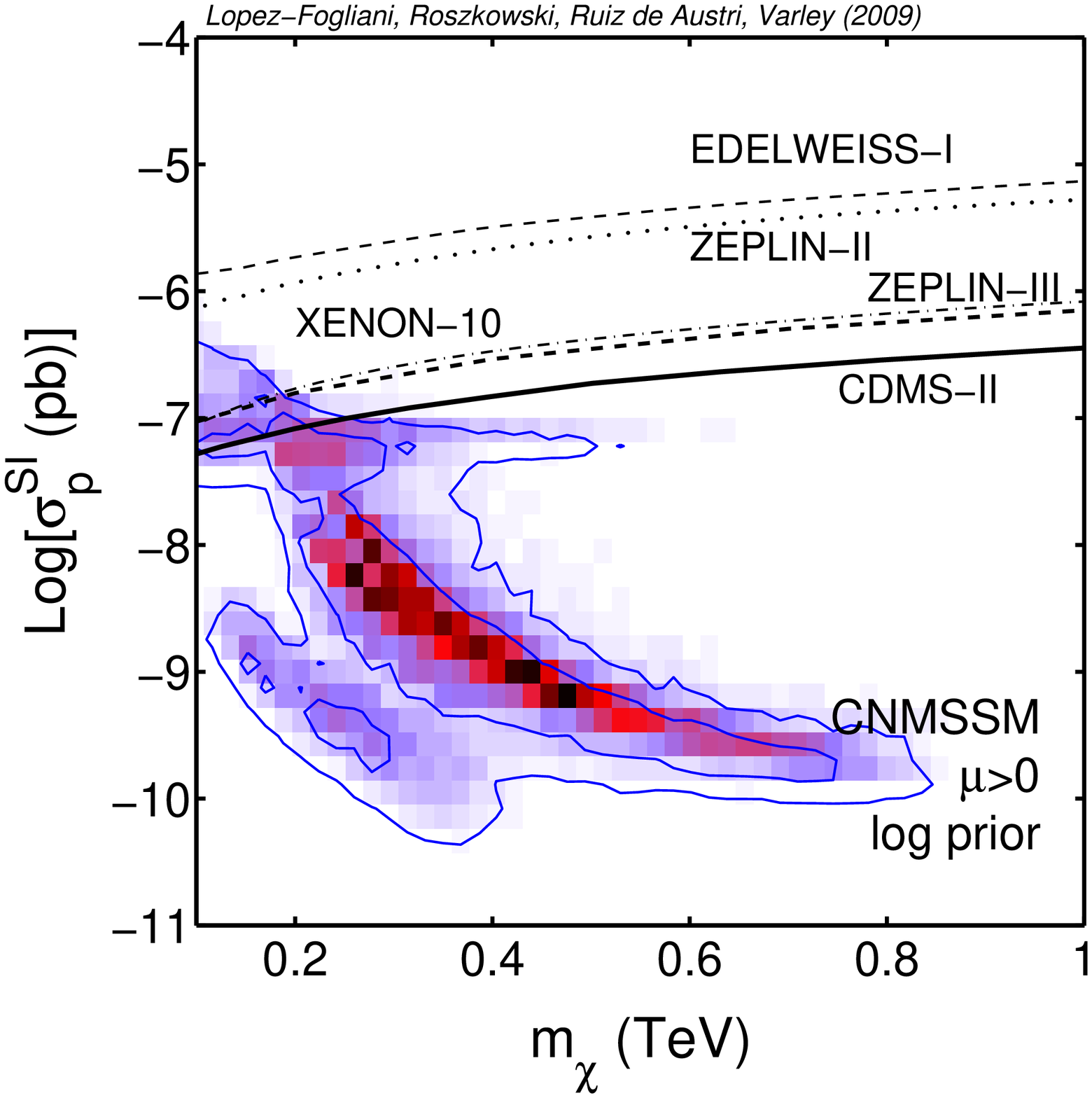}
  & \includegraphics[width=0.35\textwidth]{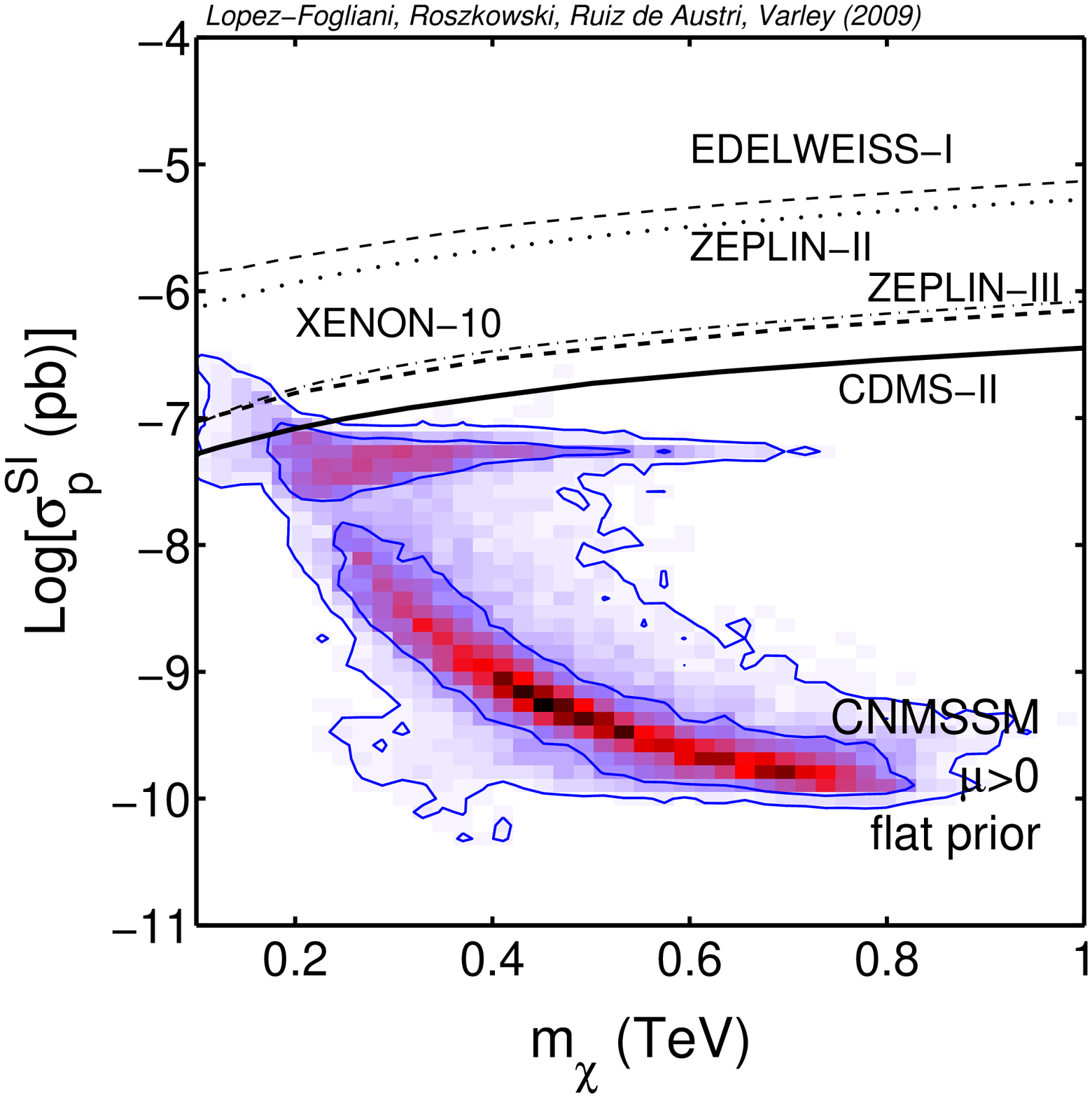}
\end{tabular}
  \includegraphics[width=0.3\textwidth]{figures/colorbar.ps}
\end{center}
\caption{\label{fig:sigsip_1d} For the dark matter spin-independent
  cross section $\sigsip$ \vs\ the neutralino mass $\mchi$ we show the
  2D relative probability density for the log prior
  (left panel) and the flat prior (right panel).   
}
\end{figure}

In Fig.~\ref{fig:sigsip_1d} we present the 2D posterior pdfs in the
usual plane spanned by the spin-independent cross section $\sigsip$
and the neutralino mass $\mchi$. The left (right) panel corresponds to
the log (flat) prior.  For comparison, some of the most stringent
90\%~CL experimental upper limits are also marked~\cite{cdms-sep05,
edelweiss-one-final, zeplin-one-final, zeplin2, zeplin3, xenon-10},
although they have not been imposed in the likelihood, as before in
our studies of the CMSSM, because of substantial astrophysical
uncertainties, especially in the figure for the local density.

Several key features can be seen in Fig.~\ref{fig:sigsip_1d}. Firstly,
the prior dependence is not very strong for both 68\% and 95\% total
probability regions, which is encouraging. It does
not affect much the banana-shape high-probability region which
corresponds to the Higgs funnel and the stau coannihilation
regions. The horizontal branch of $\sigsip\simeq 7\times 10^{-8}\pb$
is more affected because it corresponds to the focus point region of
large $\mzero$. Next, the overall shape rather closely resembles the
case of the CMSSM, see, \eg, Fig.~18 of Ref.~\cite{tfhrr1} or  Fig.~13
of Ref.~\cite{rtr1}. (The slight upwards shift in $\sigsip$ results
from changing the code from DarkSusy to Micromegas.) It does, on the
other hand, differ from the predictions of the NUHM which features an
additional higgsino-like region at $\mchi\sim 1\tev$; see Fig.~12 of Ref.~\cite{nuhm1}.

Independently of the prior, basically the whole 68\% and 95\% total
probability regions are likely to be within the planned reach of
$10^{-10}\pm$ of future 1-tonne detectors. Some of the currently
operating detectors are already probing some of the high probability
regions, and with a ``modest'' improvement down to $\sim 10^{-8}\pb$,
they will be testing some of the most likely cross sections. 

\section{Conclusions and summary}\label{sec:summary}

The Next-to-Minimal Supersymmetric Standard Model solves the
$\mu$-problem of the MSSM but, without grand unification,  both
models suffer from a large number of parameters. The constrained
versions of both models are in this respect much more
well-motivated. Because of the additional singlet superfield present
in the CNMSSM, the resulting phenomenology in the Higgs and neutralino
sectors is considerably richer. Therefore, a prior one could expect
that the models could be distinguished in experimental tests. 

The global exploration of wide ranges of CNMSSM parameters and a
Bayesian analysis show that, from the statistical point of view, this
is not the case. The coupling $\lambda$ strongly favors as small
values as possible, in other words it tends towards the decoupling
regime in which one recovers the CMSSM plus the basically decoupled
singlet Higgs and the singlino. As a result, Higgs and superpartner
mass spectra also tend to resemble those of the CMSSM, as does the
cross section for direct detection of neutralino dark matter.
Nevertheless, we have identified a limited number of cases where the
LSP is indeed singlino-dominated, but statistically they are not very
significant. 

In conclusion, the CNMSSM is, for the most part, testable at the LHC
and in dark matter searches, which is certainly encouraging. On the
other hand, should a CMSSM-like signal be detected, it is likely to be very
challenging to distinguish between the two models.


\medskip
{\bf Acknowledgments} \\ LR is partially supported by the EC 6th
Framework Programmes MRTN-CT-2004-503369 and MRTN-CT-2006-035505.
RRdA is supported by the project PARSIFAL (FPA2007-60323) of the
Ministerio de Educaci\'{o}n y Ciencia of Spain. DL-F and TV are supported by
STFC.  The use of the Iceberg computer cluster at the
University of Sheffield is gratefully acknowledged. LR would like to
thank the CERN Theory Division for hospitality during the final stages
of the project.


\end{document}